\documentclass[12pt,a4paper]{article}
\usepackage{latexsym,graphicx,psfrag,amsmath}

\newcommand{\im}{\mbox{Im}\,}
\newcommand{\re}{\mbox{Re}\,}

\newcommand{\arcsinh}{\mbox{arcsinh}}
\newcommand{\arccosh}{\mbox{arccosh}}
\newcommand{\be}{\begin{multline}}
\newcommand{\bdm}{\begin{displaymath}}
\newcommand{\edm}{\end{displaymath}}
\newcommand{\bea}{\begin{eqnarray}}
\newcommand{\eea}{\end{eqnarray}}
\newcommand{\no}{\nonumber \\}

\newcommand{\fs}{\; .}
\newcommand{\co}{\; ,}
\newcommand{\al}{\!&\!}
\newcommand{\eff}{{e\hspace{-0.1em}f\hspace{-0.18em}f}}

\newcommand{\Mbare}{m}
\newcommand{\Mp}{M}
\newcommand{\gAbare}{g}

\newcommand{\Dpi}{\Delta_\pi}
\newcommand{\DN}{\Delta_{\ind N}}

\newcommand{\GSS}{\mbox{\tiny\it GSS}}
\newcommand{\U}{u}
\newcommand{\Ubar}{\bar{u}}
\newcommand{\gpiN}{g_{\pi\hspace{-0.1em}{\scriptscriptstyle N}}}
\newcommand{\mD}{m_{\ind \Delta}}
\newcommand{\gD}{g_{\ind \Delta}}
\def\circa#1{$\!\bigcirc$\hspace{-0.7em}\raisebox{0.05em}
{\scriptsize #1}\hspace{0.2em}}

\newcommand{\cspace}{\hspace{3cm}}
\newcommand{\dspace}{\hspace{1cm}}

\newcommand{\ind}{\scriptscriptstyle}

\newcommand{\M}{M_\pi}
\newcommand{\m}{m_{\hspace{-0.1em}\ind N}}
\newcommand{\F}{F_\pi}
\newcommand{\gA}{g_{\scriptscriptstyle A}}
\newcommand{\Omq}{\Omega_q}

\newcommand{\SL}[1]{#1 \!\!\!\! /}
\newcommand{\Sl}[1]{#1 \!\!\! /}

\author{T.~Becher and H.~Leutwyler}
\date{23.~3.~01}

\begin{document}
\setcounter{equation}{0}
\setcounter{figure}{0}
\begin{flushright}{\begin{tabular}{l}
CLNS 01/1727\\ BUTP--01/11\rule{1em}{0em} \end{tabular} }\end{flushright}

\vspace{0.5cm}
\begin{center}
\renewcommand{\thefootnote}{\fnsymbol{footnote}}
\LARGE{\bf Low energy analysis of \boldmath$\pi N\rightarrow \pi
  N$} \\
\vspace{0.8cm}
{\large T.~Becher\hspace{1pt}$\phantom{}^a$ and H.~Leutwyler\hspace{1pt}$\phantom{}^b$}\\
\vspace{0.4cm} 
\normalsize $\phantom{}^a$ Newman Laboratory of Nuclear Studies, Cornell University,\\
 Ithaca, NY 14853, USA\\ \vspace{0.2cm}
$\phantom{}^b$ Institute for Theoretical Physics, University of Bern,\\
Sidlerstr.~5, CH-3012 Bern, Switzerland
\end{center}
\vspace{0.2cm}

\begin{abstract}
  We derive a representation for the pion nucleon scattering amplitude
  that is valid to the fourth order of the chiral expansion.  To
  obtain the correct analytic structure of the singularities in the
  low energy region, we have performed the calculation in a
  relativistic framework (infrared regularization). The result can be
  written in terms of functions of a single variable. We study the
  corresponding dispersion relations and discuss the problems
  encountered in the straightforward nonrelativistic expansion of the
  infrared singularities. As an application, we evaluate the
  corrections to the Goldberger-Treiman relation and to the low energy
  theorem that relates the value of the amplitude at the Cheng-Dashen
  point to the $\sigma$-term. While chiral symmetry does govern the behaviour
  of the amplitude in the vicinity of this point, the representation
  for the scattering amplitude is not accurate enough to use it for an
  extrapolation of the experimental data to the
  subthreshold region. We propose to perform this extrapolation on the
  basis of a set of integral equations that interrelate the lowest
  partial waves and are analogous to the Roy equations for $\pi\pi$
  scattering.
\end{abstract}
\thispagestyle{empty}

\newpage
\tableofcontents
\newpage
\renewcommand{\thefootnote}{\arabic{footnote}}
\section{Introduction}

\begin{center}
\it Here, perhaps an accuracy of 5 per cent is more appropriate.
\end{center}
\begin{flushright}
  \small Henley and Thirring (\cite{Henley Thirring}, 1962)  on the
   static model.
\end{flushright}

The description of the pion nucleon interaction in terms of effective
fields has quite a long history. The static model represents a
forerunner of the effective theories used today. In this model, the
kinetic energy of the nucleon is neglected: The nucleon is described
as a fixed source that only carries spin and isospin degrees of
freedom.  For an excellent review of the model and its application to
several processes of interest, we refer to the book of Henley and
Thirring \cite{Henley Thirring}.

The systematic formulation of the effective theory relies on an expansion
of the effective Lagrangian in powers of derivatives and quark masses.
Chiral symmetry implies that the leading term of this expansion is fully
determined by the pion decay constant, $F_\pi$, and by the nucleon matrix
element of the axial charge, $\gA$. Disregarding vertices with three or
more pions, the explicit expression for the leading term reads
\bea {\cal L}_{\eff}= -\frac{\gA}{2 F_\pi}\bar{\psi}\,\gamma^\mu\gamma_5
\,\partial_\mu
\pi\psi +\frac{1}{8 F_\pi^2}\bar{\psi}\,\gamma^\mu i[\pi,\partial_\mu\pi]\psi
+\ldots\nonumber\eea
The success of the static model derives from the fact that it properly 
accounts for the first term on the right hand side -- in the
nonrelativistic limit, where the momentum of the nucleons is neglected
compared to the nucleon mass.

The static model is only a model. In order for the effective theory to
correctly describe the properties of QCD at low energies, that
framework must be extended, accounting for the second term in the
above expression for the effective Lagrangian, for the vertices that
contain three or more pion fields, for the contributions arising at
higher orders of the derivative expansion, as well as for the chiral
symmetry breaking terms generated by the quark masses $m_u$, $m_d$.  A
first step in this direction was taken by Gasser, Sainio and \v{S}varc
\cite{gss}, who formulated the effective theory in a manifestly
Lorentz invariant manner. In that framework, chiral power counting
poses a problem: The loop graphs in general start contributing at the
same order as the corresponding tree level diagrams, so that the loop
contributions in general also renormalize the lower order couplings. A
method that does preserve chiral power counting at the expense of
manifest Lorentz invariance was proposed by Jenkins and Manohar
\cite{man}, who used a nonrelativistic expansion for the nucleon
kinematics. The resulting framework is called ``Heavy Baryon Chiral
Perturbation Theory'' (HBCHPT). It represents an extension of the
static model that correctly accounts for nucleon recoil, order by
order in the nonrelativistic expansion (for reviews of this approach,
see for instance refs.~\cite{EckerReview,mei2}). In
\cite{moj,Fettes:1998ud}, this formalism has been used to obtain an
$O(q^3)$ representation for the pion nucleon scattering amplitude. In
the meantime, the HBCHPT calculation has been extended to $O(q^4)$
\cite{Fettes:2000xg}.
  
As pointed out in ref.~\cite{Becher Leutwyler 1999}, the
nonrelativistic expansion of the infrared singularities generated by pion
exchange is a subtle matter. The HBCHPT representations of the scattering
amplitude or of the scalar nucleon form factor, for example, diverge in the
vicinity of the point $t=4M_\pi^2$.  The problem does not arise in the Lorentz
invariant approach proposed earlier \cite{gss}.  It originates in the fact
that for some of the graphs, the loop integration cannot be interchanged with
the nonrelativistic expansion.

The reformulation of the effective theory given in ref.~\cite{Becher
Leutwyler 1999} exploits the fact that the infrared singular part of
the one loop integrals can unambiguously be separated from the
remainder. To any finite order of the nonrelativistic expansion, the
regular part represents a polynomial in the momenta. Moreover, the
singular and regular pieces separately obey the Ward identities of
chiral symmetry.  This ensures that a suitable renormalization of the
effective coupling constants removes the regular part altogether. The
resul\-ting representation for the various quantities of interest
combines the virtues of the heavy baryon approach with those of the
relativistic formulation of ref.~\cite{gss}: The perturbation series
can be ordered with the standard chiral power counting and manifest
Lorentz invariance is preserved at every stage of the calculation. The
method has recently been extended to the multi-nucleon sector
\cite{GLPS}.

Previously, this framework has been applied in calculations of the
scalar, axial and electromagnetic form factors \cite{Becher Leutwyler
1999,Schweizer,Kubis}. In the present paper, we
use it to derive a representation for the pion nucleon scattering
amplitude which is valid to the fourth order of the chiral
expansion. After a comparison of infrared regularization with standard
dimensional regularization, we discuss some technical issues in the
evaluation of the Feynman diagrams. In sec.~\ref{nucleon mass and
coupling constant} we give the chiral expansion of the masses and of
the pion nucleon coupling constant $g_{\pi N}$.  By comparing the
quark mass dependence of the pion and the nucleon mass, we illustrate
that the infrared singularities encountered in the baryon sector are
considerably stronger than in the Goldstone sector.

In secs.~\ref{Goldberger-Treiman}--\ref{Chiral symmetry} we discuss the
constraints that chiral symmetry imposes on the scattering amplitude.
Using the result for the axial coupling constant $g_A$ obtained by
Kambor and Moj\v zi\v s \cite{Kambor:1998pi}, we show that the
Goldberger-Treiman relation is free of infrared singularities up to
and including $O(q^3)$.  Our calculation confirms a result obtained in
\cite{BKM}: The difference between the $\Sigma$-Term and the scalar
form factor at $t=2\M^2$ does not involve a chiral logarithm at order
$q^4$ -- the chiral expansion of the difference $\Sigma-\sigma(2\M^2)$
starts with a term proportional to the square of the quark masses.

In section \ref{Cuts} we discuss the analytic structure of the amplitude. There
are three categories of branch cuts in the one loop amplitude: cuts in the
variables 
$s$ and $u$ due to $\pi N$-intermediate states and cuts arising from $\pi
\pi$- and $\bar{N} N$-intermediate states in the $t$-channel. The box graph is
the only contribution to the amplitude which contains a simultaneous cut in
two kinematic variables. Since the $\bar{N} N$ cut only starts at 
$t=4\,\m^2$, far outside the
low energy region,  its contribution to the amplitude is well
represented by a polynomial. We make use of this fact to simplify the
$t$-dependence of the box graph. Dropping higher order terms suppressed by
$t/4\,\m^2$, the amplitude is given by the Born-term, a crossing symmetric
polynomial of order $q^4$ and 9 functions of a {\em single} variable, either
$s$, $t$ or $u$. These functions are given by dispersion integrals over
the imaginary parts of the one loop graphs. 

The representation in terms of dispersion integrals contains terms of
arbitrarily high order in the chiral expansion.
The expansion of the integrals can be carried out explicitly. Truncating the
series at $O(q^4)$, we obtain a simple, explicit representation of the
scattering amplitude in terms of elementary functions. The result can be
compared directly with the representation obtained in HBCHPT, where the chiral
expansion of the loop integrals is performed ab initio. In contrast to that
approach, our method allows us to examine the convergence of the chiral
expansion of the loop integrals. The matter is discussed in
detail in section \ref{Algebraic representation}, where we point out that the
infrared singularities contained in these integrals give rise to several
problems. In particular, we show that the kinematic variables
must be chosen carefully for the truncated expansion to represent a
decent approximation. This leads to considerable complications in the
analysis. Our dispersive representation of the
scattering amplitude is comparatively simple and is perfectly adequate for
numerical analysis.

The representation of the scattering amplitude to $O(q^4)$ neither
includes the cuts due to the exchange of more than two stable
particles nor the poles on unphysical sheets generated by
resonances. It accounts for these effects only summarily, through their
contribution to the effective coupling constants.  Since the
$\Delta$-resonance lies close to the physical threshold it generates
significant curvature in that region. In section \ref{Higher orders}
we discuss the role of the higher order contributions both
in the real and the imaginary part of the amplitude.

Using the example of the $S$-wave scattering lengths, we illustrate that
the one loop representation of chiral perturbation theory does not
cover a sufficiently large kinematic range to serve as a bridge
between the experimentally accessible region and the Cheng-Dashen
point. The problem arises because the one loop amplitude fulfills the
unitarity condition only up to higher orders in the chiral expansion
and the perturbative series for the amplitude goes out of control
immediately above threshold. In sec.~\ref{Total cross section} we
discuss these unitarity violations on the basis of the optical
theorem.

We conclude that a reliable extrapolation from the physical
region to the Cheng-Dashen point can only be obtained by means of
dispersive methods and propose a set of integral equations for the lowest
partial waves, similar to the Roy equations \cite{Roy} for
$\pi\pi$-scattering. Based on partial wave relations
\cite{Steiner,Koch:1986bn}, they allow for a determination of the $S$- and 
$P$-waves in the elastic region from their imaginary part at higher
energies. The structure of the amplitude that underlies these
equations matches the representation obtained in chiral perturbation
theory, but in contrast to that representation, the lowest partial waves
strictly obey the constraints imposed by  
unitarity. These equations
allow us to extend the region where the one loop approximation of chiral
perturbation theory provides a reliable description of the scattering
amplitude. In the case of $\pi\pi$ scattering, the method has been shown to
yield a remarkably accurate representation of the amplitude throughout the
elastic region \cite{ACGL,CGL}. We expect that the application of this method
to $\pi N$ scattering will lead to a reliable
determination of the pion nucleon $\sigma$-term.

\setcounter{equation}{0}
\section{Kinematics}
\begin{figure}[ht]\centering
\psfrag{q,a}[rb]{$q,a$}
\psfrag{q',a'}[lb]{$q',a'$}
\psfrag{p}[rt]{$P$}
\psfrag{p'}[lt]{$P'$}
\psfrag{s}{$s$}
\psfrag{t}{$t$}
 \includegraphics[width=2in]{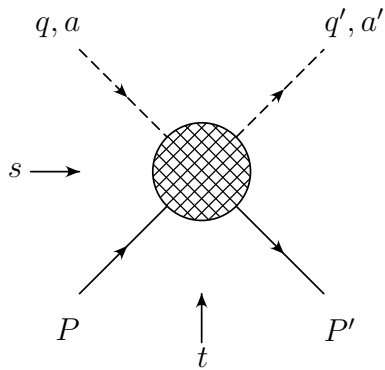}
    \caption{\label{cap:kin}Kinematics of the elastic $\pi N$ scattering
    amplitude. $P$, $q$ ($P'$,$q'$) denote the momentum of the incoming
    (outgoing) nucleons and pions, and $a$ ($a'$) stands for the isospin
    index of the incoming (outgoing) pion. }
  \end{figure}
We consider the scattering amplitude on the mass shell,
\[  
P^2={P'}^2=\m^2,\;\;\;
    q^2={q'}^2=M_\pi^2\fs
\]
In view of crossing symmetry, it is often convenient to express the energy
dependence in terms of the variable
\[ \nu=\frac{s-u}{4 \m}\fs\]
The standard decomposition involves the 
four Lorentz invariant
  amplitudes $A^\pm,B^\pm$:
\begin{eqnarray*}
T_{a'a}\al=\al \delta_{a'a}T^+ + 
\mbox{$\frac{1}{2}$}[{\bf \tau}_{a'},{\bf \tau}_a]\,T^- \no
T^{\pm}\al=\al\bar{u}'
\left\{A^{\pm} +\mbox{$\frac{1}{2}$}\,(q\hspace{-0.5em}/^{\,\prime}+
  q\hspace{-0.5em}/\hspace{0.1em} )B^{\pm}\right\} u\fs
\end{eqnarray*}
This decomposition is not suited to perform the low energy expansion
of the amplitude, because the leading contributions from $A$ and $B$
cancel.  We replace $A$ by $D\equiv A+\nu\,B$, so that the scattering
amplitude takes the form 
\bea 
T^{\pm}=\bar{u}'\left\{ D^{\pm}
-\frac{1}{4 \m} [\,q\hspace{-0.5em}/^{\,\prime},\,
q\hspace{-0.5em}/\hspace{0.15em}]B^{\pm} \right\}u \fs \nonumber
\eea
Our evaluation of the chiral perturbation series to one loop allows us
to calculate the amplitudes $D^\pm$ and $B^\pm$ to $O(q^4)$ and
$O(q^2)$, respectively.

\setcounter{equation}{0}
\section{Effective Lagrangian}
 \label{eL} The variables of the effective theory are
the meson field $U(x)\in\mbox{SU(2)}$ and the Dirac spinor $\psi(x)$
describing the degrees of freedom of the nucleon.  The quark mass matrix 
$m_{\scriptscriptstyle q}=\mbox{diag}(m_u,m_d)$ only enters the effective
Lagrangian together with
the constant $B$ that specifies the magnitude of the quark condensate in 
the chiral limit. As usual, we denote the relevant product by 
$\chi=2 B m_{\scriptscriptstyle q}$. The constant $B$ also determines 
the leading term in the chiral expansion of the square of the pion
mass, which we denote by $M^2$,
\bea M^2\equiv(m_u+m_d) B\fs\nonumber\eea
We disregard isospin breaking effects and set $m_u=m_d$.

The effective Lagrangian relevant for the sector with baryon number equal to 
one contains two parts, 
\bea {\cal L}_\eff={\cal L}_\pi +{\cal  L}_{\ind N}\fs\nonumber
\eea 
The first one only involves the field $U(x)$ and an even
number of derivatives
thereof. Using the abbreviations
\bea u^2 \al=\al
U\co\hspace{1.5em} u_\mu =i u^\dag\partial_\mu U u^\dag\co\hspace{1.5em}
\Gamma_\mu = \mbox{$\frac{1}{2}$}[u^\dag,\partial_\mu u]\co\hspace{1.5em}
\chi_\pm = u^\dag \chi u^\dag \pm u \chi^\dag u \nonumber\co\eea
this part of the Lagrangian takes the form \footnote{In 
the notation of
ref.~\cite{Gasser:1984yg}, the matrix $U$ stands for 
$U=U^0+i\,\vec{\tau}\cdot \vec{U}$. In the isospin limit,
we have $\chi_+=2\,M^2 U^0$, $\chi_-=-2\,i\,M^2\,\vec{\tau}\cdot \vec{U}$,
so that $\langle \chi_-\rangle=0$, $\langle\chi_-^2\rangle=
\frac{1}{2}\langle\chi_+\rangle^2-4\langle\chi^\dagger\chi\rangle$.}
\bea\label{Lpi}{\cal L}_\pi\al=\al\mbox{$\frac{1}{4}$}F^2\langle
u_\mu u^\mu 
+\chi_+\rangle +
\mbox{$\frac{1}{8}$}\,l_4\,\langle u_\mu u^\mu\rangle \langle \chi_+\rangle
+\mbox{$\frac{1}{16}$}\,(l_3+l_4)\,\langle\chi_+\rangle^2+O(q^6)\fs
\rule{2em}{0em}\eea
The expression agrees 
with the one used in ref.~\cite{gss}, but differs from
the Lagrangian in ref.~\cite{Gasser:1984yg} by a term proportional to the
equation of motion. 
The difference is irrelevant, because it amounts to a change of the meson 
field variables. For the physical quantities to remain the same,
this change of variables, however, 
also needs to be performed in the term ${\cal L}_{\ind N}$ -- the 
significance of the
effective coupling constants occurring in that term
does depend on the specific form used for ${\cal L}_\pi$ \cite{Ecker:1994pi
  Ecker:1996rk}.

The term ${\cal L}_{\ind N}$ is bilinear in $\bar{\psi}(x)$,
$\psi(x)$ and the derivative expansion contains odd as well as even powers of
momentum: 
\bea
{\cal L}_{\ind N} ={\cal L}_{\ind N}^{(1)} +{\cal L}_{\ind N}^{(2)}+{\cal
  L}_{\ind N}^{(3)}+\ldots\nonumber
\eea 
The leading contribution 
is fully determined by the nucleon mass $\m$, the pion decay constant
$F_\pi$ and the matrix element of the axial charge $g_A$. We denote the
values of these quantities in the chiral limit by $\Mbare$, $F$ and $\gAbare$,
respectively. 
With $D_\mu\equiv\partial_\mu+\Gamma_\mu$, the explicit 
expression then takes the form
\begin{equation} {\cal L}^{(1)}_{\ind N}=\bar \psi
\left( iD \!\!\!\!/-\Mbare\right)\psi +
\mbox{$\frac{1}{2}$}\, \gAbare\, \bar{\psi}\, u
\hspace{-0.5em}\slash\,\gamma_5 \psi\co\nonumber\end{equation}   
The Lagrangian of order $q^2$ contains
four independent coupling constants\footnote{We use the conventions of
  ref.~\cite{mei}. In this notation, the coupling constants of ref.~\cite{gss}
  are given by: $\Mbare\, c_1^{\GSS}\! =\! F^2 c_1$, $\Mbare\, c_2^{\GSS}\!=\!
  - F^2 c_4$, $\Mbare\, c_3^{\GSS}\!=\! - 2 F^2 c_3$, $\Mbare\,
  (c_4^{\GSS}-2\,\Mbare\, c_5^{\GSS})\!=\!- F^2 c_2$ (to order $q^2$, the
  terms $c_4^{\GSS}$ and $c_5^{\GSS}$ enter the observables only in this
  combination), while those of ref.~\cite{Ecker:1994pi Ecker:1996rk} 
read: $16\,
  a_1\!=\! 8\, \Mbare \, c_3+\gAbare^2$, $8\, a_2\! =\! 4\,\Mbare\,
  c_2-\gAbare^2$, $a_3\!=\!  \Mbare\, c_1$, $4\,a_5\! =\! 4\,\Mbare\,
  c_4+1-\gAbare^2$. In the numerical analysis, we work with
 $F_\pi=92.4\,\mbox{MeV}$, $g_A=1.267$,  $\m=m_p$, $M_\pi=M_{\pi^+}$.}  
\bea \label{L2}
{\cal L}^{(2)}_{\ind N}\al=\al c_1\, \langle\chi_+\rangle\,\bar \psi\, \psi
-\frac{c_2}{4\Mbare^2}\,\langle u_\mu u_\nu \rangle\, (\bar{\psi}\, D^\mu\!
D^{\nu}\!\psi + \mbox{h.c.})\,\\\al\al +\frac{c_3}{2}\,\langle u_\mu
u^\mu\rangle\,\bar{\psi}\,\psi
-\frac{c_4}{4}\,\bar{\psi}\,\gamma^\mu\,\gamma^\nu\, [u_\mu,u_\nu]\,\psi
\fs\nonumber\eea 
The terms ${\cal L}_{\ind N}^{(3)}$ and ${\cal L}_{\ind N}^{(4)}$ enter the 
representation of the scattering amplitude to $O(q^4)$ only at tree level. 
The Lagrangian of order $O(q^3)$ is taken from ref.~\cite{Fettes:1998ud}: 
\begin{multline} 
{\cal L}_{\ind N}^{(3)}
= \bar{\psi} \Big\{ - \frac{d_1+d_2}{4m} \big(
[u_\mu,[D_\nu,u^\mu]+[D^\mu,u_\nu]] D^\nu + \mbox{h.c.}  \big) \nonumber \\ 
 +\,\frac{d_3}{12m^3} \big( [u_\mu,[D_\nu,u_\lambda]] (D^\mu D^\nu D^\lambda
+ \mbox{sym.})  + \mbox{h.c.}  \big) \nonumber \\   + \frac{i\,d_5}{2m}
\big( [\chi_-,u_\mu] D^\mu + \mbox{h.c.}  \big) \nonumber \\   +
\frac{i\,(d_{14}-d_{15})}{8m} \big( \sigma^{\mu\nu} \langle [D_\lambda,u_\mu]
u_\nu- u_\mu [D_\nu,u_\lambda] \rangle D^\lambda + \mbox{h.c.}  \big)
\nonumber \\   + \frac{d_{16}}{2} \gamma^\mu \gamma_5 \langle \chi_+ \rangle
u_\mu +\,\frac{i\,d_{18}}{2} \gamma^\mu \gamma_5 [D_\mu,\chi_-]\Big\}\psi 
\fs\end{multline}
At tree level only five combinations of the coupling constants occurring here
enter the scattering amplitude. We do not give an explicit expression
for ${\cal L}_{\ind N}^{(4)}$, but characterize this part of the effective
Lagrangian through its contributions to the scattering amplitude 
(sec.~\ref{Tree graphs}).

\setcounter{equation}{0}
\section{Tree graphs}\label{Tree graphs}
The tree graph contributions to the amplitude have the following
structure
\begin{center}
\begin{tabular}{ccc}
    \includegraphics[height=2cm]{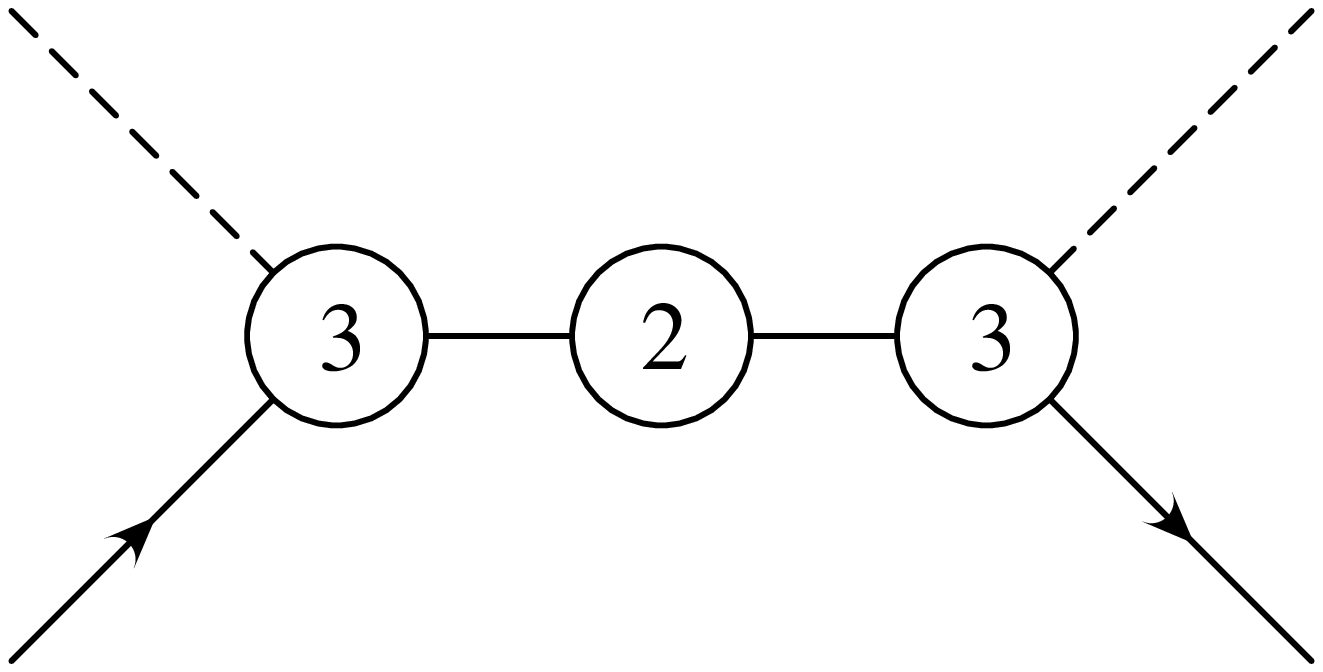} &&
    \includegraphics[height=2cm]{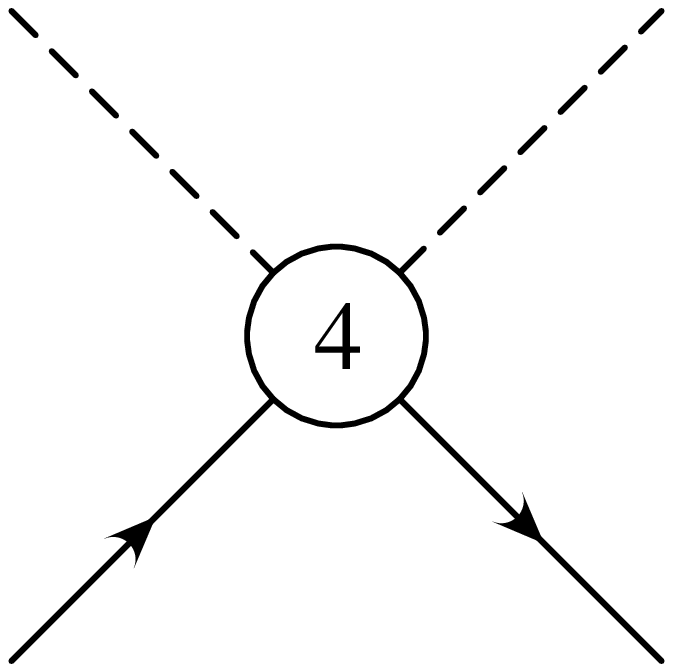} \fs
\end{tabular}
\end{center}
The vertex contributions \circa{3} involve the axial coupling $\gAbare$ and a
quark mass correction to it from the Lagrangian ${\cal L}^{(3)}_{\ind N}$. The
quark mass correction can be taken into account by replacing the coupling
constant $\gAbare$ in ${\cal L}^{(1)}$ with $g_2=\gAbare+
2M^2(2d_{16}-d_{18})$. The contributions of type \circa{2} represent mass
insertions generated by ${\cal L}^{(2)}_{\ind N}$ and ${\cal L}^{(4)}_{\ind
  N}$ at tree level.  They are taken into account by replacing the bare mass
$m$ in ${\cal L}^{(1)}_{\ind N}$ with $m_4= m-4c_1 M^2+e_1\, M^4$ (we adopt
the notation of \cite{Becher Leutwyler 1999} and denote the relevant coupling
constant in ${\cal L}^{(4)}_{\ind N}$ by $e_1$).

When evaluating the relevant diagrams, we must distinguish between
the bare and physical masses of the nucleon.
As we are working on the physical mass shell, $P^2=\m^2$,
we need to use the corresponding Dirac equation for the nucleon spinors,
$P\!\!\!\!/\;\U({ P})=\m\,\U({ P})$. The result for the sum of all
contributions of type \circa{2} and \circa{3}
reads (here, the 
amplitude $A$ is more convenient than $D$, because the representation then 
only involves functions of a single variable):
\begin{align*}
  A^\pm&=A(s)\pm A(u) &A(s)&=\frac{g_2^2\,(m_4\,+\,\m)}{4\,F^2}
  \left(\frac{s\,-\,\m^2}{s\,-\,m_4^2}\right) \\
  B^\pm&=B(s)\mp B(u)& B(s)&=\frac{g_2^2\, (s+2\,m_4\,\m+\m^2)}{4\,F^2\,
(m_4^2-s)}\fs
 \end{align*}  

The tree graphs of type \circa{4} yield polynomials in
$\nu$, $t$ and $M^2$. We only list the nonzero contributions. 
Those from ${\cal L}^{(1)}_{\ind N}$ are given by
\bea D^-=\frac{\nu}{2F^2}\co\;\;\;B^-=\frac{1}{2F^2}\co\nonumber\eea
while ${\cal L}^{(2)}_{\ind N}$ generates the terms
\bea D^+=\label{treeO2}
-\frac{4\,c_1 M^2}{F^2}+\frac{\,c_2\, (16\, \m^2\,\nu^2-t^2)}{8\,F^2\, m^2} +
\frac{c_3(2\M^2-t)}{F^2}\co\;\;
B^-=\frac{2\,c_4\,\m}{F^2}\fs\nonumber\eea
Note that the coupling constant $c_2$ gives rise to a contribution 
proportional to $t^2$ which is of $O(q^4)$. Quite generally, vertices
involving derivatives 
of the nucleon field may generate contributions beyond the order indicated by
chiral power counting. 

In the graphs from ${\cal L}_{\ind N}^{(3)}$ and ${\cal L}_{\ind N}^{(4)}$, we 
may replace
the bare constants by the physical ones, because the distinction is beyond the
accuracy of our calculation.  The term ${\cal L}^{(3)}_{\ind N}$ only shows up 
in
those amplitudes that are odd under crossing:
\begin{align*}
D^-&=\frac{2\,\nu}{\F^2}\left\{
    2( d_1+d_2+2 d_5)\, \M^2  -(d_1+d_2)\,t  + 2\, d_3\,{\nu
    }^2\right\}+O(q^5)\\ 
B^+&=\frac{4\, \nu\, \m}{\F^2}\, (d_{14}-d_{15}) +O(q^3)
\end{align*}
Finally, ${\cal L}^{(4)}_{\ind N}$ generates a polynomial that is even under 
crossing. We identify the coupling constants $e_3,\,\ldots\,,e_{11}$
with the coefficients of this polynomial:
\bea\label{eq:L4}
D^+\al=\al\frac{1}{\F^2}\left\{e_3 \M^4 + e_4\,\M^2 \nu^2 + e_5\,\M^2\,t+e_6\,
  \nu^4 + e_7\,\nu^2\,t+e_8\,t^2  \right\}+O(q^6)\no
B^-\al=\al\frac{\m}{\F^2}\left\{e_9 \M^2 + e_{10} \,\nu^2 + e_{11}\,t
  \right\}+O(q^4) 
\eea
A contribution analogous to $e_3 \M^4 +e_5\M^2\,t$ 
also occurs in the representation of the scalar 
form factor to $O(q^4)$ (in
\cite{Becher Leutwyler 1999} the corresponding contribution to $\sigma(t)$ 
is denoted by $2e_1\,\M^4+e_2\,\M^2\,t$). The coupling constants are
different, because 
the terms $\langle \chi_-^2 \rangle\,\bar{\psi}\,\psi$ and $\langle \chi_+
\rangle \langle u_\mu u^\mu \rangle \,\bar{\psi}\,\psi$ of ${\cal
  L}^{(4)}_{\ind N}$ 
do contribute to the scattering amplitude, but do not show up in 
the scalar form factor.   

\newpage
\begin{figure}[!ht]
\begin{center}
\includegraphics{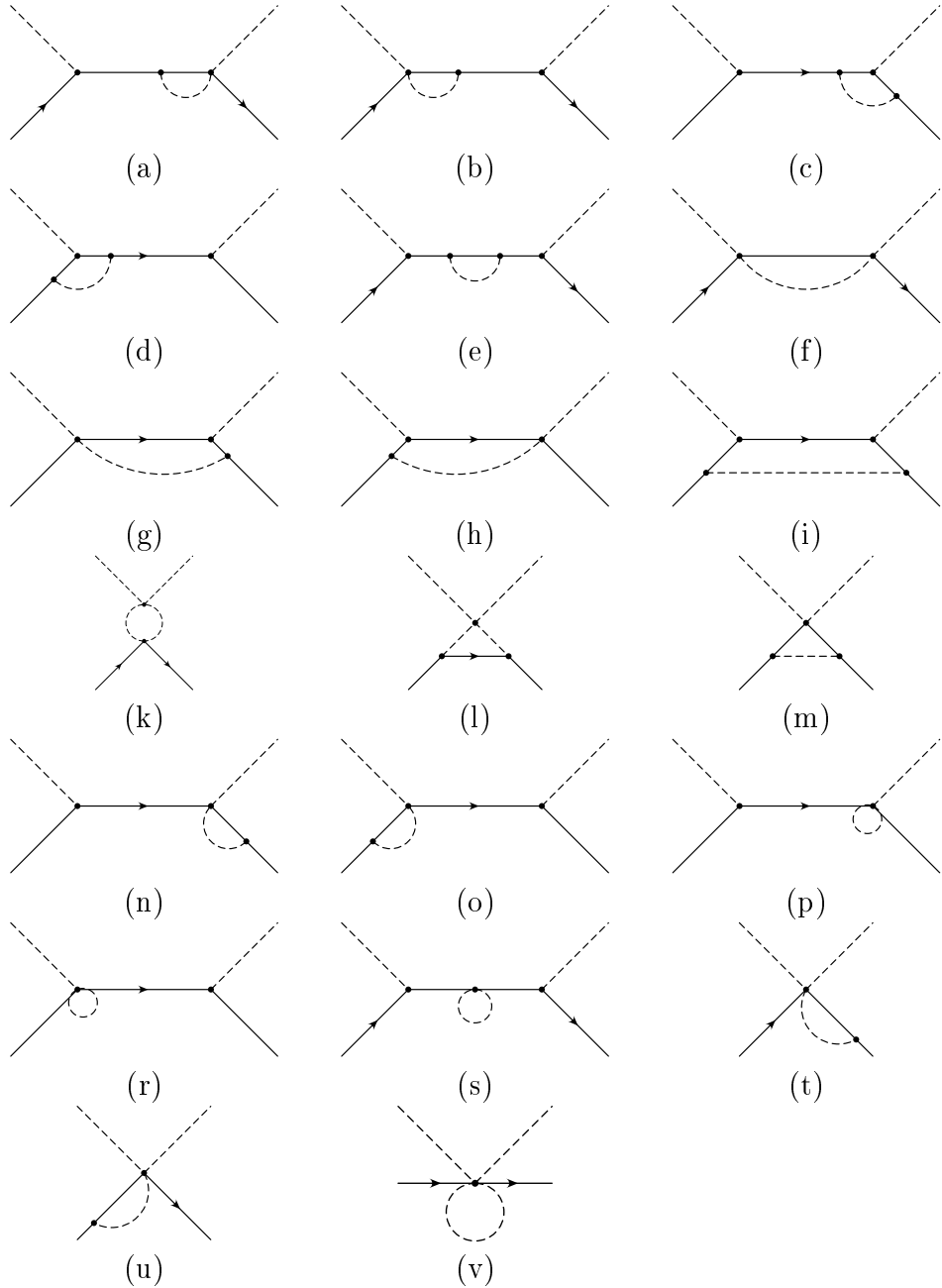}
\end{center}
\caption{One loop topologies for $\pi N$-scattering. Crossed diagrams and
  external leg corrections are not shown.  We do not display topologies
  involving closed fermion loops since they vanish in
  infrared regularization.}\label{fig:loop}
\end{figure}
\newpage

\setcounter{equation}{0}
\section{One loop graphs in infrared regularization}
\begin{figure}[htbp]
\vspace*{2em}
  \begin{center}
\includegraphics{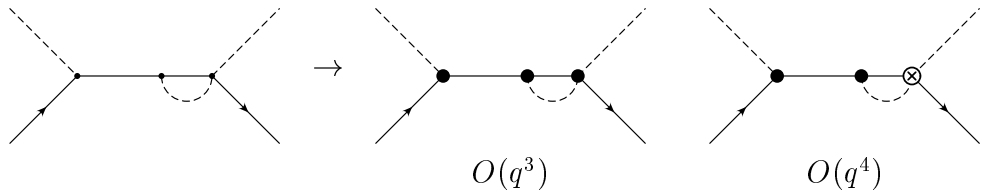}
\caption{By exclusively inserting vertices from ${\cal L}_{\ind N}^{(1)}$ 
(drawn as full circles)
into topology (a), we obtain a graph of
  $O(q^3)$. The same topology also occurs if one of the vertices 
is replaced by one of those in ${\cal L}_{\ind N}^{(2)}$, which involve
the coupling constants $c_1$, $c_2$, $c_3$ or $c_4$  
(denoted by a circled cross). The resulting graph then starts contributing at 
$O(q^4)$. Since ${\cal
  L}_{\ind N}^{(2)}$ exclusively contains vertices with an even number of
pions, only 
the rightmost vertex can be replaced.} 
  \label{fig:graphA}
  \end{center}
\vspace*{-1em}
\end{figure}

\label{Loops
  Infrared}
To obtain a representation of the scattering amplitude to order $q^3$, 
it suffices to evaluate the one loop graphs of ${\cal L}^{(1)}_{\ind N}$. 
As we wish to
evaluate the scattering amplitude to order $q^4$, we in addition 
need to consider
loops containing one vertex from ${\cal L}^{(2)}_{\ind N}$. The various
topologies 
are shown in figure \ref{fig:loop}. In this figure, we do not differentiate
between the vertices of ${\cal L}^{(1)}_{\ind N}$ and those of ${\cal
  L}^{(2)}_{\ind N}$.  
Graph (a) for instance describes two different contributions (see figure
\ref{fig:graphA}). 
Since the second order Lagrangian ${\cal
  L}_{\ind N}^{(2)}$ only involves couplings with an even number of pions, 
the topologies (c), (d), (e), (i), (l), (p), (r), (t) and
(u) do not give rise to $O(q^4)$ diagrams. 
 
The calculation of the loop diagrams generated by ${\cal
L}^{(1)}_{\ind N}$ was carried out quite some time ago by Gasser,
Sainio and \v{S}varc \cite{gss}, using a relativistic Lagrangian and
dimensional regularization. As mentioned in the introduction,
this approach does not preserve
the counting rules. In a recent paper \cite{Becher
Leutwyler 1999}, we have have set up a variant of dimensional
regularization that avoids these difficulties and which we call
``infrared regularization''.  We now briefly discuss this method.

\subsection{Infrared regularization}
\label{self energy}
\begin{figure}[ht]\centering
 \includegraphics[width=2in]{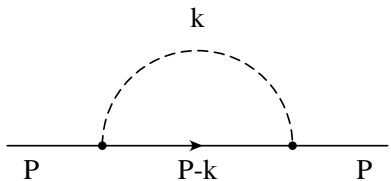}
    \caption{\label{fig:selfE} Self energy}
\end{figure}

 To explain the essence of the method, 
we consider the simplest example, the self energy graph shown in figure
\ref{fig:selfE}.
 The corresponding scalar loop integral has the form
\begin{align}\label{eq:selfE}
H(P^2)&=\frac{1}{i}\int\frac{d^d k}{(2\pi)^d}
\;\frac{1}{M^2-k^2}\;\frac{1}{m^2-(P-k)^2}
=\frac{1}{i}\int\frac{d^d k}{(2\pi)^d}\;\frac{1}{a}\;\frac{1}{b}  
\end{align}
We need to analyze the integral for external momenta in the vicinity
of the mass shell: $P=m
v+r$, where $v$ is a timelike unit vector and $r$ is a quantity of order $q$.
In the limit $M\rightarrow 0$, the integral develops an infrared singularity,
generated by small values of the variable of integration, $k=O(q)$. In that
region, the first factor in the denominator is of $O(q^2)$, while the second is
of order $O(q)$. Accordingly the chiral expansion of the integral contains
terms of order $q^{d-3}$. The order of the infrared singular part follows from
the counting rules at tree level, because it is generated by the region of
integration, where the momenta flowing through the propagators are of the same
order as in the tree level graphs.  The remainder of the integration region
does not contain infrared singularities and may thus be expanded in an ordinary
Taylor series. An evaluation of the integral at $P^2=s_+=(m+M)^2$ nicely
shows the two parts:
\begin{equation*}
H(s_+)=\frac{\Gamma(2-\frac{d}{2})}
{(4\pi)^\frac{d}{2}(d-3)}\;\Big\{\frac{\Mp^{d-3}}{(\Mbare+\Mp)}+
\frac{\Mbare^{d-3}}{(\Mbare+\Mp)}\Big\}=I+R 
\end{equation*}
The infrared singular part $I$ is proportional to $M^{d-3}$, while the
remainder $R$ is proportional to $m^{d-3}$ and does therefore not contain a
singularity at $M= 0$, irrespective of the value of $d$. Since the regular
contribution stems from a region where the variable of integration $k$ is of
order $m$, it violates the tree-level counting rules.

The explicit expression for the loop integral $H(P^2)$ in $d$ dimensions is
lengthy, but the splitting into the infrared singular part $I$ and regular
part $R$ is easily obtained in the Schwinger-Feynman representation of the
integral.
\begin{align*}
  H\;&=\;\int\frac{d^d k}{(2\pi)^d}\, \frac{1}{a}\,\frac{1}{b} = \int_0^1
  {dz}\,\int\frac{d^d k}{(2\pi)^d}\,\frac{1}{[\,(1-z)\, a+z
    \,b\,]^2}\\ \\
  &=\;\int_0^{\infty}-\int_1^{\infty}dz\,\int\frac{d^d k}{(2\pi)^d}\,
  \frac{1}{[\,(1-z)\, a+z\, b\,]^2}\; 
  = I+R \,.
\end{align*}
Schwinger-Feynman parametrizations of the infrared singular and regular parts
 of a general one loop integral are given in \cite{Becher Leutwyler
 1999}. There, the following statements are proven:
\begin{itemize}
\item To any given order, the chiral expansion of the infrared regular part
 of a one loop integral is a polynomial in the quark masses and the
 external momenta. 
\item The regular part of the one loop amplitudes is chirally symmetric.
\end{itemize}
Taken together, the two statements imply that, in a one loop calculation, one
 may simply drop the regular parts of the loop integrals because the
 contributions they generate can be absorbed in the coupling constants 
of the effective
 Lagrangian. This procedure is called ``infrared regularization'': The loop
 graph contributions are identified with the infrared singular
 parts of the corresponding dimensionally regularized integrals.

\subsection{Comparison with dimensional regularization}
The calculation of loop graphs in infrared regularization is completely
analogous to the evaluation in dimensional regularization. 
The expressions for the graphs in terms of loop integrals look the same.
In dimensional regularization, the contribution of
the graph (f) in figure \ref{fig:loop} 
to the isospin even amplitude, for instance, is given by 
\begin{multline*}
T_f^+=\frac{1}{16F^4}\Ubar^{\prime}
\Big\{ 4(s-m^2)(2m+q\hspace{-0.5em}/^{\,\prime}+
q\hspace{-0.5em}/\hspace{0.1em})H^{(1)}(s)\\ 
-(4M^2\, H(s)+\Delta_\pi-4\Delta_{\ind N})
(\,q\hspace{-0.5em}/^{\,\prime}+q\hspace{-0.5em}/\hspace{0.1em}) \Big\} {\U}
\end{multline*}
The terms $\Delta_\pi$ and $\Delta_{\ind N}$ stand for the pion and nucleon
propagators at the origin, respectively: 
\begin{align*} \Delta_\pi&=\frac{1}{i}\int_I \frac{d^dk}{(2\pi)^d}\,\frac{1}
{M^2-k^2-i\epsilon}\co\;\;\;\;\DN =\frac{1}{i}\int_I
\frac{d^dk}{(2\pi)^d}\,\frac{1} 
{m^2-k^2-i\epsilon}\fs
\end{align*}
The function $H^{(1)}(s)$ may be expressed in terms of the scalar self energy
integral $H(s)$, which we considered in the preceding subsection:
\bea H^{(1)}(s)=\frac{1}{2s}\,\left\{(s-m^2 + M^2)\,H(s) +\Dpi-\DN\right\}
\fs\nonumber\eea

To obtain the corresponding result in infrared regularization, it suffices to
replace all loop integrals by their infrared singular parts. Integrals like
$\Delta_\pi$ 
with pion propagators only do not have an infrared regular part, because they
do not involve the heavy scale $m$: In the meson sector, infrared
regularization coincides with ordinary dimensional regularization. 
Nucleon loop integrals like $\Delta_{\ind N}$, on the other hand, vanish in
infrared 
regularization, because they are infrared regular. Integrals
which involve both kinds of propagators do posses a regular and an
infrared singular part. These integrals are replaced by their infrared singular
part. The result for graph (f) in infrared regularization is thus obtained with
the substitution $H(s)\rightarrow I(s)$, $H^{(1)}(s)\rightarrow I^{(1)}(s)$ and
$\Delta_{\ind N}\rightarrow 0$. 

The result for the various graphs is given in appendix \ref{Loop graphs}. The
definitions of the pertinent loop integrals can be found in appendix \ref{Loop
integrals}. We have checked that our result for the loop graphs of ${\cal
L}^{(1)}_{\ind N}$ agrees with the one of Gasser, Sainio and \v{S}varc
\cite{gss}, if we 
evaluate our loop integrals in dimensional regularization and expand around
$d=4$. Note that we do not display the contributions proportional to loop
integrals that exclusively contain nucleon propagators, because these vanish
in infrared regularization. Also, the basis used in the decomposition of the
vector and tensor integrals differs from the one used by Gasser, Sainio and
\v{S}varc -- in our basis, the chiral power counting is more transparent.

\setcounter{equation}{0}
\section{Simplification of the {\boldmath${O(q^4) }$\unboldmath}  diagrams}
\label{Simplification}
The chiral expansion of the loop graphs in general contains terms of 
arbitrarily high order. Since we need
the amplitude only to $O(q^4)$, we can simplify the representation
by neglecting terms of $O(q^5)$ or higher. 
The example of the triangle graph (graph h
in fig.~\ref{fig:loop}) shows, however, that the chiral expansion
of infrared singularities is a subtle matter \cite{Becher Leutwyler
  1999}, which we will discuss 
in detail later on. The problem arises from the
{\it denominators} that occur in the various loop integrals -- the expansion
thereof gives rise to an infinite series of terms, which in some cases needs
to be summed up to all orders to arrive at a decent representation 
of the integral. Concerning, the {\it numerators}, 
this problem does not arise -- the corresponding expansion 
does not give rise to an infinite series of terms and hence preserves
the analytic structure of the integral, term by term.
We exploit this fact to simplify the calculation of those graphs that
contain vertices from ${\cal L}^{(2)}_{\ind N}$, which start contributing only
at $O(q^4)$ (in the loop graphs generated by ${\cal L}^{(1)}_{\ind N}$, 
we retain all contributions). 

As an illustration, we consider the part of the $O(q^4)$ graph (a) that is
proportional to $c_2$ 
\begin{multline*}
T^+=\frac{c_2\, g_A^2}{4\, F^4\,m^2}\,\frac{1}{m^2-s}\;\frac{1}{i}\int_I 
\frac{d^dk}{(2\pi)^d}\,\,\frac{1}{(M^2-k^2)(m^2-(P+q-k)^2)}\\\times
\big(k\!\cdot\! 
P'\,q'\!\cdot\! P'+k\!\cdot\! (P+q-k)\,\, q' \!\cdot\! (P+q-k)\big)  \\
\times \Ubar'\,
(m+\SL{P}+\Sl{q}-\Sl{k})\,\gamma_5\,\Sl{k}\,(m+\SL{P}+\Sl{q})\,
\gamma_5\,\Sl{q}\,\U
\end{multline*}
The numerator of the integrand involves terms 
with up to five powers of the
loop momentum. It can explicitly be expressed in terms of the
functions $I(s)$, $I^{(1)}(s)$, $\ldots\,$, that arise in the tensorial 
decomposition of the generic loop integral with one meson and one nucleon
propagator. The resulting expression is rather lengthy, however. 
As just discussed, we only retain the leading 
term in the expansion of the numerator:
\begin{multline*}
T^+=\frac{c_2\, g_A^2}{4 F^4 m^2}\,\frac{1}{m^2-s}\;\frac{1}{i}\int_I 
\frac{d^dk}{(2\pi)^d}\,\,\frac{1}{(M^2-k^2)(m^2-(P+q-k)^2)}\\ \times2 k\cdot
P'\, q'\cdot P'\,  \Ubar'\,
2m\,\gamma_5\,\Sl{k}\,(m+\SL{P}\,\,)\,\gamma_5\,\Sl{q}\,\U+O(q^5)
\end{multline*}
\begin{multline*}
\phantom{T^+}=\frac{c_2\, g_A^2}{2\,F^4\, m}\;\frac{1}{i}\int_I 
\frac{d^dk}{(2\pi)^d}\,\,\frac{1}{(M^2-k^2)(m^2-(P+q-k)^2)}\\ \times k\cdot
P'  \Ubar\,\Sl{k}\,(m-\SL{P}\,\,)\,\Sl{q}\,\U+O(q^5)\co
\end{multline*}
where we have used $P'=P+O(q)$. The relation $2 k\cdot
P'=m^2-(P+q-k)^2+O(q^2)$ then allows us to remove the nucleon propagator, so
that we arrive at 
\bea
T^+=\frac{c_2\, g_A^2}{4\,F^4\, m}\;\frac{1}{i}\int_I 
\frac{d^dk}{(2\pi)^d}\,\,\frac{1}{(M^2-k^2)}\, \Ubar'\,\Sl{k}\,
(m-\SL{P}\,\,)\,\Sl{q}\,\U+O(q^5)\fs
\nonumber\eea
The remaining integral vanishes because the integrand is antisymmetric in $k$,
so that the entire contribution is of $O(q^5)$ and can be dropped.

\setcounter{equation}{0}
\section{Wave function renormalization}\label{sec:Wave Function}
According to the rules of perturbation theory, the scattering amplitude
may be evaluated from the connected four-point-function
\bea \langle 0|\,T\,N(x')\,P_{a'}(y')\,P_a(y)\,\bar{N}(x)\, 
|0\rangle\rule[-0.3em]{0em}{0em}_c\co\nonumber\eea
where $N(x)$ and $P_a(x)$ are any operators that interpolate between the
relevant incoming and outgoing states, so that the matrix elements 
\bea \langle 0|N (x)|N\rangle =\sqrt 
Z_{\hspace{-0.1em}\scriptscriptstyle N}\, u\,e^{-i P x}\co\;\;\; 
\langle 0| P_a(x)|\pi_b\rangle =\sqrt Z_\pi\,\delta_{ab}\, e^{-i q x}\co
\nonumber\eea
are different from zero (we suppress the spin and
isospin quantum numbers of the nucleon). 
The scattering amplitude $T_{a'a}$ is determined by the residue of the poles 
occurring
in the Fourier transform of this correlation function:
\begin{multline*}
i\!\!\int\!\! d^4x' d^4y' d^4y \langle 0|\,  T\, N(x')\,{P}_{a'}(y')\,P_a(y)\,
\bar{N}(x)\,|0\rangle\rule[-0.3em]{0em}{0em}_c \,
e^{i(P' x'+q' y'- P x-q y)}\\
=\frac{\sqrt{Z_\pi}}{M_\pi^2-{q}^{\prime\, 2}}\;
\frac{\sqrt{Z_\pi}}{M_\pi^2-q^2}\;
\frac{\sqrt{Z_{\hspace{-0.1em}\scriptscriptstyle N}}}{\m^2-P^{\prime\,
    2}}\; 
\frac{\sqrt{Z_{\hspace{-0.1em}\scriptscriptstyle N}}}{\m^2-{P}^2}\;
\U'\,\Ubar\,T_{a'a} +\dots 
\end{multline*}
The result for $T_{a'a}$ does not depend on the choice of the 
interpolating fields. 

In our context, it is convenient to identify the interpolating fields
with the variables of the effective theory, $N(x)=\psi(x)$, $P_a(x)=\pi_a(x)$.
We emphasize that these are auxiliary quantities that represent the variables
of integration in the path integral -- the effective fields
do not have counterparts in QCD and do thus not represent 
objects of physical significance. The renormalization of the
coupling constants occurring in our effective Lagrangian 
only ensures that the scattering amplitude remains finite
when the regularization is removed, $d\rightarrow 4$.
A multiplicative renormalization of the effective fields,
$\psi(x)=\sqrt{Z_{\hspace{-0.1em}\scriptscriptstyle N}}\,
\psi(x)^{\scriptscriptstyle ren}$, $\pi_a(x)=\sqrt{Z_\pi}\,
\pi_a(x)^{\scriptscriptstyle ren}$, does not suffice to obtain finite 
correlation functions for these. Note also that the
correlation functions depend on the parametrization used for the matrix
$U(x)$.

It is easy to see why the correlation functions of the effective fields
do not represent meaningful quantities. The corresponding 
generating functional 
is obtained by adding source terms to the effective Lagrangian,
\bea {\cal L}_{\eff}\rightarrow {\cal
  L}_{\eff}+f^a\pi_a +\bar{g}\,\psi+ \bar{\psi}\, g\fs\nonumber \eea
This operation, however, ruins the symmetries of the effective 
theory, because the representation of the chiral group 
on $\pi_a(x)$ and $\psi(x)$ involves the pion field 
in a nonlinear manner. 
To arrive at finite correlation functions, we would need to
add extra counter terms that are not invariant under chiral rotations.
 
A more physical choice for the interpolating fields is discussed in appendix
\ref{app:sources}. This is of interest, for instance, in connection
with the matrix elements relevant for baryon decay. The corresponding
correlation functions do have a physical interpretation within QCD.
Since the relevant operators carry anomalous dimension, they depend on the
running scale of that theory, but are otherwise free from ambiguities. 
In our context, the choice of the interpolating fields is irrelevant -- the 
result obtained for the $S$-matrix by using the fields discussed in appendix
\ref{app:sources} would be the same.

The perturbative calculation of the scattering matrix can be
simplified in the familiar manner: It suffices to consider the amputated
four-point-function, obtained by (i) 
discarding all graphs that contain insertions in the 
external lines and (ii) replacing the free propagators that describe the
external lines in the remaining graphs by plane wave factors. Denoting the
amputated four-point-function by $\Gamma_{a' a}$, the scattering amplitude
is given by
\begin{equation*}
T_{a'a}=Z_{\hspace{-0.1em}\scriptscriptstyle N}\,Z_\pi\,\Ubar' \,
\Gamma_{a'a}\, \U\fs 
\end{equation*}
We have performed our calculations in the
so-called sigma parametrization
\begin{equation*}
U(x)=\sqrt{1-\frac{\vec{\pi}(x)^2}{F^2}}+i\,
\frac{\vec{\pi}(x)\cdot\vec{\tau}}{F}\fs
\end{equation*}
The wave function renormalization constants are then given by \cite{Becher
  Leutwyler 1999}
\begin{align*}
Z_\pi&=1- \frac{2\,M^2}{F^2}\left( \ell_4 + \lambda_\pi \right)+
O(M^4)\\
Z_{\hspace{-0.1em}\scriptscriptstyle N}&=1-\frac{9\,M^2 g_A^2}{2 F^2}
\left\{\lambda_\pi +
\frac{1}{48\pi^2}-\frac{M}{32\pi\,m}\right\}+
O(M^4)\co\\
\lambda_\pi&=\frac{M^{d-4}}{(4\pi)^2}\left\{\frac{1}{d-4}-\frac{1}{2}
\left(\rule{0em}{1em}\,\mbox{ln}\, 4\pi
  +\Gamma'(1)+1\right)\right\}\fs
\end{align*}   
The divergences $\propto (d-4)^{-1}$ are contained in the quantity 
$\lambda_\pi$, which involves the pion mass instead of an arbitrary scale 
$\mu$ -- we are in effect identifying the renormalization scale 
with the pion mass. This is convenient, because it simplifies the formulae: 
The divergences are always accompanied
by a chiral logarithm. Expressed in terms of the usual pole term
$\lambda$, we have
\bea \lambda_\pi=\lambda +\frac{1}{16\pi^2}\,\ln \frac{M}{\mu}\fs
\nonumber\eea
The effective coupling constants of ${\cal L}_\pi^{(4)}$,
${\cal L}_{\ind N}^{(3)}$ and ${\cal L}_{\ind N}^{(4)}$ pick up
renormalization. The relevant formulae are listed in appendix
\ref{sec:renormalization}. We have checked that if the scattering amplitude
is expressed in the renormalized couplings
$\ell_3^r(\mu),\ell_4^r(\mu),d_i^r(\mu),e_i^r(\mu)$ introduced there, it
indeed remains finite at $d\rightarrow 4$. 

\setcounter{equation}{0}
\section{Chiral expansion of {\boldmath$m_{\scriptscriptstyle N}$\unboldmath}, 
{\boldmath$ g_{\pi{\scriptscriptstyle N}}$\unboldmath} and {\boldmath$
  \gA$\unboldmath}}
\label{nucleon mass and coupling constant}

The graphs (e) and (s) in fig.~\ref{fig:loop} contain a double pole at
$s=m_4^2$ and their crossed versions exhibit the analogous singularity
at $u=m_4^2$ (we recall that the mass shifts generated by the tree
graphs of ${\cal L}_{\ind N}^{(2)}$ and ${\cal L}_{\ind N}^{(4)}$ are
included in the free Lagrangian, so that the nucleons propagating in
the loops are equipped with the mass $m_4=m-4 c_1 M^2 +e_1\,
M^4$). The double pole disappears as it should if the Born term, which
contains the pole due to one-nucleon exchange, is expressed in terms
of the physical mass of the nucleon. In the context of the low energy
expansion, the standard, pseudoscalar form of the Born term is not
convenient, because it corresponds to a scattering amplitude of order
$q^0$. We instead use the pseudovector form, defined by
\begin{align}\label{eq:pvBorn} D_{pv}^+(\nu,t)&=
  \frac{g_{\pi N}^2}{\m}\;\frac{\nu_B^2}{\nu_B^2-\nu^2}\co &
  B_{pv}^+(\nu,t)& =
  \frac{g_{\pi N}^2}{\m}\;\frac{\nu}{\nu_B^2-\nu^2} \co\\
  D_{pv}^-(\nu,t)& = \nu B^-_{pv}(\nu,t) \co &
  B_{pv}^-(\nu,t)& =\frac{g_{\pi N}^2}{\m}\,\frac{\nu_B}{\nu_B^2-\nu^2}-
  \,\frac{g_{\pi N}^2}{2 \m^2}\co\no  
\nu&  =\frac{s-u}{4\, \m}\co  & \nu_B & =\frac{t-2
  M_\pi^2}{4\,\m}\fs\nonumber\end{align}
The Born term exclusively involves the physical values of 
$\m$, $\gpiN$ and $M_\pi$. The explicit expression for $\m$ reads \cite{Becher
  Leutwyler 1999}  
\bea\label{mN} \m\al=\al m-4\,c_1 M^2
-\frac{3\,\gAbare^2 M^3}{32\pi  F^2}-\frac{3\,(\gAbare^2-8\,c_1\, m +c_2\, m 
+4\,c_3\, m)\,\lambda_\pi M^4}{2\, m\, F^2}\no
\al \al - \frac{3\,(2\,\gAbare^2-c_2\, m)M^4}{128\pi^2\, m\, F^2}+
e_1 M^4+O(M^5)\co\eea
while the one for the pion-nucleon coupling constant is given by
\bea
  \gpiN\al=\al \frac{m\gAbare}{F}\left\{1-\frac{ l_4 M^2}{F^2}-
\frac{4\,c_1 M^2}{m}+\frac{2\,(2\,d_{16}-d_{18}) M^2 }{\gAbare} -
\frac{4\,\gAbare^2\lambda_\pi \, M^2}{F^2}\right.\no \al-\al\left.
\frac{\gAbare^2 M^2}{16 \pi^2 F^2}+
 \frac{(12 +3 \,\gAbare^2-16\, c_3\, m + 32\, c_4\, m)\,M^3}
{96\pi\, m\, F^2 }+O(M^4)\right\}\,.\rule{2em}{0em}\eea
To order $M^2$, the result agrees with the one given in ref.~\cite{gss} -- 
only the terms of order $M^3$ are new.
The quantity $\gAbare$ represents the value of the axial charge in the chiral
limit. Kambor and Moj\v zi\v s \cite{Kambor:1998pi} and Schweizer
\cite{Schweizer} have calculated the physical axial charge $\gA$ to $O(q^3)$,
with the result \bea \gA\al=\al\gAbare\left\{1+\frac{4\,
    d_{16}\,M^2}{\gAbare}-\frac{2\,(2\gAbare^2+1) \,\lambda_\pi\,
    M^2}{F^2}-\frac{\gAbare^2\,M^2}{16\,\pi^2\,F^2}\right.\no
\al+\al\left.\frac{(3+3\, \gAbare^2-4\,c_3\,m+8\,c_4 \,m)\,M^3}{24 \pi
    \,m\,F^2}+O(M^4)\right\}\fs\eea

These expressions are analogous to the well-known representations that
describe the dependence of $M_\pi$ and $F_\pi$ on the 
quark masses \cite{Gasser:1984yg},
\begin{align*}
M_\pi^2&=M^2\left\{1+\frac{M^2}{F^2}\,(2\,l_3+\lambda_\pi )+O(M^4)\right\}\co\\
\F&=F\left\{1+\frac{M^2}{F^2}\,(l_4 -2\,\lambda_\pi)+O(M^4)\right\}\fs
\end{align*}
In all cases, the expansion contains contributions that are not analytic
in the quark masses, generated by the infrared singularities. In the meson
sector, only the chiral logarithm contained in $\lambda_\pi$ occurs,
but in the nucleon sector, the expansion in addition involves contributions
with $M^3\propto (m_u+m_d)^{\frac{3}{2}}$. 

\setcounter{equation}{0}
\section{Strength of infrared singularities}
\label{Strength of infrared singularities}
The fact that the low energy expansion in the baryon sector contains
even as well as odd powers of $M$ implies that the infrared singularities
occurring there are stronger than those in the meson sector.
We illustrate this with the $\sigma$-term matrix element, that is, with 
the response of the nucleon mass to a change in the quark masses:
\bea\label{FH} \sigma=m_u\frac{\partial \m}{\partial m_u}+
m_d\frac{\partial \m}{\partial m_d}\fs\eea
More specifically, we wish to discuss the dependence of this quantity on the
quark masses, which is quite remarkable. In this context, the numerical value 
of $\sigma$ is not of crucial importance. For definiteness, we use 
$\sigma=45\,\mbox{MeV}$ \cite{Gasser Leutwyler
  Sainio}. The chiral expansion of the $\sigma$-term is readily obtained
from the formula (\ref{mN}), with the result
\bea\label{sigma} \sigma=k_1 M^2 +\frac{3}{2}\,k_2 M^3 +k_3 M^4 
\left\{2\,\ln \frac{M^2}{m^2}+1\right\} + 
2\, k_4 M^4 +
O(M^5)\co\eea 
As discussed above, the term proportional to $M^3$ arises from an infrared
singularity in the self energy of the pion cloud. It lowers the magnitude of
the $\sigma$-term
by $3/2\times 15 \,\mbox{MeV} \simeq 23\, \mbox{MeV}$. 
The coefficient $k_3$ can also be expressed in terms
of measurable quantities \cite{Becher Leutwyler 1999}. Numerically the
contribution from 
this term amounts
to $- 7 \,\mbox{MeV}$, thus amplifying the effect seen at $O(M^3)$. 
Chiral symmetry does not determine all of 
the effective coupling constants entering the regular contribution 
$k_4 M^4$, which is of the same type as the correction 
$\Delta_{\scriptscriptstyle CD}= 
k_{\scriptscriptstyle CD}M^4$ to the low energy
theorem (\ref{DeltaCD}). As discussed above, corrections of this type are
expected to be very small -- we simply drop the term $k_4 M^4$.
The value of $k_1$ is then fixed by the input $\sigma=45\,\mbox{MeV}$
for the total,  so that we can now discuss the manner in which $\sigma$
changes when the quark masses are varied. 

\begin{figure}

\leavevmode   
\psfrag{x}{$M^2$}
\psfrag{y}{$\sigma$}
\psfrag{z}{\raisebox{-0.3em}{$\uparrow$}}

\centering

\includegraphics[width=8cm]{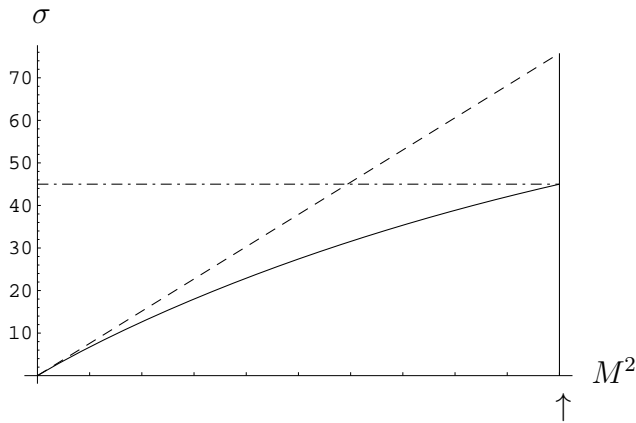}
\caption{ $\sigma$-term (in MeV) as a function of the quark
  masses. The horizontal axis gives the values of $M^2=(m_u+m_d)B$.  
The arrow corresponds to the physical value of the quark
masses.  For definiteness, 
it is assumed that the physical value of $\sigma$ is 45 MeV (dash-dotted
  line). The dashed line depicts the linear dependence that
results if the infrared singular 
contributions proportional to $M_\pi^3$ and to $M_\pi^4 \ln
M_\pi^2/\m^2$ are dropped. }
\end{figure}

At leading order, $\sigma$ is given by  $k_1 M^2$. In figure 1 
this contribution is shown as a dashed straight line. The full curve 
includes the contributions generated by the infrared singularities,
$k_2 M^3$ and $k_3 M^4\{2 \ln M^2/m^2+1\}$. 
The figure shows that the expansion of the 
$\sigma$-term in powers of the quark masses contains large
contributions from infrared singularities. These must show up in evaluations
of the $\sigma$-term on a lattice: The ratio $\sigma/(m_u+m_d)$ must change
significantly if the quark masses are varied from the chiral limit to their
physical values. Note that in this discussion, 
the mass of the strange quark mass is kept fixed at
its physical value -- the curvature seen in the figure exclusively arises
from the perturbations generated by the two lightest quark masses.

It is instructive to compare this result with the dependence of $M_\pi^2$ on
the quark masses. In that case, the expansion only contains even powers of
$M$:
\bea M_\pi^2 =M^2 +\frac{M^4}{32 \pi^2 F^2}\ln\frac{M^2}{\Lambda_3^2} + 
O(M^6)\fs\eea
The quantity $\Lambda_3$ stands for the renormalization group invariant scale
of the effective coupling constant $l_3$. The SU(3) estimate for this
coupling constant given in ref.~\cite{GL 1984} reads $\bar{l}_3\equiv -\ln
M_\pi^2/\Lambda_3^2 =2.9\pm 2.4$. The error bar is so large that the estimate 
barely determines the order of magnitude 
of the scale $\Lambda_3$.
Figure 2 shows, however, that this uncertainty does not significantly affect 
the dependence of $M_\pi^2$ on the quark masses, because the logarithmic
contribution is tiny: the range of $\bar{l}_3$ just quoted
corresponds to the shaded region shown in the figure.

\begin{figure}

\leavevmode   
\psfrag{x}{$M^2$}
\psfrag{y}{$M_\pi^2$}
\centering

\includegraphics[width=8cm]{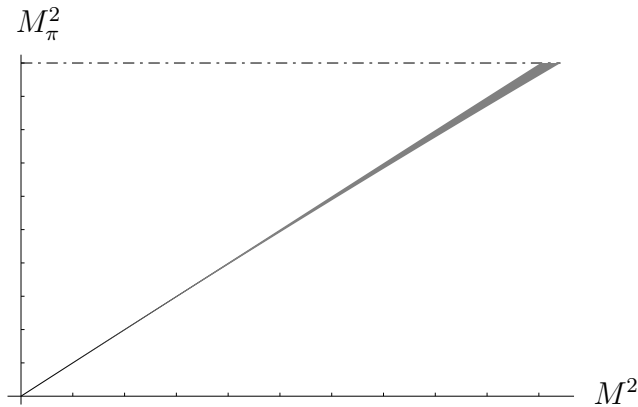}

\caption{Square of the pion mass as a function of the quark masses.
The horizontal axis gives the values of the leading order term 
$M^2=(m_u+m_d)B$. The
dash-dotted line indicates the physical value of $M_\pi^2$.}
\end{figure}

The logarithmic term occurring in the chiral expansion of $M_\pi^2$
gets enhanced by about a factor of two if
we consider the pion $\sigma$-term,  
\bea\sigma_\pi=\langle\pi|\,
m_u\,\bar{u}u+m_d\,\bar{d}d\,|\pi\rangle=
m_u\frac{\partial M_\pi^2}{\partial m_u}+ m_d
\frac{\partial M_\pi^2}{\partial m_d}\fs \eea 
We do not show the corresponding curve, because it is also nearly a straight
line. 

The main point here is that
the infrared singularities encountered in the self energy of the 
nucleon are much stronger than those occurring in the self energy of the pion.

\setcounter{equation}{0}
\section{Goldberger-Treiman relation}
\label{Goldberger-Treiman}
The above expressions for the quantities $\gpiN$, $\gA$, $\m$ and
$F_\pi$ allow us to analyze the correction that occurs in the
Goldberger-Treiman relation up to and including terms of order $M^3$.
We write the relation in the form \bea \gpiN=\frac{\gA \m} {F_\pi}\{1+
\Delta_{\scriptscriptstyle GT}\}\fs\eea If the quark masses $m_u$ and
$m_d$ are turned off, the strength of the $\pi N$ interaction is fully
determined by $\gA$ and $F_\pi$: $\Delta_{\scriptscriptstyle GT}=0$.
Inserting the formulae of the preceding section in the ratio $\gA
\m/F_\pi$ and expanding the result to order $M^3$, we find that the
coefficients of the infrared singular terms proportional to $M^2\ln
M^2/m^2$ and to $M^3$ are identical with those in $\gpiN$. Up to and
including third order, the correction $\Delta_{\scriptscriptstyle GT}$
is therefore free of infrared singularities and takes the simple form
\bea \Delta_{\scriptscriptstyle GT}=-\frac{2\,
d_{18}\,M^2}{g}+O(M^4)\fs\eea As far as the logarithmic terms of
$O(M^2)$ are concerned, this result was established a long time ago
\cite{gss}. What 
the present calculation adds is that the infrared singularities also
cancel at $O(M^3)$, in agreement with ref.~\cite{Fettes:2000xg}.

The result shows that in the case of the Goldberger-Treiman relation,
the breaking of chiral symmetry generated by the quark masses does not
get enhanced by small energy denominators.  Chiral symmetry does not
determine the magnitude of the coupling constant $d_{18}$. A crude
estimate may be obtained from the assumption that the scale of the
symmetry breaking is the same as in the case of $F_K/F_\pi$, where it
is set by the massive scalar states, $M_{\ind S}\simeq 1\,\mbox{GeV}$.
This leads to $\Delta_{\scriptscriptstyle GT}\simeq M_\pi^2/M_{\ind
S}^2\simeq 0.02$.  A more detailed analysis based on models and on the
SU(3) breaking effects seen in the meson-baryon coupling constants may
be found in ref.~\cite{Dominguez Gensini Thomas}. Recently the
Goldberger-Treiman discrepancy has also been determined from QCD sum
rules \cite{Nasrallah:2000fw}. Both evaluations confirm the
expectation that $\Delta_{\scriptscriptstyle GT}$ must be very small
-- a discrepancy of order 4\% or more would be very difficult to
understand.

Since the days when the Goldberger-Treiman relation was discovered
\cite{Goldberger Treiman},
the value of $\gA$ has increased considerably. Also, 
$F_\pi$ decreased a little, on account of radiative corrections. The main
source of uncertainty is $\gpiN$. The comprehensive analysis of $\pi N$
scattering published by
H\"ohler in 1983 \cite{Hoehler} led to 
$f^2=\gpiN^2 M_\pi^2/(16 \pi \m^2)=0.079$. 
With $\gA=1.267$ and $F_\pi=92.4 \,\mbox{MeV}$, this value yields
$\Delta_{\scriptscriptstyle GT}=0.041$.
The data accumulated since then
indicate that $f^2$ is somewhat smaller, 
numbers in the range from 0.076 to 0.077 looking more likely. This range
corresponds to $0.021<\Delta_{\scriptscriptstyle GT}<0.028$. 

We conclude that, within the current experimental uncertainties to be attached
to the pion-nucleon coupling constant, the Goldberger-Treiman relation does
hold to the expected accuracy. Note that at the level of 1 or 2 \%,
isospin breaking cannot be ignored. In particular, 
radiative corrections need to be analyzed carefully. Also, the coupling
constant relevant for the neutral pion picks up a significant contribution
from $\pi^0-\eta$ interference.

\setcounter{equation}{0}
\section{Low energy theorems for {\boldmath $D^+ $\unboldmath} and 
{\boldmath $D^- $\unboldmath} }
\label{LET}
The representation of the scattering amplitude to order $q^4$ also
allows us to analyze the corrections occurring in the well-known
low energy theorem \cite{Cheng Dashen,Brown Pardee Peccei} for the
value of the scattering amplitude $D^+(\nu,t)$ at the Cheng-Dashen
point, $\nu=0$, $t= 2 M_\pi^2$. The theorem relates the
quantity\footnote{At the Cheng-Dashen-point, the amplitude is
singular, on account of the Born term. The bar indicates that this
term is removed in the pseudovector form specified in
eq.~(\ref{eq:pvBorn}): $\bar{D}^\pm=D^\pm-D^\pm_{pv}$,
$\bar{B}^\pm=B^\pm-B^\pm_{pv}$.}  \bea\label{def Sigma} \Sigma=F_\pi^2
\bar{D}^+(0,2\M^2) \eea to the scalar form factor of the nucleon, \bea
\langle N'|\,m_u \,\bar{u}u+m_d\,\bar{d}d\,|N\rangle=
\sigma(t)\,\bar{u}' u\fs\eea The relation may be written in the form
\bea\label{CD} \Sigma= \sigma(2M_\pi^2)+\Delta_{\scriptscriptstyle
CD}\fs\label{DeltaCD}\eea The theorem states that the term
$\Delta_{\scriptscriptstyle CD}$ vanishes up to and including
contributions of order $M^2$.  The explicit expression obtained for
$\Sigma$ when evaluating the scattering amplitude to order $q^4$ again
contains infrared singularities proportional to $M^3$ and $M^4\ln
M^2/m^2$. Precisely the same singularities, however, also show up in
the scalar form factor at $t=2M_\pi^2$, so that the result for
$\Delta_{\scriptscriptstyle CD}$ is free of such
singularities\footnote{The cancellation of the terms of order $M^3$
was pointed out in ref.~\cite{gss,Pagels Pardee} and the absence of
logarithmic contributions of order $M^4$ was shown in
ref.~\cite{BKM}.}: \bea \Delta_{\scriptscriptstyle
CD}=\left(-\frac{c_2}{2\,m^2}-2\, e_1-2\, e_2+e_3 +2\, e_5+4\,
e_8\right) M^4+O(M^5) \fs\eea Equally well, we could replace the bare
coupling constants by the renormalized ones: In the combination
occurring here, the poles at $d=4$ contained in $e_1,\ldots\,,e_8$
cancel. Crude estimates like those used in the case of the
Goldberger-Treiman relation indicate that the term
$\Delta_{\scriptscriptstyle CD}$ must be very small, of order 1
MeV.

Unfortunately, the experimental situation concerning the magnitude
of $D^+$ at the Cheng-Dashen point leaves much to be desired (for a recent
review, see ref.~\cite{Sainio}). The low energy
theorem makes it evident 
that we are dealing with a small quantity here -- the object
vanishes in the chiral limit. The
inconsistencies among the various data sets available at low energies
need to be clarified to arrive at a reliable value for $\gpiN$. Only then
will it become possible to accurately measure 
small quantities such as the $\sigma$-term.

There is a low energy theorem also
for the isospin odd amplitude: In the
chiral limit, the quantity 
\bea\label{eq:defC} C=2 \,F_\pi^2\,\frac{\bar{D}^-(\nu,t)}{\nu}\, 
\rule[-1em]{0.03em}{2.5em}_{\hspace{0.3em}\nu\, =\,
0,\hspace{0.4em}t\,=\, 2M_\pi^2}\eea
tends to 1. The representation of the scattering amplitude to $O(q^4)$
yields the corrections up to and including $O(M^3)$:
\bea\label{eq:C} C \al=\al 1 -\frac{1+5\, \gA^2}{24\,\pi^2 F^2}\;M^2\, \ln \!
\frac{M}{\mu}+ k_1\,M^2+k_2\, M^3 +O(M^4)\co
\\
k_1\al=\al 16\,d_5^r(\mu)-\frac{(14-3\,\pi)+(22+3\,\pi)\gA^2}{288\,\pi^2
  F^2}\co\no
k_2\al=\al
\frac{64\,m\,c_1+2\,\gA^2(4+\gA^2)+\sqrt{2}\,\ln
  (1+\sqrt{2})\,\gA^2}{32\,\pi\,m\, F^2}\fs\nonumber\eea
Note that the scale dependence of the renormalized coupling constant
$d_5^r(\mu)$ cancels against the one of the chiral logarithm.

In contrast to the quantity $\Delta_{\ind CD}$, this expression does contain infrared singularities at the order considered here: A chiral
logarithm as well as terms of order $M^3$.  We will discuss the size
of these effects in section \ref{Chiral symmetry} -- first, we need to
establish contact with the available phenomenological representations
of the scattering amplitude.

\setcounter{equation}{0}
\section{Subthreshold expansion}
\label{Subthreshold expansion}
H\"ohler
 and collaborators \cite{Hoehler} analyze the low energy structure in
terms of an expansion of the amplitude 
around the point $\nu=t=0$. More precisely, 
the expansion concerns the difference
between the full amplitude and the pseudovector Born term, specified in 
eq.~(\ref{eq:pvBorn}). 
Crossing symmetry implies that 
the amplitudes $X\in\{D^+,D^-/\nu,B^+/\nu,B^-\}$ are even in $\nu$, so that the
expansion takes the form 
\bea X(\nu,t)=X_{pv}(\nu,t)+ x_{00}
+x_{10}\,\nu^2 +x_{01}\,t+x_{20}\,\nu^4+x_{11}\,\nu^2
t+x_{02}\,t^2+\ldots\nonumber
\eea

At leading order, only the tree graphs from ${\cal L}^{(1)}_{\ind N}$
contribute:
\bea d^+_{00}=O(M^2)\co\hspace{2em} d^-_{00}=\frac{1}{2F^2}+O(M^2)\co 
\hspace{2em}b^-_{00}=
\frac{1}{2F^2}+O(1)\fs\nonumber\eea
The first two relations may be viewed as a variant of Weinberg's predictions
for the $S$-wave scattering lengths -- the two low energy
theorems discussed in section \ref{LET} represent yet another version of these
relations. The third does not contain
significant information, because the coefficient $b^-_{00}$ picks up
contributions of the same order also from ${\cal L}^{(2)}_{\ind N}$ -- chiral
symmetry does predict the values of $d^+_{00}$ and $d^-_{00}$ for $m_u=m_d=0$, 
but does not constrain $b^-_{00}$.

The representation of the scattering amplitude to $O(q^4)$ is fully determined
by the coupling constants of the effective theory. Accordingly, we can
calculate the various coefficients in terms of these, to 
the corresponding order of the chiral expansion. 
It suffices to 
expand the loop integrals first around $\nu=t=0$ and then in
powers of $M$. For $d^-_{00}$, for instance, the calculation 
yields
\bea 
d_{00}^-\al
=\al\frac{1}{2\F^2}\left[1+\M^2\left\{\rule[-1em]{0em}{2em}\right.\! 8\,(
      d_1+d_2 + 2\,d_{5} )  +\frac{4\,
     \gA^4}{3\,F_\pi^2}\left(\lambda_\pi+ \frac{1}{32\,\pi^2}\right)\!\!
\left.\rule[-1em]{0em}{2em}\right\} \right.
\no\al\al  - 
  \left.\M^3\,\left\{ \frac{8 + 
          12\,\gA^2 + 11\,\gA^4}{64\,
          \pi \,\F^2\,\m} - 
      \frac{4\,c_1 + 
         \left( c_3 - c_4 \right) \,
          { \gA^2}}{2\,\pi \,\F^2} \
     \right\}+O(\M^4)\;\right]\fs\nonumber\eea
We have expressed the
result in terms of the physical values of $F_\pi$, $\M$ and $\gA$. 
The formula represents an exact result: Up to and including terms of 
$O(m^\frac{3}{2})$, the expansion of $d_{00}^-$ in powers of the
quark masses is fixed by the coupling constants of the effective Lagrangian
specified in section \ref{eL}. In particular, the result shows that this
expansion contains a chiral logarithm at order $\M^2$, as well as a term
of $O(\M^3)$.
Similar formulae can be given also for the other subthreshold
coefficients. These are listed in appendix \ref{Subthreshold coefficients}. 
Note that, for some of the coefficients, the
expansion starts with $\M^{-1}$, indicating that those
coefficients diverge in the chiral limit, on account of the infrared
singularities required by unitarity.

In the loop integrals, the coupling constants
$c_1,\,\ldots\,,c_4$ are needed only to leading order. As discussed in detail
in ref.~\cite{Becher Leutwyler 1999}, we may invert the  
relations for $d_{00}^+,d_{10}^+,d_{01}^+,b_{00}^-$  
and represent these couplings in terms of the subthreshold coefficients:
\bea\label{cd} c_1\al=\al
-\frac{F_\pi^2}{4M_\pi^2}\,(d_{00}^++2\,M_\pi^2\,d_{01}^+)+O(\M) 
\co\hspace{1em}
 c_2= \frac{F_\pi^2}{2}\,d_{10}^++O(\M)\co\\
c_3\al=\al -F_\pi^2\, d_{01}^++O(\M)\co\hspace{7.7em}
c_4=\frac{F_\pi^2}{2\,\m}\,b_{00}^--\frac{1}{4\m}+O(\M)\fs\nonumber\eea
Apart from the Born term, the tree graph contributions to the invariant
amplitudes represent polynomials
in the momenta. The coefficients of these polynomials involve precisely
the same combinations of effective coupling constants as the corresponding
subthreshold coefficients -- we may trade one set of parameters for the other:
Up to and including $O(q^4)$, the scattering amplitude 
may equally well be expressed in terms of subthreshold coefficients.

\setcounter{equation}{0}
\section{Chiral symmetry}
\label{Chiral symmetry}

When parametrizing the scattering amplitude in terms of the subthreshold
coefficients, part of the information is lost: Chiral symmetry imposes 
constraints on these coefficients.
In order to identify the constraints, it is useful to split the effective
Lagrangian into 
a chirally symmetric part, ${\cal L}_s$, and a remainder, ${\cal L}_{sb}$, 
that collects the symmetry breaking vertices. The leading term 
${\cal L}_{\ind N}^{(1)}$ does not contain a symmetry breaking piece. In
${\cal L}_{\ind N}^{(2)}$, the term proportional to 
$c_1 \langle \chi_+\rangle$ belongs to ${\cal L}_{sb}$, while those involving
$c_2,c_3,c_4$ survive in the chiral limit and hence
belong to ${\cal L}_s$. At order $q^3$, the coupling constants
$d_5, d_{16}, d_{18}$ represent symmetry breaking effects.
At $O(q^4)$, we did not specify the
Lagrangian, but instead wrote down the corresponding contributions to the
scattering amplitude, in eq.~(\ref{eq:L4}). Concerning $D^+$, only those terms 
in ${\cal L}^{(4)}_{\ind N}$ that contain four
derivatives of the meson field belong to the symmetric part. In view of
$2\,q\cdot q' =2M_\pi^2-t$ and $4\,\m\, \nu=(P+P')\cdot(q+q')$, 
the expressions $(t-2M_\pi^2)^2$, $(t-2M_\pi^2)\,\nu^2$ and $\nu^4$ are of
this type. Hence the coupling constants $e_6$, $e_7$, $e_8$,
belong to the chirally symmetric part of the Lagrangian, while the 
combinations $e_3-4e_8$,  $e_4+2e_7$ and $e_5+4e_8$ represent symmetry
breaking terms. For the same reason, $e_{10}$, $e_{11}$ belong to ${\cal
  L}_s$, while the combination $e_9+2e_{11}$ controls the symmetry breaking 
in the amplitude $B^-$. The coupling constants $e_1$ and $e_2$, that occur in 
the chiral expansion of the scalar form factor at first non--leading order,
also belong to the symmetry
breaking part of the Lagrangian. 

If the quark masses are
turned off, only the coupling constants of ${\cal L}_s$ contribute. 
One readily checks that, in that limit,
the subthreshold coefficients listed in appendix \ref{Subthreshold
  coefficients} represent linearly independent combinations of the symmetric
coupling constants, except for two constraints: $d_{00}^+$ vanishes and 
$d_{00}^-$ is determined by $F$ -- precisely the two low energy theorems
encountered already at leading order. We conclude that, in addition to the
Goldberger-Treiman relation, chiral symmetry only imposes
two conditions on the subthreshold coefficients. 

As noted above, we can express the amplitude in terms of subthreshold
coefficients instead of coupling constants, using the relations (\ref{cd}) to
eliminate $c_1,\ldots,c_4$ in the contributions from the loops. For the value
of $\F^2D^+$ at the Cheng--Dashen point, we then obtain
\bea\label{eq:Sigma} 
\Sigma \al=\al F_\pi^2\,(\kappa_1 \,d^+_{00}+2\,\kappa_2 \M^2
d^+_{01}+4\M^4 d^+_{02}+
\kappa_3\M^2  d^+_{10})+\Delta^+\co\no
\kappa_1\al=\al 1+\frac{19\M^2}{120\,\pi^2\F^2}-\frac{3\M^2}{64\, \pi \F^2}
\co \\
\kappa_2\al=\al 1+\frac{19\M^2}{96\,\pi^2\F^2}-\frac{3\M^2}{64\, \pi \F^2}\co\nonumber
\eea
\bea
\kappa_3\al=\al \frac{19\M^2}{960\,\pi^2\F^2}-\frac{\M^2}{128\, \pi\F^2}\co\no
\Delta^+\!\al=\al \frac{37\,\gA^2\,\M^3}{3840\,\pi\F^2}-
\frac{11\,\gA^2\M^4}{80\,\pi^2\,\m\F^2}+\frac{3\,\gA^2\M^4}{128\,\pi\,\m\F^2} + O(\M^5)
\fs\nonumber\eea
The only infrared singularity occurring in this relation is the term
proportional to $\M^3$. Since the coefficient of this term is very small,
the contributions from the loop graphs are tiny: Numerically, we find
$\kappa_1=1.003$, $\kappa_2=1.012$, $\kappa_3=-0.001$ and 
$\Delta^+=1.1\,\mbox{MeV}$.
Inserting the values of the subthreshold coefficients from
ref.~\cite{Hoehler}, the relation yields $\Sigma=61\,\mbox{MeV}$, in perfect
agreement with the value given in that reference.

We conclude that, in the present case, the expansion is rapidly convergent.
According to section \ref{LET}, the term $\Sigma$ nearly coincides with
the value of
the scalar form factor at $t=2\M^2$. The difference between $\sigma(2\M^2)$ 
and $\sigma\equiv \sigma(0)$ is well understood -- the evaluation
within chiral perturbation 
theory \cite{Becher Leutwyler 1999} confirms the result of the
dispersive calculation described in ref.~\cite{Gasser Leutwyler Sainio}:
\bea \sigma(2\M^2)-\sigma(0)=15.2\pm 0.4\,\mbox{MeV}\fs\eea 
The relation (\ref{eq:Sigma}) may thus be rewritten in the form
\bea \sigma= F_\pi^2\,(d^+_{00}+2\, \M^2
d^+_{01}+4\M^4 d^+_{02})-\sigma_1
\fs\eea
Evaluating the small correction terms with the subthreshold
coefficients of ref.~\cite{Hoehler}, we obtain
\bea \sigma_1=13\pm 2\,\mbox{MeV}\co\eea 
where the error is our estimate for the uncertainties arising
from the correction $\Delta_{\ind CD}$, as well as from the higher order
contibutions, which our calculation neglects. 

For the isospin odd amplitude $D^-$, the term analogous to $\Sigma$ is 
the quantity $C$, defined in eq.~(\ref{eq:defC}). The representation in
terms of the subthreshold coefficients is of the form
\bea\label{eq:C2} C\al=\al 2\F^2\,(d^-_{00}+2\M^2
d^-_{01})+\Delta^-\co\eea
where the correction $\Delta^-$ is given by
\bea \Delta^-\al=\al \frac{(5\gA^2-2)\,\M^2}{72\,\pi^2 \F^2}-
\frac{(\gA^2-1)\,\M^2}{96\,\pi
  \F^2}-\frac{\gA^2(3\gA^2+22)\,\M^3}{192\,\pi\,\m \F^2}\no\al\al+
\frac{\gA^2\,\M^3}{16\sqrt{2}\,\pi\,\m\,\F^2}\,\ln(1+\sqrt{2})
+ O(\M^4)
\fs\nonumber\eea
The infrared singular terms proportional to $M_\pi^3$ are small, 
of order $0.03$. Moreover, they nearly cancel against those of
$O(M_\pi^2)$, so that the net correction from the one loop graphs is tiny, $\Delta^- = - 0.003 + O(M_\pi^4)$. The contributions of order
$M_\pi^4$ are more important. In particular, the $\Delta$ resonance
generates a term of this order, which is enhanced by the third power
of the corresponding energy denominator (note that the resonance also
contributes to the coefficients $d_{00}^-$ and $d_{01}^-$): 
\bdm \Delta^-_\Delta= \frac{8 F_\pi^2
g_\Delta^2 M_\pi^4}{9\,\m (m_\Delta-\m)^3 }\fs\edm
Numerically, this term amounts to $0.02$.\footnote{ The
corresponding contribution to $\Delta^+$ is even smaller, of order 
$0.1\,\mbox{MeV}$.} The dispersive analysis of
ref.~\cite{Hoehler} confirms that the resonance contribution dominates. 
In our opinion, the estimate  
\bea\label{eq:Deltaminus} \Delta^-=0.02\pm0.01 \eea
covers the uncertainties. We conclude that the relation (\ref{eq:C2})
very accurately determines the value of the isospin
odd amplitude at the Cheng-Dashen point, in terms of the
subthreshold coefficients $d_{00}^-,d_{01}^-$.

Unfortunately, our knowledge of the coefficient $d_{00}^-$ is not
satisfactory. The result obtained for this quantity is very sensitive
to the value used for $\gpiN$: The difference $d_{00}^--\gpiN^2/2\m^2$
is relatively well known, as it is given by an integral over a total
cross section
\begin{equation}
d_{00}^- - \frac{\gpiN^2}{2\m^2}=\frac{2}{\pi}\int_{\M}^\infty
\frac{d\omega}{\omega^2}\,k\, \sigma^-(\omega)\fs 
\end{equation}
If the integral is taken known, a shift in $f^2=\gpiN^2\M^2/16
\pi\m^2$ from the value $f^2 = 0.079$ \cite{Hoehler} to $f^2 = 0.076$
\cite{SP98} lowers the result for $d_{00}^-$ by $0.08\M^{-2}$, so that
$C$ decreases by $0.07$. Indeed, the recent evaluations of the
subthreshold coefficients by Stahov \cite{Stahov} and Oades
\cite{Oades} confirm that the value of $d_{00}^-$ is lower for the
partial wave analyses of the VPI/GW group than for the Karlsruhe
analyses. For a precise determination of the value of the isospin odd
amplitude at the Cheng-Dashen point, the discrepancy in the value of
the pion nucleon coupling constant needs to be resolved.
\begin{table}
\begin{center}
\begin{tabular}{|cc|c|c|c|c|}\hline
    && $d_{00}^-$ & $d_{01}^-$ & $C$ &$16\,\M^2\, d^r_5(\mu)$ \\ \hline 
KH80&\hspace{-0.7em}\cite{Hoehler}& 1.53\phantom{0} $\pm$
0.02\phantom{0} &$ -0.134$ $\pm$ 0.005 & 1.13 $\pm$ 0.02
& $0.04$ $\pm$ 0.03 \\ \hline
KA84&\hspace{-0.7em}\cite{Stahov} & 1.510 $\pm$ 0.001 & $-0.136$ $\pm$
0.003 & 1.11 $\pm$ 0.01 & $0.03$ $\pm$ 0.01 \\
\hline SP98&\hspace{-0.7em}\cite{Stahov} & 1.468 $\pm$ 0.004 &
$-0.138$ $\pm$ 0.003 & 1.06 $\pm$ 0.01 &
$0.00$ $\pm$ 0.01 \\ \hline
KH80&\hspace{-0.7em}\cite{Oades} & 1.53\phantom{0} $\pm$
0.02\phantom{0} & $-0.14\phantom{0}$ $\pm$ 0.03\phantom{0} & 1.12 $\pm$
0.06 & 0.04 $\pm$ 0.06 \\ \hline
SP98&\hspace{-0.7em}\cite{Oades} &
 1.46\phantom{0} $\pm$ 0.01\phantom{0} & $-0.12\phantom{0}$ $\pm$ 0.01\phantom{0} & 1.09 $\pm$ 0.02 & 0.06 $\pm$ 0.02 \\ \hline
\end{tabular}
\end{center}
\caption{\label{tab:C}Value of the isospin-odd amplitude at the 
Cheng-Dashen point. The results for $C$ are obtained with 
eqs.~(\ref{eq:C2}) and (\ref{eq:Deltaminus}). 
The value of $d_5^r(\mu)$ follows from eq.~(\ref{eq:C}) and refers to the
running scale $\mu=1\, \mbox{GeV}$.} 
\end{table}

We can now check whether the phenomenological information is
consistent with the low energy theorem (\ref{eq:C}), which predicts
the value of $C$ to order $\M^3$, except for the contribution
$C_{sb}=16\,M_\pi^2\,d_5^r(\mu)$ from the symmetry breaking coupling
constant $d^r_5$.  For the reasons given earlier, we expect
contributions of this type to be very small, of the order of a few per
cent. This is confirmed by the numerical results listed in the last
column of table \ref{tab:C} which are obtained by comparing
eqs.~(\ref{eq:C}) and (\ref{eq:C2}) -- the relation (\ref{cd}) is used
to express $c_1$ in terms of subthreshold coefficients, the running
scale $\mu$ is set equal to $1\,\mbox{GeV}$ and contributions of order
$\M^4$ or higher are dropped.

As mentioned above, the Weinberg predictions for the two $S$-wave
scattering lengths represent yet another version of the two energy
theorems under discussion. These predictions only hold to leading
order of the chiral expansion, but the one loop representation of the
amplitude allows us to analyze the corresponding corrections -- it
suffices to evaluate our representation of the amplitudes
$D^\pm(\nu,t)$ at $\nu=\M$, $t=0$.  The expansion of the result in
powers of $\M$ then yields an explicit expression for the $S$-wave
scattering lengths, valid up to and including $O(\M^4)$. In fact, this
calculation has been done already, in the framework of HBCHPT
\cite{Fettes:2000xg}, and we have verified that our method yields the
same result (note that the results given in Appendix A of
\cite{Fettes:2000xg} contain typographical errors).

In the case of the scattering lengths, however, the corrections are by no
means small -- the 
first few terms of the chiral expansion 
only yield a rather poor approximation. This is in
marked contrast to the case of $\pi\pi$ scattering, where the chiral expansion
of the scattering lengths converges rather rapidly.
The main reason is that the distance between the Cheng-Dashen point and
threshold is of order $\M$ and is therefore much larger than for $\pi\pi$
scattering, where the analog of the Cheng-Dashen point occurs at
$s=u=\M^2$, $t=2\M^2$, so that the distance to threshold is of order $\M^2$.
Accordingly, the expansion parameter is larger -- we cannot
expect the chiral 
representation at threshold to have the same accuracy as in the case of
$\pi\pi$ scattering.  
The problem is accentuated by the fact that the threshold sits on top of the
branch point required by unitarity --
when considering the scattering lengths, we are in effect
analyzing the amplitude at a singular point. In this respect,
the situation is comparable with the case of $\pi\pi$ scattering, where the
$S$-wave scattering lengths also contain chiral logarithms with a large
coefficient. As shown in ref.~\cite{CGL}, the expansion of the $\pi\pi$
scattering amplitude in the subthreshold region
converges much more rapidly than at threshold. 

In the variants of the low energy theorems discussed
above, these problems did not occur, because we have been comparing the
properties of the amplitude at the Cheng--Dashen point, $s=u=\m^2$, $t=2\M^2$,
with those at $s=u=\m^2+\M^2$, $t=0$. In this region, the amplitude does not
contain branch points and the
relevant distance is small, of order $\M^2$.
This is why the corrections encountered turned out to be rather small. 

\setcounter{equation}{0}
\section{Cuts required by unitarity}
\label{Cuts}
\begin{figure}[ht]\centering
\includegraphics[width=0.8\textwidth]{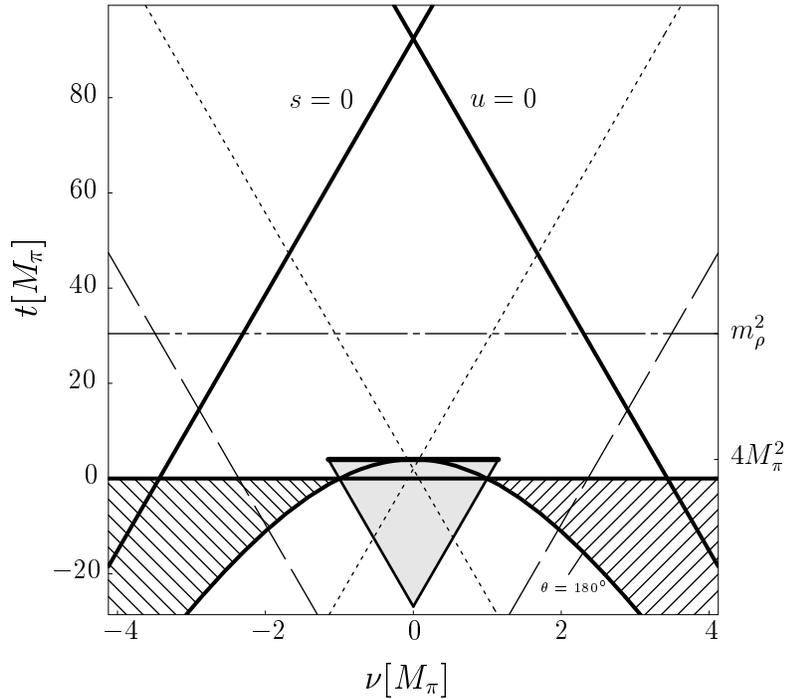}
 \caption{ Mandelstam plane. Hatched: Physical regions
   of $\pi N$ scattering. In the shaded region $s<(\m+\M)^2$,
   $u<(\m+\M)^2$ and $t<4\M^2$ the amplitude is real and analytic, once the
   poles at $s=\m^2$ and $u=\m^2$ (thin lines) are subtracted. 
   Dotted: Nucleon poles. Dashed: $\Delta$-resonance.
   Dot-dashed: $\rho$-resonance }\label{fig:mandelstam}
\end{figure}
In the following sections,
we examine the energy dependence of the loop graphs
in some detail, in order to clarify the problems encountered
when extrapolating the chiral representation to threshold or even beyond. 
For this purpose, we first study the analytic structure of the loop graphs.

In addition to the nucleon pole terms, the scattering
amplitude contains branch points 
at $s=(\m+\M)^2$, $u=(\m+\M)^2$ and $t=4M_\pi^2$. 
These correspond to the sides of the shaded triangle shown in
fig.~\ref{fig:mandelstam}.
In the chiral representation of the amplitude,
the corresponding cuts are described by the loop integrals. Crossing
symmetry relates the discontinuities in the $u$-channel to those in the
$s$-channel, so that we only need to discuss the $s$- and $t$-channel
imaginary parts.

We recall that we are absorbing the self energy
insertions due to the tree graphs of ${\cal L}^{(2)}_{\ind N}$ and  
${\cal L}^{(4)}_{\ind N}$ by replacing the bare nucleon mass with $m_4= m-4c_1
M^2+e_1\, M^4$. As far as the loop integrals are concerned, 
the difference between $m_4, M$ and the physical masses $\m, \M$ is beyond
the accuracy of our calculation. In the following, we represent
these integrals in terms of the physical masses.

\subsection[$s$-channel cuts]{{\boldmath$s$\unboldmath}-channel cuts}
The branch cut at $s> (\m+M_\pi)^2$ is an immediate consequence of the
unitarity condition,
\begin{equation}\label{eq:unitarity}
\langle f|{\bf T}|i\rangle-\langle f|{\bf T}^\dag|i\rangle=i\sum_{\{n\}}\langle
f|{\bf T}^\dag |n\rangle\langle n|{\bf T}|f\rangle \fs
\end{equation}
At one loop level only $\pi N$ intermediate states enter this
relation -- inelastic channels only show up at higher
orders.
The condition thus exclusively involves the $s$-channel partial wave amplitudes
\bea f^I_{l\pm}(s)=q^{-1}\exp i\,\delta^I_{l\pm}(s)\sin
\delta^I_{l\pm}(s)\fs\eea 
The lower index specifies the orbital angular momentum $l$ as well as the
total angular momentum $J=l\pm \frac{1}{2}$. The upper index refers to the
isospin $I=\frac{1}{2},\frac{3}{2}$ and  $q$ stands for the c.m. momentum, 
\bea q^2=\frac{1}{4 s}\,\{s-(\m+M_\pi)^2\}\{s-(\m-M_\pi)^2\}\fs\eea
In this normalization, elastic unitarity takes the form 
\bea\label{elastic unitarity}
\im\,f_{l\pm}^{I}(s)= q\, | f_{l\pm}^{I}(s) |^2 \fs\eea

The low energy expansion of the partial waves starts at $O(q)$.
The relation (\ref{elastic unitarity}) implies that an 
imaginary part starts showing up only at $O(q^3)$ -- the 
tree graphs generated by ${\cal L}_{\ind N}^{(1)}$ and
${\cal L}_{\ind N}^{(2)}$ thus suffice to evaluate the 
imaginary parts of the amplitude to $O(q^4)$. In agreement with the
chiral counting rules, the imaginary parts generated by the square of the
contributions due to ${\cal L}^{(2)}_{\ind N}$ are of $O(q^5)$ and we can
drop these.

We have checked that our representation of the scattering amplitude indeed
obeys elastic unitarity. In this respect, there is no difference
between infrared and dimensional regularization -- the $s$-channel
discontinuities are identical \cite{Becher Leutwyler 1999}. 
The real parts are different, however. To any order in the low energy
expansion, the difference is a polynomial, but the degree of the polynomial
increases with the order.

\subsection[$t$-channel cuts]{{\boldmath$t$\unboldmath}-channel cuts}
Some of the graphs contain a branch cut for $t>4M_\pi^2$.
The corresponding discontinuity is determined by the extended unitarity
relation
\begin{equation}\label{t-channel unitarity}
\im\,f^J_\pm(t)=\left\{1-4M_\pi^2/t\right\}^{\frac{1}{2}}\;
t_{J}^{I}(t)^\star\; 
f^J_\pm(t)\co\;\;\;\;4\M^2<t<16\M^2 \fs
\end{equation}
The quantities $f^J_\pm(t)$ represent the $t$-channel partial waves 
of the $\pi N$ scattering amplitude. The quantum number $J$ stands for the
total angular momentum, while the lower index again refers to the spin
configuration. There is no need to in
addition specify the isospin quantum number, because it is
determined by the total angular momentum:  $J$ even $\rightarrow I=0$,
$J$ odd $\rightarrow I=1$. The functions
\bea t_{J}^{I}(t)=\left\{1-4M_\pi^2/t\right\}^{-\frac{1}{2}}
\exp i\, \delta^I_J(t)\,\sin\delta^I_J(t)\nonumber\eea 
denote the partial wave projections of the
$\pi\pi$ scattering amplitude. 

The low energy expansion of the right
hand side of (\ref{t-channel unitarity}) again starts at $O(q^3)$. 
In our context, the evaluation of this
condition is particularly simple, because the $\pi\pi$ scattering amplitude
is needed only to leading order, where it is a polynomial. 
Hence only the partial waves with $J=0,1$ contribute. These
are given explicitly in appendix \ref{t-channel partial waves}.

We have checked that, in the region $t<4\m^2$, the imaginary parts of
our loop integrals do obey the relation (\ref{t-channel
unitarity}). Note that this relation only accounts for the $\pi\pi$
intermediate states.  States with three or more
pions do not occur at one loop, but outside the low energy region, the
nucleon-antinucleon states yield an additional contribution, which
does show up in the box graph and in the triangle graph,
fig.~\ref{fig:loop}m, through a branch cut for $t>4\m^2$. Because this
singularity is absent at any fixed order of the chiral expansion of
the loop integrals, it is not preserved by infrared
regularization. Indeed one finds that the regular part of the
corresponding loop integrals has a branch cut for $t>4\m^2$ so that
the imaginary part of the amplitudes in infrared regularization agrees
with the one obtained in dimensional regularization only for
$t<4\m^2$.

\subsection{Box graph}\label{Box graph}
The analytic structure of the Box graph (i) is more complicated: The
corresponding loop integrals contain a simultaneous cut in the variables
$s$ and $t$ and thus contribute to the Mandelstam double spectral
function (at one loop level, only the box graph generates such a contribution).

Let us consider the relevant scalar loop integral. In view of the four
energy denominators, this integral does not require
regularization.
It may be represented in terms of the double dispersion
relation 
\bea H_{13}(s,t)\al =\al\frac{1}{\pi^2}\! 
\int_{s_+}^\infty\hspace{-0.5em} ds'
\int_{t_{\ind min}(s')}^\infty \hspace{-2em}dt'\hspace{1em}
\frac{\rho_2(s',t')}{(s'-s-i\epsilon)\,
(t'-t-i\epsilon)}\;\;\fs\nonumber\eea 
The boundary of the integration region is given by
\bea s_\pm=(\m\pm \M)^2\co\hspace{1em}
t_{\ind min}(s)\al=\al\frac{4\,(s-\m^2-2\M^2)\,
(\m^2\, s-(\m^2-\M^2)^2)}{(s-s_+)\,(s-s_-)}
\co\nonumber\eea
and the explicit expression for the double spectral density 
reads\footnote{The expression for $\rho_2(s,t)$
may be read off from the one given for the imaginary
part $\mbox{Im}_s\,H_{13}(s,t)=\mbox{Im}_s\,I_{13}(s,t)$ in appendix
\ref{Imaginary parts}. The square root occurring in that expression becomes
imaginary for $t>t_{\ind min}$. The double spectral density is the 
discontinuity across the cut generated by this square root.}

\bea \rho_2(s,t)\al=\al\frac{1}{8\sqrt{\,t\,
(t-t_{\ind min}(s))\, (s-s_+)\,(s-s_-)}}\;\;\nonumber\fs\eea

In view of $t_{\ind min}(s)> 4 \m^2$, we may expand the integral in a Taylor
series in powers of $t$. Because the singularity
in the variable $t$ lies far outside the low energy region, the higher
order 
terms in $t$ are strongly suppressed -- these contributions are similar to
those we dropped when simplifying the integrands of the 
loop integrals in section \ref{Simplification}. 
The statement also holds in infrared regularization and for the
tensor integrals associated with the box graph.

Using the explicit expressions for the box graph in appendix \ref{Loop
  graphs}, we have determined the order to which the $t$-dependence of the
integrals $I_{13}(s,t)$, $I_{13}^{(1)}(s,t)$, $I_{13}^{(2)}(s,t)$
is needed -- because of cancellations between the various terms occurring in
the representation of the invariant amplitudes in terms of 
these, the order needed cannot 
directly be read off.
Within the accuracy
of our calculation, the box-amplitudes of $B^\pm$ are given by their value at
$t=0$ and the amplitudes $D^\pm$ can be replaced by their first order
expansion in $t$, so that we are allowed to replace the box integrals by their
Taylor expansion in $t$ to first order.

The expanded box integrals can be represented in terms of a spectral function
with a single variable. In the case of the full scalar integral, for instance,
the explicit representation reads: \bea
H_{13}(s,t)\al=\al\frac{1}{\pi}\int_{s_+}^\infty
ds'\,\frac{\rho_1(s')}{s'-s-i\epsilon}\,\left(1+\frac{2\,t}{3\,t_{\ind min}(s)}+O(t^2)\right)\;\co\no
\rho_1(s)\al=\al\frac{\sqrt{(s-s_-)\,(s-s_+)}}{16\pi (s-\m^2-2\M^2)\, (\m^2
  s-(\m^2-\M^2)^2)}\;\;\fs\nonumber \eea
The formula nicely displays the suppression of the $t$-dependent part. Since
$t_{\ind min}(s)$ is a quantity of order $q^0$ it is also evident that the
$t$-dependent part is an $O(q^2)$ correction to $H_{13}(s,0)$.

This shows that, at low energies, the contributions 
generated by the Mandelstam double spectral function can be described
in terms of single variable dispersion integrals. 

\setcounter{equation}{0}
\section{Dispersive representation}
\label{Dispersive representation}
With the above simplification of the box
graph contribution and after the 
reduction of the numerators for the diagrams discussed in section
\ref{Simplification}, all of the loop integrals are of a similar structure:
The discontinuities across the $s$- and $u$-channel cuts are
linear in $t$, while those across the $t$-channel cut are linear in $\nu$. Up
to higher order corrections, the $s$-channel imaginary parts thus have the
form 
\begin{align*}
\mbox{Im}_sD^\pm(s,t)&=\mbox{Im}D^\pm_1(s)+t\,\mbox{Im}D^\pm_2(s)+O(q^5)\co \\
\mbox{Im}_sB^\pm(s,t)&=\mbox{Im}B^\pm_1(s)+O(q^3)\co
\end{align*}  
while the $t$-channel discontinuities have the structure (explicit
expressions for the functions $\mbox{Im}D^+_3(t)$, $\mbox{Im}D^-_3(t)$,
$\mbox{Im}B^-_2(t)$ are given in appendix
\ref{t-channel partial waves})
\begin{align*}
\mbox{Im}_tD^+(\nu,t)&= \mbox{Im}D^+_3(t)+O(q^5)\co &
\mbox{Im}_tD^-(\nu,t)&=\nu\, \mbox{Im}D^-_3(t) +O(q^5)\co\\
\mbox{Im}_tB^+(\nu,t)&=O(q^3)\co &
\mbox{Im}_tB^-(\nu,t)&=\mbox{Im}B^-_2(t)+O(q^3)\fs
\end{align*}

We now make use of the fact that an analytic function that
  asymptotically only grows with a power of the argument is fully
  determined by its singularities, up to a polynomial. This allows us
  to establish an alternative representation of the scattering
  amplitude that also holds to first nonleading order and has two
  advantages as compared to the one in terms of the explicit
  expression for the loop integrals: It exhibits the structure of the
  amplitude in a more transparent manner and only involves the
  expressions for the imaginary parts. Since unitarity determines the
  latter in terms of the tree level amplitude, the dispersive
  representation does not involve any loop integrals.

Explicitly, the dispersive representation reads
\bea D^+(\nu,t)\al=\al
  D^+_{pv}(\nu,t)+D^+_1(s)+D^+_1(u) 
+t\,D^+_2(s)+t\,D^+_2(u)\no
\al\al\hspace{8em}+\;D^+_3(t)+D^+_p(\nu,t)+O(q^5)\co\no
\label{representation to O4}
D^-(\nu,t)\al =\al D^-_{pv}(\nu,t)+D^-_1(s)-
D^-_1(u)+t\,D^-_2(s)-t\,D^-_2(u)\\
\al\al\hspace{8em}+\;\nu\,D^-_3(t)+D^-_p(\nu,t)+O(q^5)\co \no
B^+(\nu,t)\al =\al  B^+_{pv}(\nu,t)+B^+_1(s)-
B^+_1(u)+B^+_p(\nu,t)+O(q^3)\co \no
B^-(\nu,t) \al=\al  B^-_{pv}(\nu,t)+B^-_1(s)+
B^-_1(u)+B^-_2(t)+B^-_p(\nu,t)+O(q^3)\fs\nonumber
\eea
The functions $D^\pm_{pv}(\nu,t),B^\pm_{pv}(\nu,t)$ represent the pseudovector
Born terms of eq.~(\ref{eq:pvBorn}).
The contributions generated by the cuts in the $s$-, $t$- and $u$-channels are
described by the functions $D^+_1(s), D^+_1(u),\,\ldots\,, B^-_2(t)$. 
The dispersion integrals that determine these in terms of the corresponding
imaginary parts are given below. Once the singularities are removed,
the low energy representation of the invariant amplitudes reduces to a 
polynomial. We denote the corresponding contributions by
$D^\pm_p(\nu,t),B^\pm_p(\nu,t)$. 

A representation of the above type was used already by Gasser, Sainio and
\v{S}varc \cite{gss,Gasser:1986pc}.  These authors noted that if the masses
occurring in the loop integrals are identified with the physical masses rather
than the bare quantities occurring in dimensional regularization, the infrared
singularities generated by the cuts are correctly accounted for.  As they did
not make use of any approximations for the loop integrals, the $t$-dependence
of their $s$- and $u$-channel cuts was however more involved. The 
representation (\ref{representation to O4}) shows that, to the accuracy of a
one loop calculation, the  
amplitude may be described in terms of nine functions of a single variable. 
An analogous representation
also holds for the $\pi\pi$ scattering amplitude \cite{Stern}. 
In that case three 
functions of a single variable suffice and the representation is valid even
to two loops, that is modulo contributions of $O(q^8)$.

\setcounter{equation}{0}
\section{Subtracted dispersion integrals}
\label{Dispersion integrals}
We now specify the functions $D^\pm_1(s),\,\ldots\,, B^\pm_2(t)$ as suitably
subtracted dispersion
integrals over their imaginary parts. For this purpose, it is convenient to
replace the variable 
$s$ by the lab energy $\omega$, which is linear in $s$
and independent of $t$ (for $t=0$, $\nu$ coincides with $\omega$), 
\bea
\omega=\frac{s-\m^2-M_\pi^2}{2 \m}\fs\eea 
The number of subtractions is
determined by the asymptotic behavior of the imaginary parts 
and is related to the superficial degree of divergence
$\omega(G)$ of the relevant diagram. This quantity is obtained by adding up the
powers of the loop
momentum $k$ in the region where it is large compared to the masses and to the
external momenta:
$\omega(G)=4+\delta-I_{\ind N}-2\,I_\pi$, where $\delta$ is the number of
derivatives at the vertices and $I_{\ind N}$, $I_\pi$ stand for the number of
nucleon and pion propagators occurring in the loop, respectively. The
vertices from ${\cal L}_{\ind N}^{(1)}$ contain one derivative, those of
${\cal L}_\pi^{(2)}$ contain two. In ${\cal L}_{\ind N}^{(2)}$, there are up
to four derivatives (vertex proportional to $c_2$).  It is important here that
we are considering only diagrams with at most one vertex from ${\cal
  L}^{(2)}_{\ind N}$ -- if the unitarity relation is evaluated with the full
square of the tree level amplitude, the imaginary part grows more rapidly.
Note also that the simplifications discussed in section \ref{Simplification}
reduce the degree of divergence.

The imaginary parts occurring in the tensorial
decomposition of the loop integrals are listed explicitly in appendix
\ref{Imaginary parts}. These expressions show that the
quantities $\mbox{Im}D^\pm_1(s)$ grow with $s^3\ln s$, while
for the parts linear in
$t$, we have $\mbox{Im}D^\pm_2(s)\propto s\ln s$. The leading
contribution stems from the graphs (g) and (h) and is proportional to $c_2$
(if only the loops generated by ${\cal L}^{(1)}_{\ind N}$ are considered, we
instead obtain $\mbox{Im}D^\pm_1(s)\propto s^2$ and
$\mbox{Im}D^\pm_2(s) \propto s$). The asymptotic behavior of the
$B$-amplitudes is given by $\mbox{Im}B^+_1(s)\propto c_2\, s\ln s$,
$\mbox{Im}B^-_1(s)\propto s $.
We subtract at $\omega=0$ and write the $s$-channel dispersion relations in
the form
\bea
\label{disp Ds}
 D^\pm_1(s)\al=\al\frac{\omega^{5}}{\pi}\int_{M_\pi}^\infty
\frac{d\omega'\,\mbox{Im}D^\pm_1(s')} 
{\omega^{\prime\, 5}\,(\omega'-\omega-i\epsilon)}\no
 D^\pm_2(s)\al=\al\frac{\omega^{3}}{\pi}\int_{M_\pi}^\infty
\frac{d\omega'\,\mbox{Im}D^\pm_2(s')} 
{\omega^{\prime\, 3}\,(\omega'-\omega-i\epsilon)}\\
B_1^\pm(s) \al=\al
\frac{\omega^3}{\pi}\int_{M_\pi}^\infty \frac{d\omega'\,\mbox{Im} B_1^\pm(s')}
{\omega^{\prime\, 3}\,(\omega'-\omega-i\epsilon)}
 \fs\nonumber\eea 
The representation (\ref{representation to O4}) is manifestly crossing
symmetric: The contributions generated by the cuts in the $u$-channel are 
are obtained from those due to the $s$-channel discontinuities 
by the substitution $s\rightarrow u$. 

The contributions due to the $t$-channel cuts may be characterized in
the same manner. The dispersion relations take the form
\bea \label{disp Dt}D_3^+(t)\al=\al \frac{t^3}{\pi}\int_{4M_\pi^2}^\infty 
\frac{dt'\;\mbox{Im} D^+_3(t')} {t^{\prime\,
    3}\,(t'-t-i\epsilon)}\co\no 
D^-_3(t)\al=\al
\frac{t^2}{\pi}\int_{4M_\pi^2}^\infty \frac{dt'\;\mbox{Im} D^-_3(t')}
{t^{\prime\, 2}\,(t'-t-i\epsilon)}\co \\ B_2^-(t)
\al=\al\frac{t}{\pi}\int_{4M_\pi^2}^\infty \frac{dt'\;\mbox{Im} B^-_2(t')}
{t^{\prime\, }\,(t'-t-i\epsilon)}\fs\nonumber \eea In the terminology
used in ref.~\cite{Gasser:1984yg}, the contributions associated with
the cuts in the $s$-, $t$- and $u$-channels represent ``unitarity
corrections''.  Together with the explicit expressions for the
imaginary parts of the various graphs given in the appendix, these
relations unambiguously specify the contributions generated by the
cuts.

By construction, the cuts do not
contribute to the first few terms in the Taylor series expansion 
of the amplitude in powers of
$\nu$ and $t$, that is to the leading coefficients of the subthreshold
expansion, which we discussed in section \ref{Subthreshold expansion}. 
The polynomials $D^\pm_p(\nu,t)$ and $B^\pm_p(\nu,t)$ account for the following
terms of this expansion:
\bea D^+_p(\nu,t)\al=\al d_{00}^++d_{10}^+\,\nu^2 +d_{01}^+\,t+d_{20}^+\,\nu^4
+d_{11}^+\,\nu^2 t +d_{02}^+\,t^2\co\no
B^+_p(\nu,t)\al=\al b_{00}^+\,\nu\co \label{eq:Tpoly} \\
D^-_p(\nu,t)\al=\al d_{00}^-\,\nu +d_{10}^-\,\nu^3 +d_{01}^-\,\nu\,t\co\no
B^-_p(\nu,t)\al=\al b_{00}^-+b_{10}^-\,\nu^2 +b_{01}^-\,t\fs\nonumber\eea
At the order of the chiral expansion under consideration, 
the higher coefficients exclusively receive contributions from the
parts of the amplitude that describe the cuts. 

To complete the construction of the dispersive representation, we incorporate
the constraints imposed by chiral symmetry. These are:  
\begin{enumerate}\item[1.] The strength of the
$\pi N$ interaction is determined by $\gA$, up to corrections from the
symmetry breaking part of the effective Lagrangian.
\item[2.] In view of the relations (\ref{cd}),
chiral symmetry and unitarity fix the imaginary parts 
in terms of the subthreshold coefficients. \item[3.] Finally, 
as shown in section \ref{Chiral symmetry}, chiral symmetry subjects the
subthreshold coefficients to the two constraints (\ref{eq:Sigma}) and
(\ref{eq:C2}).\end{enumerate}
With these supplements, the dispersive representation of the
amplitude becomes equivalent to the one in terms of infrared regularized loop
integrals. In effect, we made use of these integrals only when analyzing the
corrections to the Goldberger-Treiman relation and to the two low energy
theorems referred to in point 3.

Note that the dispersive representation does not rely on an expansion of the
infrared singularities 
in powers of the momenta --
as mentioned above, this is a subtle matter. In particular, the
result depends on the choice of the two independent  
kinematic variables that are kept fixed when performing the expansion.
The above definition of the
cut contributions in terms of their absorptive parts precisely serves
the purpose of avoiding such problems at this stage. We will give the 
explicit algebraic representations for the functions $D_1^\pm(s),\,\ldots\,,
B^\pm_2(t)$ that result, for instance, if the expansion is performed by 
keeping the ratio $\omega/M_\pi$ fixed and discuss the changes occurring 
if we replace the lab energy $\omega$ by the c.m.~energy. 
Also, we will discuss the range of validity of such
representations in some detail, but first we wish to describe the result
of our calculation without invoking expansions of this sort.

\setcounter{equation}{0}
\section{Dispersion relations versus loop integrals}
\label{DR versus loops}
The dispersive representation
of the loop integrals may be viewed as yet
another regularization of the loop graphs that is consistent with power
counting. We add a few remarks concerning the comparison of this 
representation with the one based on infrared regularized loop integrals. 
For this purpose, we first again consider the
scalar self energy integral, for which both of these representations can be
given explicitly (compare section \ref{self energy}). In dimensional
regularization, the relevant integral is given by 
$H(s)$, while in infrared regularization, only the infrared singular
part thereof, $I(s)$, is retained. On the other hand, the dispersive
representation of the self 
energy integral, which we denote by $D(s)$, is determined by the imaginary
part of $I(s)$. The relevant 
dispersion relation involves the minimal number of subtractions required by
the asymptotic behaviour:
\bea D(s)=\frac{\omega}{\pi}\int_{\M}^\infty\frac{d\omega'\,\mbox{Im}I(s')}
{\omega'(\omega'-\omega-i\epsilon)}\fs\nonumber\eea
On the interval of interest, $s>(\m+\M)^2$, 
we have $\mbox{Im}I(s)=\mbox{Im}H(s)$. While the function $H(s)$ is analytic
except for the discontinuity on this interval, the function $I(s)$ in addition
contains fictitious 
singularities far outside the low energy region: a pole at $s=0$, as well as
a cut along the negative $s$-axis. Since by construction, the dispersive
representation only contains the right hand cut, $D(s)$ can differ from the
function $H(s)$ only by a polynomial. In fact, the asymptotic behaviour shows
that the polynomial is a constant. Hence the explicit expression for $D(s)$ 
reads 
\bea D(s)=H(s)-H(\m^2+\M^2)\fs\nonumber\eea
In view of $H(s)=I(s)+R(s)$, the difference between the two representations we
wish to compare may be represented as
\bea D(s)-I(s)=R(s)-R(\m^2+\M^2)-I(\m^2+\M^2)\fs\nonumber\eea 
Since the regular part of the self energy integral, $R(s)$,
does not contain any infrared
singularities, we may replace it by the first few terms in the
chiral expansion, which in the present case amounts to a simultaneous
expansion in powers of  $\omega$ and $\M$. The resulting representation for
$R(s)$ is a polynomial in 
$\omega$, with coefficients that are polynomials in $\M^2$. 
We conclude that, to any finite order in
the chiral expansion, $D(s)-I(s)$ is a polynomial in $\omega$ or,
equivalently, in $s$. 
Moreover, since all of its coefficients except the constant term only pick up
contributions from $R(s)$, they
are polynomials in $\M^2$. The constant term, on the other hand, does contain
infrared singularities, from $I(\m^2+\M^2)$. 

The same line of reasoning also applies to the other
graphs. When forming the sum of all contributions, the difference
also amounts to a polynomial and only the coefficients listed 
in eq.~(\ref{eq:Tpoly}) can pick up infrared singularities, from the analogues
of the term $I(\m^2+\M^2)$ encountered above. As we are explicitly
supplementing the dispersion integrals with the relevant polynomial, the
result agrees with the representation in terms of infrared regularized loop 
integrals, up to terms that are beyond the accuracy of a one loop calculation.

\setcounter{equation}{0}
\section{Chiral expansion of the amplitude}
\label{Algebraic representation}

The dispersive representation given above includes terms of arbitrarily high
order in the chiral expansion. It is suitable for numerical
analysis, but since the expressions for the imaginary parts
are lengthy, an approximate representation that 
replaces these integrals by elementary functions is useful. 
Such a representation may be
obtained by performing the chiral
expansion in the integrand of the loop integrals. The result 
is very similar to the one obtained in HBCHPT, where the unitarity cuts are 
described in terms of simple functions such as $\arccos(x)$.
In \cite{Becher Leutwyler 1999}, we have proven that
this expansion converges for all of the integrals encountered here, 
at least in part of the low energy region.
In the present section, we first briefly comment on the chiral expansion of
the integrals associated with the $t$-channel cut, where the discussion
given in ref.~\cite{Becher Leutwyler 1999} can be taken over without
significant complications. Then we analyze the analogous issue for the
$s$-channel and discuss the infrared singularities associated with the
corresponding cut in some detail.

\subsection[$t$-channel cut]{{\boldmath$t$\unboldmath}-channel cut}
Only the diagrams belonging to the topologies (i), (k), (l) and (m)
\begin{figure}[htb]
\begin{center}
\includegraphics{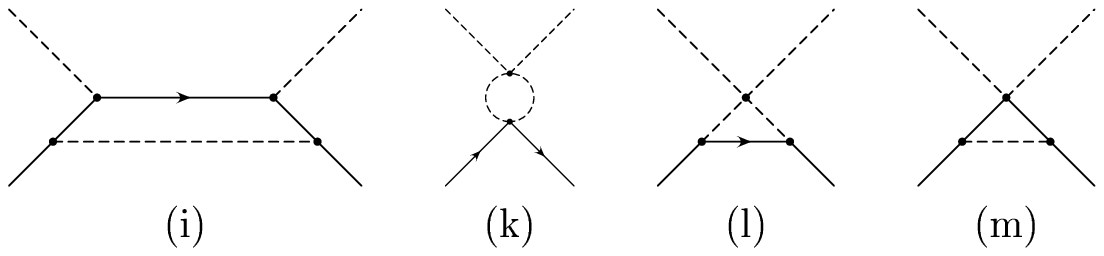}
\vspace*{-2em}
\end{center}
\end{figure}
develop an imaginary part for $t>4M_\pi^2$. For graphs (i) and (m), the
$t$-channel cut only starts at $t=4\m^2$. As discussed in section
\ref{Cuts}, these do not show up at finite orders of the chiral expansion, 
so that the diagrams (i) and (m) do not contribute. The integrals arising from
the graphs (k) and (l) are polynomials in $\nu$. 
It is convenient to replace the pion mass and the momentum transfer by the two
dimensionless variables
\bea \alpha=\frac{\M}{\m}\co\hspace{2em}\tau=\frac{t}{\M^2}\fs\nonumber\eea
Graph (k) may be expressed in terms of the scalar loop integral with two
meson propagators, $J(t)$. We remove the divergence by subtracting
at $t=0$. The remainder, $\bar{J}(t)=J(t)-J(0)$, only depends on the ratio
$\tau$ and the explicit expression 
reads
  \bea \bar{J}(\tau)=\frac{1}{8 \pi^2}
\left\{1-\sqrt{\frac{4-\tau}{\tau}}\,
\arcsin\,\frac{\sqrt{\tau}}{2}\right\}\fs\nonumber\eea
The triangle graph (m) involves integrals with two meson propagators and one 
nucleon propagator. The chiral expansion of these functions
does not cover the entire low
energy region: It breaks down in the vicinity of the point $t=4\M^2$. The
  matter is discussed in detail in ref.~\cite{Becher Leutwyler 1999}, where it
  is shown that the elementary function
\begin{multline*} g(\tau)
= \frac{ 1}{32\,\pi \,{\sqrt{\tau}}} \,\ln  \frac{2\, + 
{\sqrt{\tau}}}{2\, -\sqrt{\tau}} -  \frac{1}{32\,\pi } \,\ln \left\{ 1 +
\frac{\alpha}{{\sqrt{4 - \tau}}}\right\} \\
+\frac{\alpha}{32\,\pi^2} \,\left\{1 +  \frac{\pi}{\sqrt{4  - \tau}}
 +  \frac{2\,\left( 2 - \tau \right)}
{\sqrt{\,\tau\left( 4 - \tau \right) }}\,\arcsin \frac{\sqrt{\tau}}{2}
\right\}   
\end{multline*}
does provide a representation that is valid throughout the low energy region.

\subsection[$s$-channel cut]{{\boldmath$s$\unboldmath}-channel cut}
The dispersive integrals associated with the 
$s$-channel cut are polynomials in $t$ with coefficients
that depend only on $s$. The $u$-channel cut is described by the same 
integrals.
The effective theory concerns the region where $s-\m^2=O(q)$. 
We may for instance consider the expansion of the loop integrals
in powers of $\alpha=\M/\m$ at a fixed value of the ratio $r=(s-\m^2)/\M$.
That expansion, however, necessarily diverges in the vicinity of the
branch point $s=(\m+\M)^2$, because this singularity corresponds to a value
of $r$ that depends on the expansion parameter $\alpha$.
A better choice of the variable to be held fixed
is the ratio of the pion lab.~energy $\omega$ to the pion mass,
which we denote by
\begin{align*}
\Omega=\frac{s-\m^2-\M^2}{2\,\m\,\M}=\frac{\omega}{\M}\fs
\end{align*}   
The $s$-channel cut starts at $\Omega=1$ and thus stays put
when $\alpha$ varies. Indeed, we have shown 
\cite{Becher Leutwyler 1999} that the convergence domain of
the expansion of the self energy integral $I(s)$ at fixed $\Omega$ is
$0<s<2\m^2+2\M^2$ and thus covers exactly the Mandelstam triangle. 

Numerically,
the convergence is rather slow, however, already at the threshold. The explicit
expression for the self energy integral $I(s)$ involves the factor
$m^2/s=(1+2\,\alpha\,\Omega+\alpha^2)^{-1}$ and some of the loop integrals
even contain the third power thereof. 
At threshold, the expansion of this factor in powers of $\alpha$ 
is roughly a geometric series with expansion parameter
$2\,\Omega\,\alpha \simeq \frac{2}{7}$. 
When the factor appears squared or cubed, the
lowest orders do therefore not give a decent approximation.

A more suitable choice is the c.~m.~energy of the pion
\begin{equation}
\Omq=\frac{s-\m^2-\M^2}{2\M\sqrt{s}}=\frac{\omega_q}{\M}\co
\end{equation}
which maps the interval $0<s<\infty$ onto $-\infty<\Omq<\infty$ and thus
pushes the singularity at $s=0$ away from the threshold at $\Omq=1$. 
Expressed in terms of this variable, the infrared singular part of the self
energy integral is given by
\begin{equation}
I(s)=\frac{\M}{16\,\pi^2\,\sqrt{s}}\left\{-2\sqrt{1-\Omq^2}
\arccos(-\Omq)+\Omq\left(1-32\,\pi^2\,\lambda_\pi\right)\right\}\fs
\end{equation}
At fixed $\Omq$, only the factor
$\sqrt{s}=\sqrt{1+\alpha^2\left(\Omq^2-1\right)}+\alpha\,\Omq$ gets expanded.
It has a branch point at $\Omega_q^c=\pm i\sqrt{1-\alpha^2}/\alpha$. 
The expansion
is convergent for $|\Omq|<|\Omega_q^c|$, which corresponds to the range
\[
(3-2\sqrt{2})(\m^2-\M^2)<s<(3+2\sqrt{2})(\m^2-\M^2)\fs
\] 
On the left, the convergence range is slightly smaller than the one obtained
for the variable $\Omega$, but on the right, the convergence is
drastically improved. The point at which the expansion at fixed $\Omq$ breaks
down is six times farther away from threshold than the edge of the Mandelstam
triangle, where the expansion at fixed $\Omega$ starts to fail.
\begin{figure}[htb]
\psfrag{x}{$\omega_q$}
\psfrag{y}[b]{$\mbox{Re} \,D^+(\omega_q)$}
\begin{center}
\begin{tabular}{c}
 \includegraphics[width=0.6\textwidth]{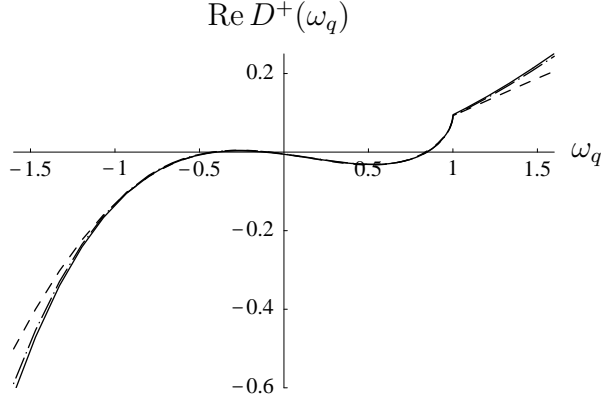}
\end{tabular}\caption{\label{figDp}Chiral expansion of the contribution from
diagram (f) to the real part of the amplitude $D^+$ 
(infrared regularization, scale $\mu=\m$). Full: relativistic
amplitude, dot-dashed: expansion at fixed $\Omq$, 
dashed: expansion at fixed $\Omega$. In all of the figures, 
dimensionful quantities are given in units of $\M$.}
\end{center}
\vspace*{-1em}
\end{figure}

We illustrate this with the contribution of graph (f) to the amplitude
$D^+$ at $t=0$. The corresponding Feynman diagram is shown in
fig.~\ref{fig:loop} and the formulae in appendix \ref{Loop graphs}
explicitly specify this contribution in terms of the function $I(s)$.
In fig.~\ref{figDp}, the full line shows the exact result for the
renormalized amplitude at $t=0$, obtained with infrared regularization
(for definiteness, we have identified the running scale with the
nucleon mass).  The dashed line corresponds to the expansion to order
$\alpha^4$ at fixed $\Omega$, while the dash-dotted curve is obtained if
we instead keep the variable $\Omq$ fixed. The figure shows that the
chiral expansion at fixed $\Omq$ describes the result quite
accurately.

\begin{figure}[thb]
\vspace*{-1em}
\begin{center}
\begin{tabular}{c}
\psfrag{x}{$k$} \psfrag{y}{\hspace{4em}\raisebox{-2em}{$n^+(k)\, \mbox{Im}_s
  D^+(k)$}} 
\includegraphics[width=0.6\textwidth]{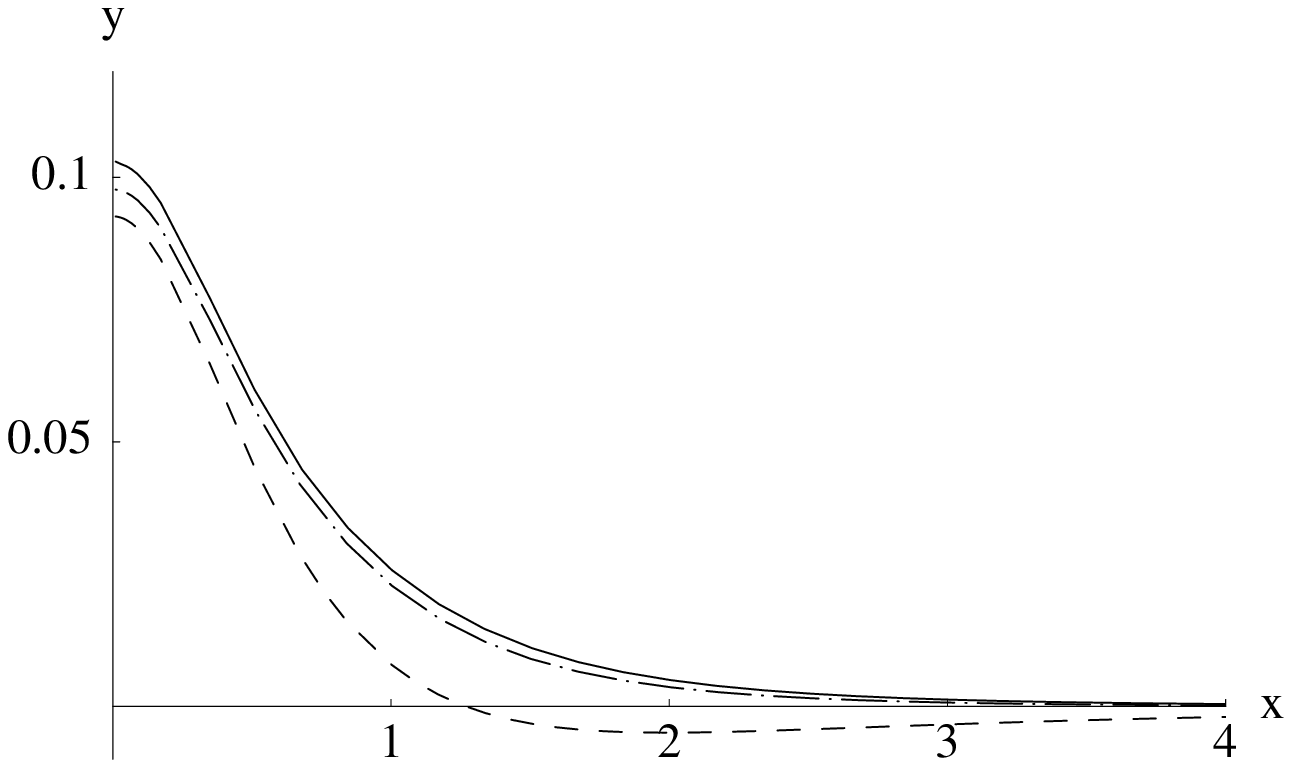}\\ 
\vspace*{-2em}
\psfrag{x}{$k$} \psfrag{y}{\hspace{4em}\raisebox{-2em}{$n^-(k)\,\mbox{Im}_s
  D^-(k)$}} 
\includegraphics[width=0.63\textwidth]{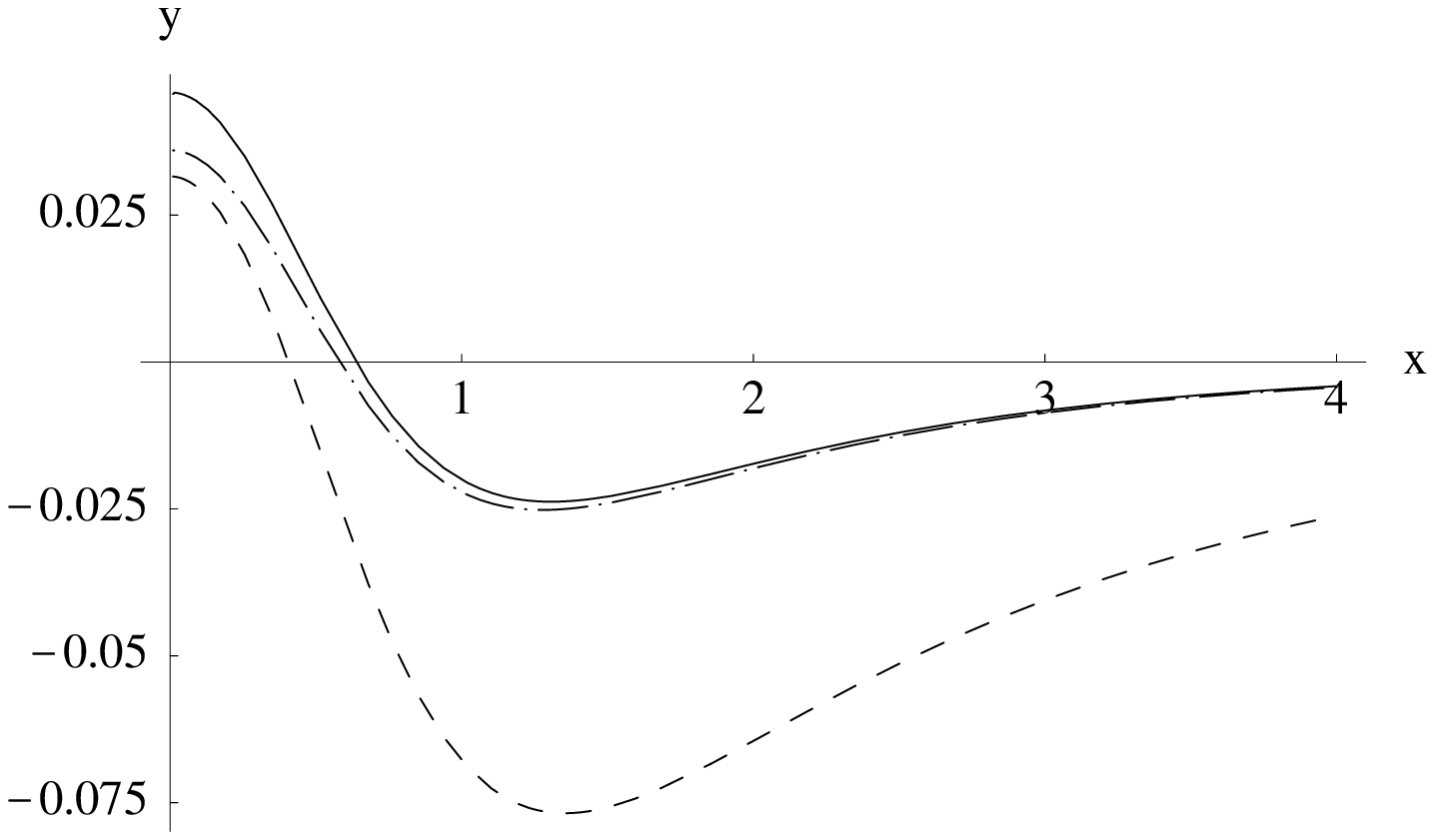}\rule{0.9em}{0em}
\end{tabular}
\end{center}
\vspace{0.3em}
\caption{Imaginary part of the $D$-amplitudes at $t=0$. The normalization
 factor $n^\pm(k)$ is chosen such that the area under the curve represents
the contribution to the
value of the amplitude at threshold.
Full: relativistic
  amplitude, dot-dashed: expansion at fixed $\Omq$, dashed: expansion at fixed
  $\Omega$.}\label{fig:imagD}
\end{figure}

In fig.~\ref{fig:imagD}, we compare the imaginary parts of the loop integrals
with the chiral expansion thereof, for the sum of all graphs. 
Near threshold, these are suppressed by phase space. To make the behavior
visible there, we plot the integrand relevant if the variable in the 
dispersion integrals is identified with the lab momentum 
$k=\sqrt{\omega^2-\M^2}$ rather than the lab energy
$\omega$.
The integrand of the dispersion relation for the value of the amplitude
at threshold, $D^\pm(M_\pi,0)=4\,\pi\,(1+\alpha)\,a^\pm_{0+}$, is then given
by $ n^\pm(k)\, \mbox{Im}D^\pm_1(k)$, with
\bea n^\pm(k)\al=\al\frac{\M^5}{\pi\,\omega^5}\;\frac{d\omega}{dk}\;
\left\{\frac{1}{\omega-\M}\mp
\frac{1}{\omega+\M}\right\}=\frac{2M_\pi^5}{
\pi\,k\,\omega^6}\left\{\hspace{-0.4em} 
\begin{array}{c} M_\pi\\ \omega\end{array}\right.\nonumber\eea
In view of the difference in the weight, the dispersion relation for $D^-$ is
more sensitive to the high energy region than the one for $D^+$.

\subsection{Chiral expansion of the dispersion integrals}
\begin{figure}[ht]\centering
\psfrag{x}{$\omega$}
\psfrag{y}{$\mbox{Re} \,D^+_1(s)$}
 \includegraphics[width=4in]{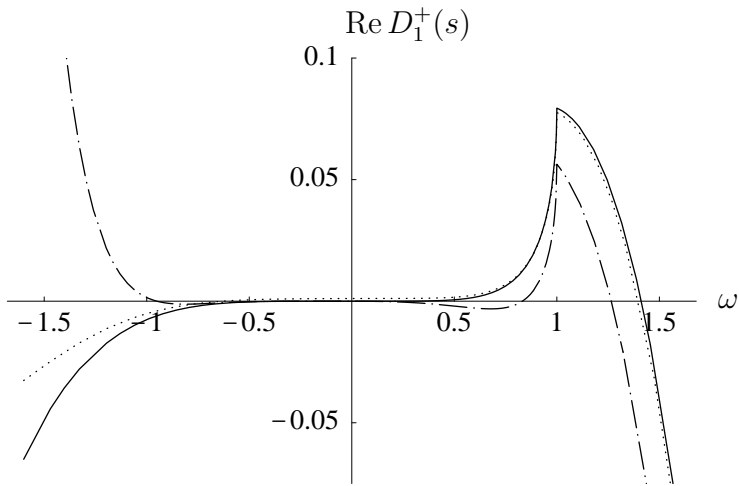}
    \caption{\label{figDs}Chiral expansion of the contribution from
      diagram (f) to the subtracted dispersion integral $D^+_1(s)$.}
  \end{figure}

The situation changes drastically, however, if we consider the
corresponding contribution to the subtracted dispersion
integral
$D^+_1(s)$, which may be evaluated as follows. We first note that
the discontinuity across the right hand cut of the function $I(s)$ 
coincides with the one of the corresponding dimensionally regularized 
object, $H(s)$. Moreover, $H(s)$ does not have a left hand cut.
Hence the expression obtained by simply replacing $I(s)$ with $H(s)$ does
have the same singularities as the dispersive integral we are looking for --
the difference is a polynomial in $\omega$. The subtracted dispersion integral
is obtained from $H(s)$ by removing the Taylor series expansion 
to order $\omega^4$. 
In fig.~\ref{figDs}, the full line represents the
exact contribution to $\mbox{Re}D^+_1(s)$, while the dot-dashed one depicts
the chiral expansion thereof at fixed $\Omq$, 
to $O(\alpha^4)$. Evidently, the truncation of the
chiral expansion at $O(\alpha^4)$ very strongly distorts the result. 

The reason for this disaster is that the 
chiral expansion of the Taylor coefficients that specify the subtraction
polynomial converges only extremely slowly. The contributions
from the $s$-channel loop graphs to the coefficient $d_{20}^+$, for instance,
are obtained by evaluating the fourth derivative of the corresponding
contributions to $D^+$ at $\omega=0$. These quantities are represented very 
poorly by the leading terms in their chiral expansion.
The problem disappears if we keep the Taylor coefficients fixed when 
performing the
chiral expansion (the representation that then results is indicated by
the dotted line).
When representing the amplitude in the form (\ref{representation to O4}),
we have absorbed the subtraction polynomial in the coefficients of the
subthreshold expansion. This implies that the chiral expansion of the 
subthreshold coefficients in powers of the quark 
masses -- in particular the one of $d_{20}^+$ -- converges only very 
slowly. 

The physics underneath the phenomenon is readily identified. The
coefficient $d_{20}^+$ may be expressed as an integral over the total
cross section. The integral is dominated by the threshold region.
If the quark masses are sent to zero, this integral diverges:
The explicit formula for
$d_{20}^+$ given in appendix \ref{Subthreshold coefficients} shows that this 
constant contains a term that is inversely proportional to $M_\pi$. 
The formula may be viewed as a low 
energy theorem: The leading term in the expansion of $d_{20}^+$ in powers of
the quark masses is determined by $\gA$ and $F_\pi$. At the same time,
however, the above discussion shows that this expansion converges only very
slowly, because the terms of higher order pick up important contributions from
the same one loop integrals that determine the leading
coefficient. 

We emphasize that the infrared singularities seen here are perfectly well 
described by the relativistic loop integrals occurring in our framework.
They generate problems only if we attempt to replace the loop integrals
by their chiral expansion, as it is done ab initio in Heavy Baryon Chiral
Perturbation Theory. The dotted curve in fig.\ref{figDs}
indicates that a decent representation of this type may be obtained if (a) the
kinematic variables are carefully chosen and (b)
the polynomial part of the amplitude is dealt with properly.
We do not elaborate further on the chiral expansion of the loop integrals, 
but now turn to the contributions
from the terms of $O(q^5)$ or higher, which our
representation does not account for.

\setcounter{equation}{0}
\section{Higher orders}
\label{Higher orders}
The representation of the scattering amplitude to $O(q^4)$ incorporates the
poles and cuts generated by the exchange of one or two stable particles
(nucleons or pions), but
accounts for all other singularities only summarily, through their
contributions to the effective
coupling constants. 

The first resonance in the $s$- and $u$-channels is the 
$\Delta(1232)$. In the $t$-channel, the $\rho(770)$ plays an analogous
role. These states generate poles on unphysical sheets of the scattering
amplitude. Since we did not include corresponding dynamical degrees of freedom
in our effective Lagrangian, we are in effect replacing these singularities
by the first few terms of the Taylor series expansion 
in powers of $\nu$ and $t$.

The $\rho$-meson pole occurs at $t\simeq 30\M^2$. In the region 
$|t|\,\mbox{\raisebox{-0.3em}{$\stackrel{<}
{\sim}$}}\,10 M_\pi^2$, the contribution from this state 
is well represented by a polynomial, so that the description in terms
of effective coupling constants is adequate. The range of validity
of this approximation is about the same as the one encountered in 
$\pi\pi$ scattering. 

The situation is different for the $\Delta$-resonance.  The curvature caused
by this state in the low energy region may be described in terms of a low
order polynomial only crudely.  A better description of the effects generated
by the singularities at $s\simeq \mD^2$, $u\simeq \mD^2$ is obtained from an
effective Lagrangian that includes the degrees of freedom of the $\Delta$
among the dynamical variables, a procedure followed in many papers on baryon
chiral perturbation theory. For details
concerning this approach, we refer to the literature \cite{Small scale}.  In
the following, we make use of the explicit representation for the tree graph
contribution to the scattering amplitude associated with $\Delta$ exchange
given in appendix D of ref.~\cite{Becher Leutwyler 1999}. We denote this
contribution by $D^\pm_{\ind \Delta}(\nu,t)$, $B^\pm_{\ind \Delta}(\nu,t)$. It
involves two parameters: the mass $\mD$ and the coupling constant $\gD$. The
$\Delta$-resonance starts contributing at $O(q^2)$. In a calculation to order
$q^4$ the coupling constants of ${\cal L}_{\ind N}^{(2)}$ enter the loop
diagrams, so 
that part of the $\Delta$ contribution is absorbed by the loop graphs with one
insertion of a vertex from ${\cal L}_{\ind N}^{(2)}$.

For the amplitude $D^+$ at $t=0$ the situation is simple, because the
contribution of the ${\cal L}^{(2)}_{\ind N}$-couplings to the dispersion
integral 
$D_1^+(s)$ is of $O(q^5)$ and numerically negligible. We can thus estimate
the remaining $\Delta$ contribution by subtracting from the $\Delta$ tree
graph the contribution to the polynomial part $D^+_{p}$. At threshold, the
remainder is given by
\begin{equation*}
D^+_\Delta =\frac{64\,\gD^2\,\m^5\,\M^6\,
     \left( \left( \m + m_\Delta \right)^2 - \M^2
       \right) }{9\,m_\Delta^2\,
     \left( m_{\Delta }^2 - \m^2 - \M^2 \right)^5}=
O\left(\frac{\M^6}{(m_\Delta-\m)^5}\right)\fs
\end{equation*}
It is formally of $O(q^6)$, but numerically enhanced by the small denominator.
Inserting the value $\gD=13.0\,\mbox{GeV}^{-1}$ \cite{Hoehler}, we obtain 
$\F^2 D^+_\Delta=2\,\mbox{MeV}$, indicating that for a determination of 
the $\Sigma$-term, corrections of this type are not entirely negligible.  
The corresponding
contribution to the scattering length is small: 
$a^+_{0+\Delta}=0.0022\,\M^{-1}$. The effective
range of the $S$-waves or the scattering lengths of the $P$- and $D$-waves
involve derivatives of the amplitude at threshold, so that the curvature due
to the $\Delta$ generates more significant effects.

Note that the singularities generated by the $\Delta$ do not affect
the Goldberger-Treiman relation. In the case of the low energy theorem
that relates the scattering amplitude at the Cheng-Dashen point to the
$\sigma$-term, the higher order terms due to the $\Delta$ amount to a
correction of order $\M^6/(M_\Delta-\m)^3$. The effect is negligibly
small: It increases the value of $\Delta_{\scriptscriptstyle CD}$ by
0.08 MeV.  Also, if the effective theory is used only in the threshold
region, for instance to study the properties of pionic atoms
\cite{Gasser:1999td}, the contributions generated by the $\Delta$ are
adequately described by a polynomial -- it suffices to replace the
Taylor series of the crossing even quantities $D^+$, $B^+/\nu$,
$D^-/\nu$, $B^-$ in powers of $\nu$ by an expansion in powers of
$\nu^2-M_\pi^2$. The effective value of $d_{00}^+$ relevant in that
context is the value of $D^+$ at threshold rather than the one at $\nu=0$, and
analogously for the other coefficients occurring in the polynomial
part of our representation. The straightforward identification of
these terms with the coefficients of the subthreshold expansion yields
an accurate representation in the vicinity of $\nu=t=0$. It is
adequate also at the Cheng-Dashen point, but at threshold, the higher
order terms of the subthreshold expansion do become important, quite
irrespective of the contributions generated by the branch point that
occurs there.

\section{Values of the coupling constants of {\boldmath${\cal L}_{\ind
    N}^{(2)}$\unboldmath}}
\label{Values of the couplings}
In the representation of the amplitude
specified in section \ref{Dispersive representation}, the coupling constants
$c_1$, $c_2$, $c_3$, $c_4$ only enter through the imaginary parts 
of the dispersion integrals.
In section \ref{Subthreshold expansion}, we have 
fixed these with the leading coefficients of the 
subthreshold expansion. Numerically, using the values of the subthreshold
coefficients of ref.~\cite{Hoehler}, this leads to
\begin{align}\label{eq:cHoehlernum}
c_1 &= -0.6\, \m^{-1} \co& 
c_2 &= 1.6\,\m^{-1} \co& 
c_3 &= -3.4\, \m^{-1}\co& 
c_4 & = 2.0\,\m^{-1}\fs &
\end{align}
If we use these values to evaluate the scattering
amplitude at threshold, the result differs significantly from
what is observed.  For the loop integrals to  
describe the strength of the $s$- and
$u$-channel threshold singularities, it might be better to fix the
couplings of ${\cal L}_{\ind N}^{(2)}$ there, 
using the values of the amplitude and their first derivatives at
threshold as input rather than those at $\nu=0$. The difference is typical of
the higher order effects that our calculation neglects and we now briefly 
discuss it numerically.

At tree level, the coupling constants $c_1$, $\ldots\,,$ $c_4$ 
only occur in $D^+$ and $B^-$. The values and first
derivatives of these amplitudes
at threshold are related to the scattering lengths of the $S$-waves,
  $a_{0+}^\pm$, to their effective
ranges, $b_{0+}^\pm$, and to the scattering lengths of the $P$-waves,
$a_{1\pm}^{\pm}$ (see appendix \ref{notation}).
Expanding the tree graph contributions from ${\cal L}^{(1)}_{\ind N}$ and  
${\cal L}^{(2)}_{\ind N}$ around threshold,
we thus obtain a representation of the threshold parameters in terms
of the effective coupling constants, valid at tree level. 
Inverting the relation, we arrive at a corresponding  
representation for the coupling constants:
\bea\label{eq:cthreshold} 
c_1\al=\al \pi\,\F^2\left\{-(4+2\,\alpha+\alpha^2)\,\frac{a_{0+}^+}{4\M^2}
+b^+_{0+}-3\,\alpha\,a_{1+}^+\right\}+ \alpha^2\,c_{\ind B}\co\no
 c_2\al=\al \frac{2\,\pi\,\F^2}{1+\alpha}\left\{\frac{a_{0+}^+}{2\,\M\,\m}
+b_{0+}^++2 a_{1+}^++a_{1-}^+\right\}-8\,c_{\ind B}\co\\
c_3\al=\al\frac{2\,\pi\,\F^2}{1+\alpha}\left\{(1-\alpha)\,\frac{
a_{0+}^+}{4\,\m^2}+
\alpha\,b_{0+}^+ -(2+3\,\alpha+3\,\alpha^2)\,a_{1+}^+-a_{1-}^+\right\}+16\,
c_{\ind B}\no
c_4\al=\al-\frac{1}{4\,\m}+4\,\pi\,\F^2\left\{\frac{a_{0+}^-}
{4\,\m^2}-a_{1+}^-+a_{1-}^-\right\}-\alpha^2\,(4-\alpha^2)\,c_{\ind  B}\co\no
c_{\ind B}\al=\al\frac{\gpiN^2\,\F^2}{4\,(4-\alpha^2)^2\,\m^3}\fs\nonumber\eea
With Koch's values for the threshold parameters 
\cite{Koch:1986bn}, the above representation yields
\begin{align}\label{eq:cthresholdnum}
 c_1 &= -0.9\, \m^{-1} \co& 
c_2 &= 2.5\,\m^{-1} \co& 
c_3 &= -4.2\, \m^{-1}\co& 
c_4 & = 2.3\,\m^{-1}\fs &
\end{align}
Although, algebraically, the formulae (\ref{cd}) and 
(\ref{eq:cthreshold}) differ only through terms of
$O(q^2)$,  the comparison of the corresponding numerical values in 
(\ref{eq:cHoehlernum}) and (\ref{eq:cthresholdnum}), shows that the 
difference
\bea \delta c_i=c_{i}\,\rule[-0.5em]{0.04em}{1em}_{\,\nu=\M}-
c_{i}\,\rule[-0.5em]{0.04em}{1em}_{\,\nu=0}\eea 
is quite significant. It mainly
arises from the curvature due to the $\Delta$. In leading order of the
small scale expansion \cite{Small scale}, the shift in the values of 
$c_2$, $c_3$ and $c_4$ generated by this state is given by
\bea
\delta c_2^{\ind \Delta}=4\,\gamma\co\hspace{2em}
\delta c_3^{\ind \Delta}=-2\,\gamma\co\hspace{2em} 
\delta c_4^{\ind \Delta}=\gamma\co\hspace{2em}
\gamma=\frac{2\,\gD^2\,\F^2\,\M^2}{9\,(\mD-\m)^3}
\co\nonumber
\eea
and thus involves the third power of the small energy denominator $\mD-\m$.
Inserting the value $\gD=13\,\mbox{GeV}^{-1}$ \cite{Hoehler}, we obtain
$\gamma\simeq 0.23 \,\m^{-1}$, so that $c_2$
is increased by about one unit, while $|c_3|$ and $c_4$ pick up
one half and one quarter of a unit, respectively. 
In $c_1$, the singularity does not show up equally
strongly: At leading order in the small scale expansion, we have
$c_1^{\ind\Delta}=-(\mD-\m)/\m \gamma\simeq -0.07 \m^{-1}$.
The comparison with the
numbers given above confirms the claim that the $\Delta$ is responsible
for the bulk of the difference.

Concerning the $t$-channel imaginary parts, the role of the coupling 
constants $c_1,c_2,c_3,c_4$ and the contributions from the 
$\Delta$ are discussed in detail in ref.~\cite{Becher Leutwyler 1999}.
These are not sensitive to the manner in which the coupling constants
of ${\cal L}^{(2)}_{\ind N}$ are fixed. As an illustration, we consider the
quantity $\Delta_\sigma=\sigma(2\M^2)-\sigma(0)$, relevant for a 
measurement of the $\sigma$-term on the basis of $\pi N$ data. 
In ref.~\cite{Becher Leutwyler 1999}, we pinned the coupling constants 
down with the subthreshold expansion and obtained 
$\Delta_\sigma=14.0\,\mbox{MeV}$. If we use the same
formulae, but fix the values of the coupling constants with the scattering
lengths, this number changes to
$\Delta_\sigma=15.9\,\mbox{MeV}$.
Both of these values agree quite well with the result of the dispersive 
analysis, $\Delta_\sigma=15.2\,\mbox{MeV}$ \cite{Gasser Leutwyler Sainio}. 

\setcounter{equation}{0}
\section{\boldmath{$S$}--wave scattering lengths}
\label{Scattering lengths}
At leading order of the chiral perturbation series, the
 scattering amplitude is described by the tree graphs from 
${\cal L}^{(1)}_{\ind N}$. This leads to Weinberg's prediction for the two
scattering lengths: \cite{Weinberg:1966kf},
\bea
a_{0+}^+=O(q^2)\co\hspace{2em}  
a_{0+}^-=\frac{\M}{8\,\pi\,(1+\alpha)\,F_\pi^2}+O(q^3)\fs
\nonumber\eea 
Numerically, 
Weinberg's formula yields 
$4\,\pi\,(1+\alpha)\,a_{0+}^-=1.14 \,\M^{-1}$, remarkably close
to the experimental value. 

The one loop representation of the scattering amplitude accounts for the
corrections up to and including $O(q^4)$. According to eq.~(\ref{representation
  to O4}), the scattering lengths are given by
\begin{align*}
4\,\pi(1+\alpha)\,a_{0+}^+&=D^+_{pv} + d_{00}^+ + \M^2\, d_{10}^+ + \M^4\, 
d_{20}^+ +
D_s^+\co\\
4\,\pi(1+\alpha)\,a_{0+}^-&=D^-_{pv} + \M\,d_{00}^- + \M^3\, d_{10}^-
  +D^-_s\co
\end{align*}
where $D^+_s$ and $D^-_s$ collect the contributions from the
dispersion integrals at $\nu=\M$, $t=0$.

In view of the fact that the value of the coupling constant $\gpiN$ plays a
significant role, it is not a simple matter to combine the information
obtained by different authors \cite{Sainio}.  An account of recent work on the
low energy parameters may be found in ref.~\cite{Oades Stahov Pavan}.  For the
purpose of the present discussion, the precise values of the various
experimental quantities do not play a significant role, however -- we simply
stick to 
the KA84 solution \cite{KA84}. The experimental values of the various terms
occurring in the above decomposition are listed in the table (all quantities
in units of $\M$).

\begin{center}
\begin{tabular}{|c|r|r|r|r|r|r|}
\hline
&$D_{pv}\hspace{0.4em}$&$d_{00}\hspace{0.5em}$&$d_{10}\hspace{0.6em}$&$d_{20}$
&$D_s\hspace{1em}$&total\hspace{0.2em}\\ 
\hline
$4\,\pi\,(1+\alpha)\,a^+_{0+}$
&$-0.15$&$-1.49$&1.17&0.2&0.15\phantom{5} &$-0.12$\\
$4\,\pi\,(1+\alpha)\,a^-_{0+}$
&0.01&1.51&$-0.17$&$-\hspace{0.3em}$&$-0.035$&1.32\\
\hline
\end{tabular}
\end{center}

\noindent
The values for $D_s^\pm$ quoted there represent the full 
dispersion integral rather than the one loop approximation to it,
so that the sum of the terms is the value of the physical amplitude at 
threshold. Evaluating these quantities in one loop approximation, we instead
obtain $D_s^+=0.08\,\M^{-1}$ and $D_s^-=-0.035\,\M^{-1}$: While the value for
$D^-_s$ agrees with the experimental result, the one for $D_s^+$ is too small
by a factor of two. Although we are discussing small effects here, they do
matter at the level of accuracy needed to extract the $\Sigma$-term:
In the case of $D^+_s$, the experimental value is $\F^2 \,D_s^+=
9.3\,\mbox{MeV}$, while the loop graphs yield $\F^2 \,D_s^+=4.9\,\mbox{MeV}$.

We conclude that the one loop representation of chiral perturbation theory
does not cover a sufficiently large kinematic region to serve as a bridge 
between the experimentally accessible region and the Cheng--Dashen point. 
A meaningful extrapolation of the data into the subthreshold region can be
achieved only by means of dispersive methods -- we will outline one such
method in section \ref{sec:Roy}. First, however, we wish to identify the 
origin of
the problem, by studying the difference between the full imaginary parts and
the one loop representation thereof. We focus on the quantities
$D_s^\pm$, which represent dispersion integrals over total cross sections. 

\setcounter{equation}{0}
\section{Total cross section}
\label{Total cross section}
The optical theorem relates the imaginary part of the amplitude $D(s,t)$ 
at $t=0$ to the total cross section:
\begin{equation}
\mbox{Im}_s D(s,0)=k\,\sigma_{tot}\fs
\end{equation}
To the order of the low energy expansion we are considering here,
the scattering is elastic, so that $\sigma_{tot}$ is the integral over the
differential cross section, given by the square of the scattering amplitude.
In the c.~m.~system, the explicit expression may be written in
the simple form
\bea
\frac{d\sigma}{d\Omega}\al=\al\frac{\m^2}{16\,\pi^2\,s}\Big\{D\, D^\ast\,
(1-\frac{t}{4\,m^2})\\
\al\al\hspace{3.5em}+\, (D\, B^\ast+B\, D^\ast)\,\frac{t\,\nu}{4\,m^2} -
B\,B^\ast\, 
\frac{t\,(t-4\,M^2+4\,\nu^2)}{16\,\m^2}\Big\}\fs
\nonumber\eea

\begin{figure}[!ht]
\begin{center}
\begin{tabular}{c}
\psfrag{om}[b]{\small$\nu$ [$\M$]}\psfrag{M}[B]{\small $\sigma\rule[-0.5em]{0em}{0em}^{\frac{1}{2}}\rule{0em}{0em}_{tot}$ [$\M^{-2}$]}
\includegraphics[width=0.8\textwidth]{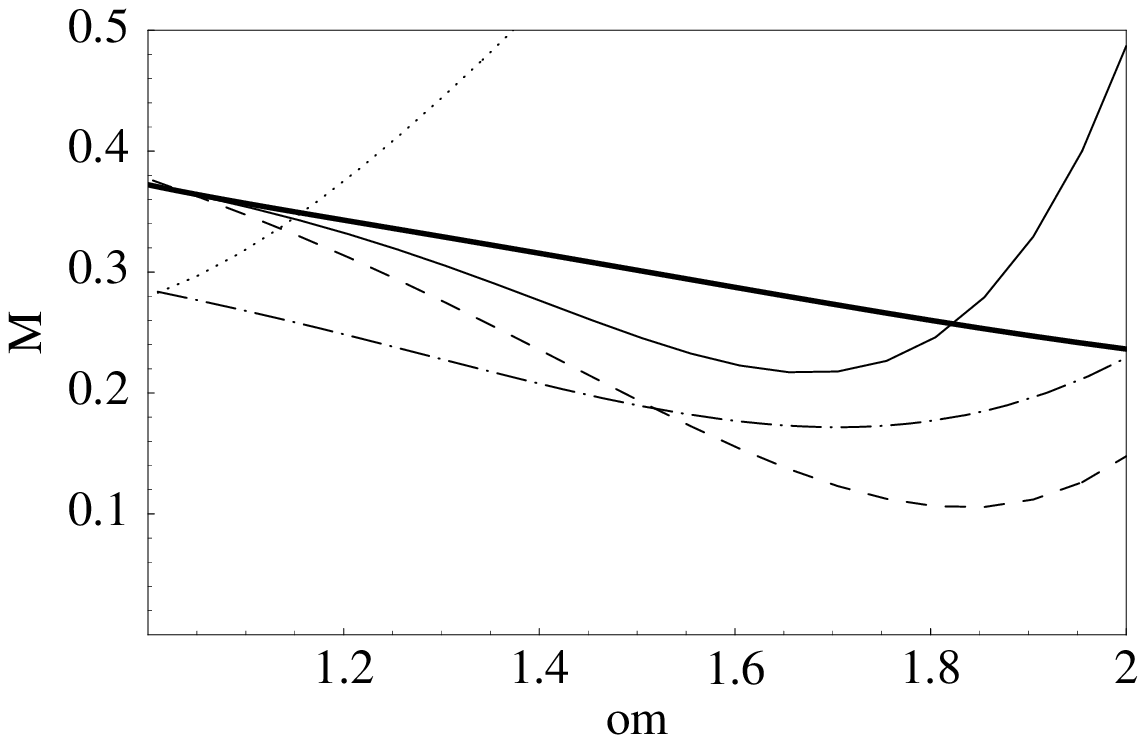} \\
\\
\psfrag{om}[b]{$\nu$ [$\M$]}\psfrag{M}[B]{$\phantom{abc}\sigma\rule[-0.5em]{0em}{0em}^{\frac{3}{2}}\rule{0em}{0em}_{tot}$ [$\M^{-2}$]}
\hspace{-0.4em}\includegraphics[width=0.8\textwidth]{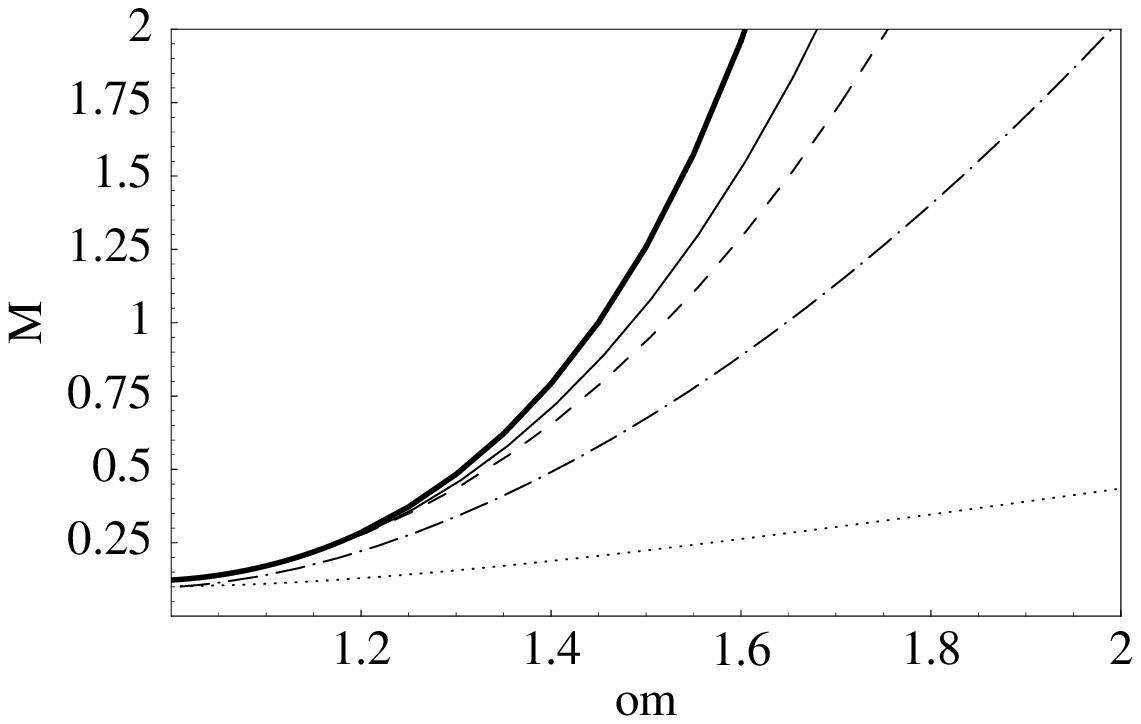}
\end{tabular}
\end{center}
\caption{Total cross sections.  Dotted: Current algebra. Dot-dashed: 
Tree graphs from ${\cal L}^{(1)}_N+{\cal L}^{(2)}_N$.
Dashed: CHPT $O(q^4)$, all couplings fixed at $\nu=t=0$.
Thin: CHPT $O(q^4)$, all couplings fixed at threshold. 
Thick: Experimental values \cite{Koch:1986bn}.}
\label{fig:sigmatot} 
\end{figure}

If the chiral perturbation series of the scattering amplitude is
truncated at $O(q^3)$,
the imaginary part of the loop integrals 
is given by the square of the tree graphs generated by ${\cal
L}^{(1)}_{\ind N}$ (current algebra approximation).  Although current
algebra does describe the $S$-wave scattering lengths remarkably well
it does not yield a realistic picture for the imaginary parts for two
reasons: (a) Since we are now dealing with the square of the
amplitude, the deficiencies of the current algebra approximation
become more visible and (b) the effects due to the $\Delta$ are more
pronounced than in the subthreshold region. This is illustrated in
fig.~\ref{fig:sigmatot}, where the dotted lines correspond to
the current algebra approximation. The full lines show the behaviour of the
experimental total cross section, reconstructed from the results of
Koch \cite{Koch:1986bn} for the amplitudes in the threshold region.

The discrepancy seen near threshold is directly related to the
difference between the experimental values of the scattering lengths and
the current algebra result:
There, the cross section is the square of the scattering length,
\begin{align*}
  \sigma\rule[-0.5em]{0em}{0em}^{\frac{1}{2}}\rule{0em}{0em}_{tot} &=
{4\,\pi}\, (a_{0+}^+ +2 a_{0+}^-)^2\co &
\sigma\rule[-0.5em]{0em}{0em}^{\frac{3}{2}}\rule{0em}{0em}_{tot}
&= {4\,\pi}\,
      (a_{0+}^+ - \,a_{0+}^-)^2\fs
\end{align*}
These expressions are dominated by $a_{0+}^-$ -- the term $a^+_{0+}$ 
vanishes at leading order. Although Weinberg's prediction represents
a very decent approximation, the difference to the observed scattering lengths
generates an effect of
order 25 \% in the cross section -- this explains why, in the vicinity of
threshold, the dotted lines are 
on the low side. 

Since the corrections from ${\cal L}^{(2)}_{\ind N}$ do not enter the
  large scattering length $a^-_{0+}$, the sum of the contributions
  from ${\cal L}^{(1)}_{\ind N}$ and ${\cal L}^{(2)}_{\ind N}$
  (dot-dashed line) is close to the current algebra approximation at
  threshold. At higher energies the second order terms are however
  sizeable: The couplings of ${\cal L}^{(2)}_{\ind N}$ pick up large
  values because of the presence of the $\Delta$ resonance. Although
  the corrections improve the energy dependence, the one loop
  representation underestimates the cross section, in particular, for
  $I=\frac{1}{2}$. In the dispersion integral for $D_s^-$, which
  involves the difference
  $\sigma\rule[-0.5em]{0em}{0em}^{\frac{1}{2}}\rule{0em}{0em}_{tot}-
  \sigma\rule[-0.5em]{0em}{0em}^{\frac{3}{2}}\rule{0em}{0em}_{tot}$,
  the deficiencies happen to cancel, so that the one loop
  approximation for this quantity agrees with the full result, but for
  $D_s^+$, where the imaginary part is given by
  $\sigma\rule[-0.5em]{0em}{0em}^{\frac{1}{2}}
  \rule{0em}{0em}_{tot}+2\,
  \sigma\rule[-0.5em]{0em}{0em}^{\frac{3}{2}}\rule{0em}{0em}_{tot}$,
  the miracle does not happen -- this is why the one loop
  approximation is too small by a factor of two in that case.

The remaining curves show the cross section obtained with the one loop
representation of the scattering amplitude: The dashed lines result if
all combinations of the coupling constants $c_i, d_i, e_i$ that enter
the amplitudes are determined with the subthreshold coefficients,
while for the thin lines, these are evaluated from the scattering
lengths.  Since in the second case the input parameters are all fixed at
threshold, this option yields a somewhat better description at higher
energies, but in either case the one-loop results start deviating
from the experimental values already at rather low energies. 
The fact that the
dispersive part of the one loop representation is determined by the
square of the tree level amplitudes rather than by the square of the
one-loop approximation means that this representation obeys unitarity
only up to contributions of order $q^5$.  The difference between the
dash-dotted and thin lines shows that the violation of
unitarity is numerically quite important, particularly in the case of
$I=\frac{1}{2}$.  The figure also shows that the perturbative
expansion of the amplitude $D^{\frac{1}{2}}$ goes out of control
immediately above threshold: The higher order ``corrections'' exceed
the ``leading'' order contribution.

The approach used in refs.~\cite{Fettes:1998ud,Fettes:2000xg} is
different. There, the main goal is an analysis of the $\pi N$
scattering amplitude in the region above threshold. The coupling
constants are determined by fitting unitarized one loop partial waves
to the data and a rather decent description is obtained -- we expect
that the corresponding results for the total cross section would come
quite close to the experimental curves in the above figures.  The
backside of the coin is that the representation then unavoidably fails
in the subthreshold region, so that it is not possible to establish
contact with the low energy theorems of chiral symmetry.
\section{Analog of the Roy equations}
\label{sec:Roy}
The preceding discussion makes it evident that the chiral representation of
the $\pi N$ scattering amplitude to
$O(q^4)$ can provide a decent
approximation only in the subthreshold region -- a reliable
determination of the $\sigma$-term from the data in the physical region
cannot be performed on this basis.  Even in the case of $\pi\pi$
scattering, where the chiral representation does converge rapidly enough
for the one--loop approximation to yield a decent description of the
amplitude in part of the physical region, the Roy equations provide a more 
accurate framework, which moreover 
also covers significantly higher energies \cite{ACGL,CGL}. 
In the present section, we briefly
outline the steps required to extend the Roy equation analysis to the case 
of $\pi N$ scattering. 

As shown by Roy \cite{Roy}, analyticity and crossing symmetry imply a set of
integral equations for the partial waves of $\pi\pi$ scattering.  
The case of $\pi\pi$ scattering is special
in that the $s$-, $t$- and $u$-channels correspond to the same physical
situation, so that the partial wave decomposition is the same in all three
channels. The Roy equations represent the real parts of the
partial waves in terms of integrals over their imaginary parts and 
two subtraction constants. In the elastic region, unitarity then
converts this system into a set of coupled integral equations.
At low energies, the angular momentum barrier suppresses the higher
partial waves, so that the amplitude 
is dominated by the $S$- and $P$-waves. As the pions are spinless and carry
isospin 1, Bose statistics implies that there are
three such waves, $t^0_0(s)$, $t^1_1(s)$, $t^2_0(s)$ (the upper index denotes
the total isospin of the $\pi\pi$ state, while the lower one specifies its 
total angular momentum).  The effects due to the higher partial waves can be
accounted for in the so--called driving terms. In the elastic region,
however, these are very small and can just as well be dropped. For a detailed
discussion of the properties of the Roy equations, we refer to \cite{ACGL}.

Two complications arise when extending this framework to $\pi N$ scattering:
The proton carries spin $\frac{1}{2}$ and, more importantly, 
there are now two different categories of unitarity cuts that matter at low
energies:  $\pi N$ states, relevant for the discontinuities
in the $s$- and $u$-channels, and states of the type $\pi\pi$ or 
$\bar{N}N$, which generate the discontinuities in the
$t$-channel. Accordingly, we need 
to invoke two different partial wave decompositions. We use the notation
of \cite{Hoehler} and denote the $s$-channel partial waves by
$f_{\ell\pm}(s)$. For $\ell=0$ and $\ell=1$, there are altogether 6 such waves:
\bea\label{eq:fs} f_1^\pm(s)\al\equiv\al f_{0+}^\pm(s)\co\hspace{2em}f_2^\pm(s)
\equiv f_{1-}^\pm(s)
\hspace{2em}f_3^\pm(s)\equiv f_{1+}^\pm(s)\fs\eea
The upper index refers to isospin -- it is convenient to use
the $t$-channel isospin basis, where $f^+$ and $f^-$ correspond to 
$I_t=0$ and $I_t=1$, respectively.

In the $t$-channel, $\bar{N}N$ states of angular momentum
$\ell$ carry parity $(-1)^{\ell+1}$. They can thus only couple to 
$\pi\pi$ states of total angular momentum $J=\ell \pm1$. In the notation of
ref.~\cite{Hoehler}, the corresponding $S$- and $P$-waves are denoted by
$f^J_\pm(t)$, 
where the upper index specifies the total angular momentum $J$, while the
lower one labels the two independent partial waves occurring for $J\geq 1$.
In the $t$-channel, the isospin quantum number is fixed by the total angular
momentum: $J$ even implies $I=0$, $J$ odd corresponds to $I=1$. Ignoring terms
with $J\geq 2$, the partial wave decomposition in the isospin
even channel contains a single wave, while the isospin odd channel 
contains two: 
\bea\label{eq:ft} f_4^+(t)\al\equiv\al f^0_{+}(t)\co\hspace{2em}
 f_4^-(t)\equiv f^1_{-}(t)
\co\hspace{2em} f_5^-(t)\equiv f^1_{+}(t)\fs\eea

Altogether, the discontinuities generated by the $S$- and $P$-waves
thus involve 9 functions of a single variable, to be compared with the
3 functions $t_0^0(s)$, $t_1^1(s)$, $t^2_0(s)$ that occur in the case
of $\pi\pi$ scattering. A closed system of integral equations for
these functions is proposed in appendix \ref{Royappendix}.  It is of
the same form as the Roy equations: 
\bea\label{integral equation}
f_i^+(x)\al=\al f_i^+(x)\rule[-0.2em]{0em}{0em}_{\ind B}+
k_i^+(x)+\sum_{k=1}^4 \int_{s_k}^\infty
dy\,K^+_{ik}(x,y)\,\mbox{Im}\,f_k^+(y) \co\no f_i^-(x)\al=\al
f_i^-(x)\rule[-0.2em]{0em}{0em}_{\ind B}+k_i^-(x)+ \sum_{k=1}^5
\int_{s_k}^\infty dy\,K^-_{ik}(x,y)\,\mbox{Im}\,f_k^-(y) \co\\
s_k\al=\al \left\{\begin{array}{ll}(\m+\M)^2\co&\hspace{2em}k=1,2,3\\
\rule{0em}{1.2em}4\M^2\co&\hspace{2em}k=4,5\end{array}\right.\nonumber
\eea 
The first term on the right represents the partial wave
projection of the pseudovector Born term, while the second contains
the subtraction constants. The kernels $K_{ik}^\pm(x,y)$ are
explicitly known kinematic functions.  The main term on the diagonal
is the familiar Cauchy kernel, 
\bea
K_{ik}^\pm(x,y)=\frac{\delta_{ik}}{\pi(y-x-i\epsilon)}+\ldots
\nonumber 
\eea 

In principle, only one subtraction constant is
required, but additional subtractions may be introduced in order to
arrive at a system of equations that is less sensitive to the
contributions from higher partial waves or higher energies.  In the
case of $\pi\pi$ scattering, it is advantageous to work with two
subtraction constants, although, in principle, one of these is
superfluous: The Olsson sum rule relates a combination of the two
subtraction constants to an integral over total cross sections. The
specific form of the equations proposed in appendix \ref{Royappendix}
for the case of $\pi N$ scattering involves four subtraction
constants.  One of these can be represented as an integral over total
cross sections, while the remaining three are to be determined with
the experimental information available at low energies. In particular, the
beautiful data obtained from pionic atoms \cite{pionic atoms} subject these
constants to a stringent constraint. 

The above system of equations may be viewed as a unitarization of the
low energy representation provided by chiral perturbation theory: The
dispersion integrals describe the low energy singularities that
necessarily accompany the tree graph contributions, on account of
unitarity.  There are two differences to the one loop representation
of the unitarity corrections in eqs.~(\ref{disp Ds}), (\ref{disp Dt}):
(i) That representation accounts for the imaginary parts only to first
nonleading order of the chiral expansion, while the one underlying the
above set of integral equations retains the full imaginary parts of
the $S$- and $P$-waves.  (ii) The chiral representation involves an
oversubtracted set of dispersion integrals. The extra subtractions are
needed, because in that representation, the energy dependence is
expanded in a Taylor series. The problem does not occur in the above
framework, because unitarity implies that the partial waves tend to
zero at high energies.

In the case of $\pi\pi$ scattering, the integral equations can be written in
exact form, where all of the partial waves are accounted for. 
The equations for the $S$- and $P$-waves then contain corrections from
the higher angular momenta, collected in the driving terms mentioned above. 
We do not know of a
corresponding set of exact equations for the case of $\pi N$ scattering -- 
driving terms would have to be added to convert (\ref{integral equation}) 
into an exact system. The size of these corrections depends on the number
of subtractions made and on the range of energies considered. 
With four subtractions, they should be
negligibly small in the elastic region, because the bulk of the contributions 
from the partial waves with $\ell\geq 2$ is then absorbed in the subtraction 
constants. The available detailed phenomenological information about the
imaginary parts of the higher waves \cite{Hoehler} allows an explicit 
evaluation of their contributions, so that this can be checked.

Alternative unitarizations are proposed in the literature
\cite{unitarization,Leisi}. The advantage of the present proposal is that the
resulting representation of the scattering amplitude does not introduce any
fictitious singularities. In particular, once the
system of equations is solved and the subtraction constants are determined,
the representation yields a formula for the value of the scattering amplitude
at the Cheng--Dashen point:
\bea \Sigma\al=\al\F^2\left\{d_{00}^++2\M^2 d_{01}^++64\M^4\int_{4\M^2}^\infty
\!\!  dt\;\frac{\mbox{Im}\,f_4^+(t)}{t^2\,(t-2\M^2)\,(4\m^2-t)}\right\}\co\no
C\al=\al 2\F^2\left\{d_{00}^-+12\M^2 \int_{4M^2}^\infty \!\!dt\;\frac{8\m
\mbox{Im}\,f_5^-(t)-\sqrt{2}\,t\,\mbox{Im}\,f_4^-(t)}{t\,(t-2\M^2)\,(4\m^2-t)}
\right\}\fs\eea

The approach outlined above is closely related to the one used by
R.~Koch \cite{Koch:1986bn}, who relies on fixed $\nu$ dispersion
relations to study the $t$-dependence.  Also, a similar method was
used in ref.~\cite{GLLS} to extract the value of the $\sigma$-term
from the data available at the time.  For more recent work in this
direction, we refer to \cite{Anant Buettiker}, where hyperbolic
dispersion relations \cite{hite} are used to supplement the partial
wave relations \cite{Hoehler,Steiner} for $K\pi$ scattering.

\setcounter{equation}{0}
\section{Summary and conclusion}
\renewcommand{\labelenumi}{\arabic{enumi}.}
\begin{list}{\bf \arabic{enumi}.}{\usecounter{enumi}
    \setlength{\labelwidth}{0.8cm} \setlength{\leftmargin}{0cm}
    \setlength{\labelsep}{0.2cm}\setlength{\rightmargin}{0cm}
    \setlength{\itemindent}{1cm}}

\item We have derived a representation of the $\pi N$ scattering amplitude
  that is valid to the fourth order of the chiral expansion. The infrared
  singularities occurring in that representation are stronger than for
  $\pi\pi$ scattering. 
  The difference also manifests itself in the expansion of the
  nucleon mass or of the $\sigma$-term in powers of the quark masses, which
  contain contributions of $O(\M^3)$ that are quite important numerically. The
  Goldberger-Treiman 
  relation and the low energy theorem that relates the $\sigma$-term to the
  value of the scattering amplitude at the Cheng-Dashen point, on the other
  hand, do not contain infrared singularities, at the order considered.

\item We have shown that the Mandelstam double spectral function only
  contributes beyond the order of the low energy expansion examined in
  the present paper. The dependence of the box graph on the
  momentum transfer, for instance, can be expanded in powers of
  $t/4\m^2$ -- the cut due to $N\bar{N}$
  intermediate states in the $t$-channel does not manifest itself at finite
  orders of the low energy expansion. Up to and
  including $O(q^4)$, the scattering amplitude can be 
  written in terms of functions of a single variable, either $s$ or $t$ or
  $u$. These functions describe the singularities associated with the
  unitarity cuts. In the framework of the effective theory, they arise from
  the one loop graphs.  We have given an explicit, dispersive representation
  for these functions, which accounts for all terms
  to $O(q^4)$ and sums up those higher order contributions that
  are generated by the threshold singularities.
  
\item The chiral expansion of the dispersion integrals reduces these to
  algebraic expressions involving elementary functions such as
  arccos($-\omega/\M$). On general grounds, the result of this expansion 
  must agree with what is
  obtained in HBCHPT, where the loop integrals are expanded ab initio. That
  expansion, however, is a subtle matter, because the integrals
  contain infrared singularities. In particular, the choice of the
  kinematic variable to be kept fixed
  plays an important role.  While the expansion at fixed $\omega/\M$ does not
  lead to a decent description of the full loop integrals, the one at fixed
  $\omega_q/\M$ does yield an approximate 
  representation for the relativistic integrals, even above
  threshold.
  
\item While the expansion at fixed $\omega_q/\M$ solves one problem,
  it generates another. To absorb the divergences of the loop
  integrals in the coupling constants of the effective Lagrangian, the
  integrals must be expanded in powers of $\M$ at fixed $\omega/\M$,
  so that an expansion of the subtraction terms is required, which
  converges only slowly.  Algebraically, the accuracy of the
  representation does not depend on the splitting between the
  polynomial part and the cut contribution, but numerically, the
  amplitude is very sensitive to this choice. We find it more
  convenient not to invoke an expansion of the integrals -- the
  problem does not occur in the relativistic representation that forms
  the basis of our work.

\item We find that the dependence of the amplitude on the momentum
  transfer $t$ is adequately described by the contributions from the
  one loop graphs, so that the extrapolation from the Cheng-Dashen
  point to $t=0$ does not pose a significant problem. The
  representation obtained for the dependence on the variables $s$ and
  $u$, however, has a very limited range of validity. Higher order
  effects, such as those due to the $\Delta$ do matter.  The one loop
  representation does not provide the bridge needed to connect the
  value of the amplitude at the Cheng-Dashen point to the physical
  region.

\item We conclude that dispersive methods are required to obtain a
  reliable description of the scattering amplitude at low energies.
  With this in mind, we propose a system of integral equations that
  is analogous to the Roy equations for $\pi\pi$ scattering and 
  interrelates the lowest partial wave amplitudes associated with the
  $s$-, $t$- and $u$-channels. The structure of the amplitude that
  underlies this system is very similar to the
  dispersive representation mentioned above, but the constraints imposed
  on the partial waves by unitarity are now strictly obeyed.  It remains to be
  seen whether or not this set of equations provides a useful framework for
  the analysis of the low energy structure, 
  in particular for a reliable determination of the $\sigma$-term. 
\end{list} 

\vspace{2em}
\noindent{\Large\bf Acknowledgment}

\vspace{0.5em}\noindent We greatly profited from J\"urg Gasser's
knowledge of the subject. This work would not have been possible
without his help and cooperation. We thank P.~B\"uttiker and J.~Stahov
for making the KA84 solution available to us and G.~Oades for
communicating his results for the subthreshold coefficients prior to
publication. Also, we acknowledge useful discussions and
correspondence with B.~Ananthanarayan, G.~Ecker, R.~Kaiser,
U.G.~Meissner, H.~Neufeld and U.~Raha. This work was supported by the
Swiss National Science Foundation, and by TMR, BBW--Contract
No. 97.0131 and EC--Contract No. ERBFMRX--CT980169 (EURODA$\Phi$NE).

\newpage

\begin{appendix}
\section{Notation}\label{notation}
We use the notation of H\"ohler's ``Collection
of Pion-Nucleon Scattering Formulas''. (Appendix of \cite{Hoehler}.)
\subsection*{Kinematics}
\begin{flalign*}
&P\co q &&\text{four momenta of the incoming nucleon and pion}\\
&P'\co q' &&\text{four momenta of the outgoing nucleon and pion}\\
\end{flalign*}\vspace{-1cm}
\begin{flalign*}
&s=(P+q)^2\co\;\; t=(q-q')^2\co\;\; u=(P-q')^2\co\;\;s+t+u=2\M^2+2\m^2& \\ 
\end{flalign*}
\begin{flalign*}
&\omega_q=\M\,\Omq=\frac{s-\m^2+\M^2}{2\,\sqrt{s}} && \text{c.~m.~pion
  energy}\\ 
&E=\frac{s+\m^2-\M^2}{2\,\sqrt{s}} && \text{c.~m.~nucleon
  energy}\\ 
&q=\sqrt{\omega_q^2-\M^2} &&\text{c.~m.~momentum} \\ 
&\cos\theta=1+\frac{t}{2q^2} && \text{scattering angle (c.~m.~system)}\\&&&\\
&\omega=\M\,\Omega=\frac{s-\m^2-\M^2}{2\,\m} && \text{lab.~energy of the
  incoming pion}\\
&k=\sqrt{\omega^2-\M^2} &&\text{lab.~momentum of the incoming pion}\\ &&&\\
&\nu=\frac{s-u}{4\m}=\omega+\frac{t}{4\m} && \\
&\nu_B=\frac{1}{4\m}(t-2\M^2) && 
\end{flalign*}

\subsection*{Isospin}
We denote the amplitudes for the reactions $\pi^\pm p\rightarrow \pi^\pm p$ by
$A_\pm$, the charge exchange reaction amplitude $\pi^- p\rightarrow \pi^0 n$
by $A_0$.  The amplitudes of the isospin eigenstates are denoted by $A^I$,
$I=\frac{1}{2},\frac{3}{2}$. The relations to the isospin odd and even
amplitudes $A^\pm$ are
\begin{flalign*}
A_+&=A^{\frac{3}{2}} =A^+-A^- \co &
A_-&=\mbox{$\frac{1}{3}$}(2\,A^{\frac{1}{2}}+A^{\frac{3}{2}}) =A^+ + A^-\\
A_0&=\mbox{$\frac{\sqrt{2}}{3}$}(A^{\frac{3}{2}}-A^{\frac{1}{2}})=-\sqrt{2}\,A^-\co &
A^\frac{1}{2}&=\mbox{$\frac{1}{2}$}(3A_- - A_+)=A^++2\,A^- \\
A^+&=\mbox{$\frac{1}{3}$}(A^{\frac{1}{2}}+2\,A^{\frac{3}{2}})\co&
A^-& =\mbox{$\frac{1}{3}$}(A^{\frac{1}{2}} - A^{\frac{3}{2}})&
\end{flalign*}

\subsection*{Partial waves}
For the partial wave decomposition, the amplitude
\bea C=A+\frac{\nu}{1-t/4\m^2}\,B =D+\frac{\nu \,t}{4\m^2-t}\,B\nonumber\eea 
is more convenient to work with than $A$ or $D$. 

In the $s$-channel, the decomposition reads
\bea\label{schannelPW} 
C\al=\al \frac{2\pi\sqrt{s}}{p_-^2}\sum_{\ell=0}^\infty 
(\ell+1)f_{\ell
    +}(s)\left\{(E+\m)P_\ell(z)-(E-\m)P_{\ell+1}(z)\right\}\no
\al \al + \frac{2\pi\sqrt{s}}{p_-^2}\sum_{\ell=1}^\infty
\ell f_{\ell -}(s)\left\{(E+\m)P_\ell(z)-
(E-\m)P_{\ell-1}(z)\right\}\co\no
B\al=\al \frac{4\pi}{q^2}\sum_{\ell=0}^\infty f_{\ell
    +}(s)\left\{-(E+\m)P_\ell'(z)+(E-\m)P_{\ell+1}'(z)\right\}\\
\al \al +\frac{4\pi}{q^2}\sum_{\ell=1}^\infty f_{\ell -}(s)
\left\{(E+\m)P_\ell'(z)-(E-\m)P_{\ell-1}'(z)\right\}\co\no
t\al=\al 2q^2(z-1)\co\hspace{2em}p_-=\sqrt{\m^2-\mbox{$\frac{1}{4}$}\,t}
\fs\nonumber\eea

The $t$-channel partial waves are defined by
\bea\label{tchannelPW} C\al=\al\frac{8\pi}{p_-^2}
\sum_{J=0}^\infty (J+\mbox{$\frac{1}{2}$})\,(p_- q_-)^J P_J(Z) f_+^J(t)\co\no
 B\al=\al 8\pi\sum_{J=1}^\infty\frac{J+\frac{1}{2}}{\sqrt{J(J+1)}}
\,(p_-q_-)^{J-1} P_J'(Z) f_-^J(t)\co\\
\nu\al=\al\frac{Z\,p_-q_-}{\m}\co\hspace{2em} 
p_-=\sqrt{\m^2-\mbox{$\frac{1}{4}$}\,t}\co\hspace{2em}
q_-=\sqrt{\M^2-\mbox{$\frac{1}{4}$}\,t}\fs\nonumber\eea 

\subsection*{Threshold expansion}
We write the threshold expansion of the amplitude $X\in \{D^+, D^-,
B^+, B^-\}$ in the form
\begin{equation*}
\re X(q^2,t)=X_{00}+X_{10}\,q^2+X_{01}\,t+X_{20}\,q^4+X_{11}\,q^2\,t+
X_{02}\,t^2+\dots
\end{equation*}
The coefficients $X_{nm}$ of this expansion are related to the scattering
lengths and effective ranges. For the threshold expansion of the real part of
the partial wave amplitudes
\begin{equation*}
T_{l\pm}=q\,f_{l\pm}=\frac{1}{2\,i}\,(\eta_{\pm} e^{2\,i\,\delta_{l_\pm}}-1)
\end{equation*}
we use the notation
\begin{equation*}
\re T_{l\pm}=q^{2l+1}(a_{l\pm}+b_{l\pm}\,q^2+c_{l\pm}\,q^4+\ldots)\fs
\end{equation*}
The lowest coefficients of the threshold expansion of the amplitudes $D$
and $B$ are obtained from the expansion of the partial wave amplitudes via 
\begin{align*}
D_{00}&= 4 \,\pi\,\left( 1 +  \alpha\right)\,a_{0+} \\
D_{10}&= 4\,\pi \left( 1 + \alpha \right) \,
\Big\{  \frac{a_{0+}}{2\,\alpha\, \m^2}  + b_{0+}  + 
 a_{1-} + 2\,a_{1+} \Big\}   \\
D_{01} &= 2\,\pi\Big\{
\frac{a_{0+} }{4\,\m^2}+ a_{1-}  + a_{1+}\,\left( 2 +  3\,\alpha \right)  
\Big\}\\
D_{20}&=\frac{2\,\pi\,(1+\alpha)}{\alpha\, \m^2}\,\Big\{
-(1-\alpha+\alpha^2)\,\frac{a_{0+}}{4\,\alpha^2\m^2}+b_{0+}+
2\,\alpha\,\m^2\,c_{0+}+a_{1-} \\
&\hspace{3em}  \rule[-1em]{0em}{1em}+\,2\, \alpha\,\m^2\,b_{1-}+
2\,a_{1+}+4\,\alpha\,
\m^2\,b_{1+}+4\,\alpha\,\m^2\,a_{2-}+6\,\alpha\,\m^2\,a_{2+}\Big\}
\end{align*}
\begin{align*} 
D_{11}&=\frac{\pi}{2\,\m^2}\Big\{-\frac{a_{0+}}{4\m^2}+b_{0+}+a_{1-}+
4\,\m^2\,b_{1-}+2\,(3+4\,\alpha)\,\frac{a_{1+}}{\alpha} \\
&\hspace{2.5em}  \rule[-1em]{0em}{1em}+4\,(2+3\,\alpha)\,\m^2 \,b_{1+}
+12\,(2+\alpha)\,\m^2\,a_{2-}+12\,(3+4\,\alpha)\,\m^2\,a_{2+}\Big\}\\
D_{02}&=\frac{3\,\pi}{4\,\m^2}\,\Big\{a_{1+}+4\,\m^2\,
a_{2-}+2\,(3+5\,\alpha)\,\m^2 \,a_{2+}\Big\}\\
B_{00}&= 8\,\pi\,\m\Big\{\frac{ a_{0+}}{4\m^2} +  a_{1-} -  a_{1+}
\Big\}\\
B_{10}&=\frac{2\,\pi}{\m}\,\Big\{-\frac{a_{0+}}{4\,\m^2}+b_{0+}+a_{1-}+
4\,\m^2\,b_{1-}+2\,a_{1+}\\
&\hspace{14em}  \rule[-1em]{0em}{1em}-4\,\m^2\,b_{1+}+12\, \m^2\, a_{2-}-
12\,\m^2\,a_{2+}\Big\}\\
B_{01}&=\frac{3\,\pi}{\m}\,\Big\{a_{1+}+4\,\m^2\,a_{2-}-4\,\m^2\,a_{2+}\Big\}\fs
\end{align*}

\section{Off-shell {\boldmath$\pi N$\unboldmath}-amplitude}
\label{app:sources}
As discussed in the text, the off-shell amplitude of the effective theory
is without physical meaning. In the underlying theory, however, 
an unambiguous off-shell extrapolation 
can be constructed from correlators of currents
with the appropriate quantum numbers. For the pion 
field, the operator of lowest dimension is the pseudoscalar density $P^a(x)$
\begin{equation*}
P^a(x)=\bar{q}(x)\,i\gamma_5 \tau^a\, q(x)\co\;a=1,2,3
\end{equation*}
which couples to the pion with strength $G_\pi$,
\begin{equation*}
\langle 0|\,{P}^a(0)\,|\,\pi^b \rangle = G_\pi\,\delta^{ab}\fs
\end{equation*}
To generate nucleons, we need a three-quark operator with the correct
quantum numbers. At lowest dimension, there are two independent such 
operators:
\begin{align*}
{\bf N}_1(x)&=\Big(q_\alpha(x)\,i{\tau}^2\,{\cal C}\, q_\beta(x)\Big) 
\gamma_5 q_\gamma(x)
\epsilon^{\alpha\beta\gamma}\\
{\bf N}_2(x)&=\Big(q_\alpha(x)\,i{\tau}^2\,{\cal C}\,\gamma_5 q_\alpha(x)\Big)
q_\gamma(x)\,\epsilon^{\alpha\beta\gamma} 
\end{align*}
In $q_\alpha(x)=(u_\alpha(x),d_\alpha(x))$ we have collected the quarks with
color index $\alpha$. 
The brackets indicate a sum over Dirac and flavor indices, $i\tau^2$ stands
for the antisymmetric tensor in SU(2) flavor space and ${\cal
  C}=-i\gamma_0\gamma_2$ is the charge conjugation matrix. These operators
  couple to the proton with coupling strength $G_{1,2}$. 
\begin{equation*}
\langle 0|\,{\bf N}_{1,2}(x)\,|\,P,s \rangle = G_{1,2}(\mu)\, {\bf u}(P,s)
\end{equation*}
Since these operators carry anomalous dimension, their matrix elements 
depend on the running
scale of QCD, but are otherwise free from ambiguities. Their Green functions
contain contact terms that make it difficult to handle the corresponding
generating functional -- the perturbation generated by the source terms
is not renormalizable. We may, however, restrict ourselves to Green functions
involving only two nucleon fields, so that a finite number of counter 
terms suffices. The pion nucleon scattering amplitude can be extracted
e.~g.~from the Green function
\begin{equation*}
\langle 0|\, {\bf T}\Big[{\bf
  N}_1(x')\,{P}^{a'}(y')\,P^a(y)\,\bar{\bf N}_1(x)\,\Big]|0\rangle \fs
\end{equation*}
Under a chiral transformation $q(x)\rightarrow\{
\frac{1}{2}(1+\gamma_5)V_R+\frac{1}{2}(1-\gamma_5)V_L\}q(x)$ the operator ${\bf
  N}_1$ transforms as
\begin{equation*}
{\bf N}_1\rightarrow \{\frac{1}{2}(1+\gamma_5)V_R-\frac{1}{2}(1-\gamma_5)V_L
\} {\bf N}_1
\end{equation*}
In our effective theory, it is easy to identify an object that
transforms in the same way, namely $\Psi=u\, \psi_R-u^\dag\,\psi_L$.
If we would add a source term $\bar{\eta}_1\,{\bf N}_1+\bar{\bf N}_1\,\eta$ to
the Lagrangian of QCD, this source would couple to the field $\Psi$ in our
effective Lagrangian and we would be able to compute the above Green
function with an extended version of our effective Lagrangian. Since the two
fields $\Psi(x)$ and $\psi(x)$ do generate the same on-shell amplitudes we can
also use the field $\psi(x)$ to extract these, as we did in our calculation.
  
\section{One-loop integrals}\label{Loop integrals}
\setcounter{equation}{0}
In this appendix, we list the integrals that occur in the evaluation
of the scattering amplitude to one loop. Throughout, we use 
infrared regularization (for a detailed discussion of that method, we
refer to \cite{Becher Leutwyler 1999}).
We perform the tensor decomposition of the integrals with respect to the
 sums and the differences of the external momenta,
\bea
\Sigma^\mu\al=\al (P+q)^\mu=(P'+q')^\mu \nonumber\\
Q^\mu\al=\al (P'+P)^\mu \nonumber\\ 
\Delta^\mu\al=\al(q'-q)^\mu=(P-P')^\mu \nonumber\fs\eea
and put all external legs on their mass shell,
\bea P^2=P^{\prime 2}=m^2\co\hspace{2em} q^2=q^{\prime 2}=M^2\fs
\nonumber \eea
The coefficients represent Lorentz invariant functions of the Mandelstam
variables. The squares of the vectors
$\Sigma$, $\Delta$ and $Q$ are given by
\bea \Sigma^2\al= \al s\co\hspace{2em} \Delta^2=t \co 
\hspace{2em}Q^2=4 m^2-t \nonumber \eea
\subsubsection*{1 meson:\; $\Dpi=I_{10}$}
\bea \Dpi =\frac{1}{i}\int_I 
\frac{d^dk}{(2\pi)^d}\,\frac{1}
{M^2-k^2}= 2M^2\,\lambda_\pi
\nonumber\eea
\subsubsection*{1 nucleon:\; $\DN=I_{01}$}
\bea \DN =\frac{1}{i}\int_I \frac{d^dk}{(2\pi)^d}\,\frac{1}
{m^2-k^2}=0\nonumber\eea 
Note that integrals formed exclusively with nucleon propagators
vanish in infrared regularization: $I_{0n}=0$.
\subsubsection*{2 mesons:\; $J=I_{20}$}
\bea \al\al \{J\,,\,J^\mu\,,\,J^{\mu\nu}\}=
\frac{1}{i}\int_{I}\frac{d^dk}{(2\pi)^d}\,
\frac{\{1\,,\,k^\mu\,,\,k^\mu k^\nu\}}
{(M^2-k^2)(M^2-(k-\Delta)^2)}\rule{4em}{0em}\nonumber \\
\al\al J^\mu=\mbox{$\frac{1}{2}$}\,\Delta^\mu\,J(t)\rule{0em}{1.7em}\nonumber\\
\al\al J^{\mu\nu}=
(\Delta^\mu \Delta^\nu-g^{\mu\nu} \Delta^2)\,J^{(1)}(t)+\Delta^\mu 
\Delta^\nu\,J^{(2)}(t) \rule{0em}{1.7em}\nonumber \eea
\subsubsection*{1 meson, 1 nucleon:\; $I=I_{11}$}
\bea\al\al \{I,\,I^\mu,I^{\mu\nu}\}=
\frac{1}{i}\int_{I}\frac{d^dk}{(2\pi)^d}\,
\frac{\{1\,,\,k^\mu\,,\,k^\mu\,k^\nu\,\}}{(M^2-k^2)(m^2-(\Sigma-k)^2)}
\nonumber\\
\al\al I^\mu=\,\Sigma^\mu\,I^{(1)}(s)\rule{0em}{1.7em}\nonumber\\
\al\al I^{\mu\nu}=\, g^{\mu\nu} I^{(2)}(s)+\,\Sigma^\mu \Sigma^\nu I^{(3)}(s)
\rule{0em}{1.7em}\nonumber\eea 
\subsubsection*{2 mesons, 1 nucleon:}
\bea \al\al\{I_{21}\,,\,I_{21}^{\;\;\mu}\,,\,I_{21}^{\;\;\mu\nu}\}=
\frac{1}{i}\!\int_I\frac{d^d
  k}{(2\pi)^d}\frac{\{1\,,\,k^\mu\,,\,k^\mu\,k^\nu\,\}} 
{(\Mp^2-k^2)\,
(\Mp^2-(k-\Delta)^2)\,(\Mbare^2-(P-k)^2)} \nonumber \\
\al\al I_{21}^{\;\;\mu}=\,Q^\mu\,I_{21}^{(1)}(t)+
\mbox{$\frac{1}{2}$}
\,\Delta^\mu I_{21}(t)\co\rule{0em}{1.7em}\nonumber \\
\al\al I_{21}^{\;\;\mu\nu}=\,g^{\mu\nu} I_{21}^{(2)}(t)+Q^\mu Q^\nu
I_{21}^{(3)}+ \Delta^\mu \Delta^\nu I_{21}^{(4)}(t)+ (\Delta^\mu Q^\nu+Q^\mu
\Delta^\nu)\mbox{$\frac{1}{2}$} I_{21}^{(1)}(t) \co\rule{0em}{1.7em}\nonumber
\eea
\subsubsection*{1 meson, 2 nucleons: $I_{12}$, $I_A$, $I_B$.}
\bea \al\al\{I_{12}\,,\,I_{12}^{\;\;\mu}\,,\,I_{12}^{\;\;\mu\nu}\}=
\frac{1}{i}\!\int_I\frac{d^d k}
{(2\pi)^d}\frac{\{1\,,\,k^\mu\,,\,k^\mu\,k^\nu\,\}} {(\Mp^2-k^2)\,
(m^2-(P_1-k)^2)\,(m^2-(P_2-k)^2)}\nonumber \eea With $Q= P_1+P_2$ and
$\Delta= P_1-P_2$, the tensorial decomposition reads \bea\al\al
I_{12}^{\;\;\mu}=\,Q^\mu\,I_{12}^{(1)}+ \,\Delta^\mu I_{12}^{(2)} \co
\nonumber\rule{0em}{1.7em} \\ \al\al I_{12}^{\;\;\mu\nu}=\,g^{\mu\nu}
I_{12}^{(3)}+Q^\mu Q^\nu I_{12}^{(4)}+\Delta^\mu \Delta^\nu
I_{12}^{(5)}+ (\Delta^\mu Q^\nu+Q^\mu \Delta^\nu) I_{12}^{(6)}
\fs\rule{0em}{1.7em}\nonumber \eea  
Several different topologies give rise to this integral: the
graphs (c), (d), (g), (h) and (m) in fig.~\ref{fig:loop}. The values
of $P_1$ and $P_2$ differ from graph to graph. In all cases, however,
one of the two momenta is on the mass shell -- we set $P_1^2=m^2$. The
coefficients of the tensor decomposition then only depend on two
variables: 
\bea I^{(n)}_{12}=I^{(n)}_{12}(s,t)\co\hspace{2em}
P_1^2=m^2\co \hspace{2em} P_2^2=s\co \hspace{2em}t= (P_1-P_2)^2\fs
\nonumber\eea 
In the case of the graph (m), we have $P_1=P$, $P_2=P'$,
so that we are dealing with the special case $s=m^2$. In the case of
(c), (d), (g) and (h) on the other hand, the difference $P_1-P_2$
represents the momentum of an on-shell pion, so that $t=M^2$.  To
simplify the notation, we introduce invariant functions that
correspond to these two special cases and only depend on a single
variable: \bea I_A(t)=I_{12}(m^2,t)\co\hspace{2em}
I_B(s)=I_{12}(s,M^2)\fs\nonumber \eea 
The coefficients of the
decomposition of the corresponding tensorial integrals are \bea 
\al\al I_{A}^{\;\;\mu}(t)=\,Q^\mu\,I_{A}^{(1)}(t) \co
\nonumber\rule{0em}{1.7em}\\ \al\al I_{A}^{\;\;\mu\nu}(t)=\,g^{\mu\nu}
I_{A}^{(2)}(t) + Q^\mu Q^\nu I_{A}^{(3)}(t) + \Delta^\mu \Delta^\nu
I_{A}^{(4)}(t) \rule{0em}{1.7em}\nonumber \\ \al\al I_B^{(n)}(s) =
I^{(n)}_{12}(s,M^2) \fs\nonumber\rule{0em}{1.7em}
\eea

\subsubsection*{1 meson, 3 nucleons:} 
\bea\al\al\hspace{-1em} \{I_{13}\,,\,I_{13}^{\;\;\mu}\}=\no
\al\al\hspace{1em} \frac{1}{i}\!\int_I\!\frac{d^d k}
{(2\pi)^d}\frac{\{1\,,\,k^\mu\,\}}{(\Mp^2-k^2)
(m^2-(P-k)^2)(m^2-(\Sigma-k)^2)(m^2-(P'-k)^2)} \nonumber \\
\al\al I_{13}^\mu(s,t)=\, Q^\mu\,I_{13}^{(1)}(s,t)+
(\Delta+2q)^\mu\,I_{13}^{(2)}(s,t)\rule{0em}{1.7em} \nonumber
\eea

\section{Results for the loop graphs}\label{Loop graphs}  

In the present appendix, we list the contributions from the various
topologies shown in fig.~\ref{fig:loop}. Note that 
the amplitude in addition contains the contributions
obtained from the graphs (a)--(i) and (n)--(s) 
through crossing,
\begin{align}
  A_{{\scriptscriptstyle G}\;tot}^\pm(s,t)&=
A_{\scriptscriptstyle G}^\pm(s,t) \pm A_{\scriptscriptstyle G}^\pm(u,t) \no
  B_{{\scriptscriptstyle G}\;tot}^\pm(s,t)&=
B_{\scriptscriptstyle G}^\pm(s,t) \mp B_{\scriptscriptstyle G}^\pm(u,t) 
\hspace{3em}\mbox{\small\it G}\in(a,\,\ldots \,,i,n,\,\ldots\,,s)\no
  D_{{\scriptscriptstyle G}\;tot}^\pm(s,t)&=
D_{\scriptscriptstyle G}^\pm(s,t) \pm D_{\scriptscriptstyle G}^\pm(u,t) 
\nonumber\end{align}

\subsection[Loop graphs of ${\cal L}^{(1)}$]{Loop graphs of {\boldmath${\cal
      L}^{(1)}$\unboldmath}}\label{graphsDrei} 
The contribution of the loops formed exclusively with vertices of the lowest
order Lagrangian ${\cal L}^{(1)}$ starts at order $O(q^3)$. We give the exact
result for the ${\cal L}^{(1)}$ loops in terms of the amplitudes $A(s,t,u)$,
$B(s,t,u)$, since the corresponding expressions can be written in a more
compact form. Note that the amplitude $A(s,t,u)$ does not obey the chiral power
counting rules.

We use the following abbreviation:
\begin{align}
F(s) &= -2\, M^2\, I(s) + (s-m^2)\,  I^{(1)}(s) \fs  \nonumber
\end{align}
\subsubsection*{graphs a+b} 
\begin{flalign} 
A^+_{ab} &= \frac{g_A^2\, m\, F(s)}{2\, F^4}  &
B^+_{ab} &= -\frac{g_A^2}{2\, F^4} \Big\{ {\frac{2\, m^2\, F(s)}{s-m^2}} + M^2
  I(s) + F(s)\Big\}   \nonumber \\ &&\nonumber  \\ 
A^-_{ab} &= A^+_{ab} &
B^-_{ab} &= B^+_{ab}  \nonumber
\end{flalign}
\subsubsection*{graphs c+d} 
\begin{flalign} 
A^+_{cd} &= \frac{g_A^4\, m}{8\, F^4} \Big\{-2\, \Delta_\pi + (s-m^2)  
I^{(1)}(s)  - 8\, m^2\left(M^2\, {I_B}(s) - (s-m^2) \, I_B^{(2)}(s)\right)
\Big\}  & \nonumber \\ 
B^+_{cd} &= {\frac{g_A^4}{8\, F^4}} \Big\{-M^2\, I(s) + (s-m^2) \left(
I^{(1)}(s) - 4
 m^2\, I_B^{(1)}(s)\right) & \nonumber \\ 
&\hspace{2.5cm}+ {\frac{4\, m^2}{s-m^2}} \left[\Delta_\pi + (3\, m^2 + s) 
\left (M^2\,
{I_B}(s) - (s-m^2)\,  I_B^{(2)}(s)\right)\right]\Big\} & \nonumber \\ &&
\nonumber  \\  
A^-_{cd} &= A^+_{cd} \cspace  B^-_{cd} = B^+_{cd} & \nonumber
\end{flalign}
\subsubsection*{graph e} 
\begin{flalign} 
A^+_{e} &= {\frac{3\, g_A^4\, m}{16\, F^4}} \Big\{ {\frac{4\, m^2\, F(s)}
{m^2-s}} - \left(2\,  M^2\, I(s) + 3\, F(s)\right) \Big\}  & \nonumber \\ 
B^+_{e} &= {\frac{3\, g_A^4}{16\, F^4}} \Big\{F(s) + M^2\, I(s) - {\frac{4\,
    m^2}{m^2-s}} \left(M^2\, I(s) + 2\, F(s)\right)  + {\frac{8\, m^4\, F(s)}
{{{(m^2-s)
        }^2}}}\Big\}   \nonumber & \\  &&\nonumber  \\  
A^-_{e} &= A^+_{e} \cspace 
B^-_{e} = B^+_{e}& \nonumber
\end{flalign}
\subsubsection*{graph f} 
\begin{flalign}
A^+_{f} &= {\frac{m\, (s-m^2)\,  I^{(1)}(s)}{2\, F^4}}  &
B^+_{f} &= {\frac{1}{8\, F^4}}  \Big\{-4\, M^2\, I(s) - \Delta_\pi + 4\, 
(s-m^2) \,  I^{(1)}(s)\Big\}
  \nonumber \\ &&\nonumber  \\  
A^-_{f} &= {\frac{A^+_{f}}{2}} &
B^-_{f} &= {\frac{B^+_{f}}{2}}  \nonumber
\end{flalign}
\subsubsection*{graphs g+h}
\begin{flalign}
A^+_{gh} &= {\frac{g_A^2\, m}{2\, F^4}}\, (s-m^2)   \Big\{-2\, I(s) + 
I^{(1)}(s) + 8\,
m^2\, I_B^{(1)}(s)\Big\}   & \nonumber \\ 
B^+_{gh} &= {\frac{g_A^2}{4\, F^4}}  \Big\{-M^2\, I(m^2) - 2\, \Delta_\pi - 
8\, m^2
M^2\, {I_B}(s) & \nonumber \\
&\hspace{2.5cm} + 2 (m^2-s)\left [I(s) - I^{(1)}(s) - 4\, m^2 \left(I_B^{(1)}(s) +  I_B^{(2)}(s)\right) \right]
\Big\} & \nonumber \\  &&\nonumber  \\   
A^-_{gh} &= 0 \cspace
B^-_{gh} = 0 &  \nonumber
\end{flalign}
\subsubsection*{graph i} 
\begin{flalign}
A^+_{i} &= {\frac{3\, g_A^4\, m}{16\, F^4}} \Big\{2\, M^2\, \left(I(m^2) - 
I(s)\right)  + (s-m^2)
\left(2\, I(s) + I^{(1)}(s)\right) & \nonumber \\
&\hspace{0.1cm} + 8\, m^2\Big[-M^2\, {I_A}(t) + 4\, m^2\, I_A^{(1)}(t) 
   -  (s - u)\,
I_A^{(3)}(t) &
\nonumber \\&\phantom{=} + M^2 \left({I_B}(s) - I_B^{(1)}(s) - I_B^{(2)}(s)\right)  - (m^2 + 3 s)\,
I_B^{(1)}(s) - (s-m^2)\,  I_B^{(2)}(s)\Big]   & \nonumber \\
&\hspace{1cm}+ 32\, m^4\,  (s-m^2)\,
I_{13}^{(1)}(s,t) \Big\} 
 &\nonumber \\ & &\nonumber \\ 
B^+_{i} &= {\frac{3\, g_A^4}{16\, F^4}} \Big\{ \left(3\, m^2 + s\right)  I(s) 
+ 4\, m^2  I^{(1)}(m^2) -
\left(m^2 + s \right)  I^{(1)}(s) & \nonumber \\
&\hspace{0.1cm} + 4\, m^2\, \Big[ M^2\, {I_A}(t) - 2\, I_A^{(2)}(t) + 2\, M^2\,  
{I_B}(s)  - 2\, (3\, m^2 + s)
I_B^{(1)}(s) & \nonumber \\
&\phantom{=} + 2\, (m^2-s) \,  I_B^{(2)}(s)\Big]  + 16\, m^4\, \Big[-M^2\, 
{I_{13}}(s,t) + 2\, (s-m^2) \,  I_{13}^{(2)}(s,t)\Big] \Big\}  & \nonumber \\
&&\nonumber  \\    
A^-_{i} &=- {\frac{1}{3}}\,A^+_{i} \cspace
B^-_{i} = - {\frac{1}{3}}\, B^+_{i} & \nonumber 
\end{flalign}
\subsubsection*{graph k} 
\begin{flalign}
A^+_{k} &=
B^+_{k} =A^-_{k} = 0 \cspace 
B^-_{k} = \frac{t\, J^{(1)}(t)}{F^4} & \nonumber
\end{flalign}
\subsubsection*{graph l} 
\begin{flalign}
A^+_{l} &= \frac{g_A^2\, m}{2\, F^4}\Big\{2 M^2\, I(m^2) - \left(M^2 - 2\,
  t\right) \left (J(t) -  4\, m^2\, 
I_{21}^{(1)}(t)\, \right) \Big\} \hspace{1.2cm}   B^+_{l}=0 &&  \nonumber \\ 
A^-_{l} &= -\frac{4\, g_A^2\, m^3}{F^4}\, (s - u)\, I_{21}^{(3)}(t)
\hspace{2cm} B^-_{l} = -\frac{g_A^2}{F^4}\Big\{ t\, J^{(1)}(t) + 4 m^2 I_{21}^{(2)}(t)\Big\}&&
 \nonumber 
\end{flalign} 
\subsubsection*{graph m} 
\begin{flalign}
A^+_{m} &=
B^+_{m} = 0  & \nonumber \\ 
A^-_{m} &= -\frac{g_A^2\, {m^3}}{F^4}\, (s - u) \, I_A^{(3)}(t)  & \nonumber
\\  
B^-_{m} &= -\frac{g_A^2}{8\, F^4} \Big\{\Delta_\pi - 4\, m^2\,
\left(I^{(1)}(m^2) +  M^2\, {I_A}(t) - 2\, I_A^{(2)}(t)\right) \Big\} &
\nonumber  
\end{flalign}
\subsubsection*{graphs n - v}
\begin{flalign}
A^\pm_{no} &= -\frac{g_A^2\, m}{F^4} M^2 I(m^2) &
B^\pm_{no} &= A^+_{no} \left({\frac{2\, m}{m^2-s}} - {\frac{1}{4\, m}}\right)
\nonumber \\ \nonumber \\ 
A^\pm_{pr} &= \frac{g_A^2 m \Delta_\pi}{2 F^4}  &
B^\pm_{pr} &= A^+_{pr} \left({\frac{2\, m}{m^2-s}} - {\frac{1}{2\, m}}\right)
\nonumber \\ \nonumber \\ 
A^\pm_{s} &= 0  &
B^\pm_{s} &= 0 \nonumber \\ \nonumber \\ 
A^+_{tu} &= -\frac{g_A^2\, m\, M^2\, I(m^2)}{F^4}  &
B^+_{tu} &= 0  \nonumber \\ 
A^-_{tu} &= 0  &
B^-_{tu} &= {\frac{A^+_{tu}}{2\, m}}  \nonumber \\ \nonumber \\ 
A^+_{v} &= 0  &
B^+_{v} &= 0  \nonumber \\ 
A^-_{v} &= 0 &
B^-_{v} &= {\frac{5\, \Delta_\pi}{8\, F^4}}  \nonumber
\end{flalign} 

\subsection[Loop graphs of ${\cal L}^{(2)}$]{Loop graphs of {\boldmath${\cal L}^{(2)}$\unboldmath}}\label{graphsVier}
The loops which, besides vertices from ${\cal L}^{(1)}$, 
also involve one vertex
from ${\cal L}^{(2)}$, start contributing at order $q^4$. As exemplified in
section (\ref{Simplification}), the fact that we
only need the leading order of the expansion of these diagrams simplifies the
calculation considerably. The results are given in terms of the amplitudes
$\{D(s,t)$, $B(s,t)\}$ to order $\{O(q^4)$, $O(q^2)\}$. We define
\begin{equation*}
Q(t)=\frac{(m^2-s)^2 + 2\, m^2\, (t-2\, M^2) }{8\, m^3}
\end{equation*}

\subsubsection*{graphs a+b}
 \begin{flalign} 
D^+_{ab} & =  Q(t)\; B^+_{ab}  & 
B^+_{ab} & =  \frac{8\, g_A^2\,{m^3}}{F^4\, (m^2-s) } \big(c_3 - c_4\, (d-2)
\big)\,  I^{(2)}(s)  & \nonumber \\  
D^-_{ab} & =  D^+_{ab}  &
B^-_{ab} & =  B^+_{ab}  & \nonumber
\end{flalign} 

\subsubsection*{graph f}
 \begin{flalign} 
D^+_{f} & = B^+_{f}= B^-_{f}= 0  & \nonumber \\ 
D^-_{f} & =  \frac{1}{2\, F^4} \Big\{ 4\, c_1\, M^2\, (m^2-s)\,  I(s) - \big[8
c_1\, m^2 \, 
  M^2 - (c_2 + c_3)\,  (m^2-s)^2 \big]\,  I^{(1)}(s) & \nonumber \\
&\cspace - 2\, (c_2 + c_3)\,(m^2-s)\,  (I^{(2)}(s) + m^2 I^{(3)}(s))\Big\} & \nonumber
\end{flalign} 

\subsubsection*{graphs g+h}
 \begin{flalign} 
D^+_{gh} & =  -D^-_{gh} - \frac{8\, g_A^2\, c_4\, (d - 2)\, m^3}{F^4} 
 \,Q(t)\,   I_{B}^{(3)}(s)  & \nonumber \\ 
B^+_{gh} & =  -B^-_{gh} - \frac{8\, c_4\, g_A^2\, m^3}{F^4} \Big\{(d-4)\,
  I_{B}^{(3)}(s)+ 2\,m\,Q(0)\,
  I_{B}^{(5)}(s)\Big\}   & \nonumber \\ & & \nonumber \\ 
D^-_{gh} & =  -{\frac{8\, g_A^2\, m^3}{4\, F^4}} \Big\{ 4\, c_3 \,Q(t)\,
 I_{B}^{(3)}(s)  
  + \,Q(0)\, \Big[-8\, c_1\, M^2\, I_{B}^{(2)}(s)& \nonumber \\
&\cspace + 
  \big(\frac{c_2}{m^2} {{(m^2-s) }^2} + 2\, c_3\, (2 M^2 - t) \big)  
  I_{B}^{(5)}(s) \nonumber \\
&\cspace\cspace- 4\, (c_2 + c_3) \, (m^2-s)\,  I_{B}^{(6)}(s)\Big] \Big\}   
  & \nonumber \\ 
B^-_{gh} & =  \frac{8\, c_3\, g_A^2\, {m^3}\, I_{B}^{(3)}(s)}{F^4}  
  & \nonumber
\end{flalign} 

\subsubsection*{graph k}
 \begin{flalign} 
D^+_{k} & =  {\frac{1}{2\, F^4}} \Big\{(M^2 - 2 t)  \left(4\, c_1\, M^2 +
  c_3\,  (t - 2 
  M^2) \right)\,  J(t) & \nonumber \\
& + 2 \Big(4\, c_1\, M^2 - {\frac{2\, c_2\, M^2}{d}} - c_3\, (M^2 
  + 2\, t) \Big)  \Delta_\pi - 2\, c_2\, t\, (2\, t - M^2)\, J^{(1)}(t) \Big\}
  & \nonumber  
  \\ 
B^+_{k} & = D^-_{k}= 0  & \nonumber \\ 
B^-_{k} & =  \frac{4\, c_4\, m\, t\, J^{(1)}(t)}{F^4}  & \nonumber
\end{flalign} 

\subsubsection*{graph m}
 \begin{flalign} 
D^+_{m} & =  -\frac{3\, g_A^2}{8\, F^4} \Big( c_2\, (s-u)^2 
   -32\, c_1\, m^2\, M^2 + 8\, c_3 \, m^2 (2\, M^2 - t) \Big) \nonumber \\
& \cspace\cspace \times  \Big( (d-1 ) \, I_{A}^{(2)}(t) + t \, 
  I_{A}^{(4)}(t)\Big)   & \nonumber \\ 
B^+_{m} & =  0  \cspace D^-_{m} =  0  & \nonumber \\ 
B^-_{m} & =  \frac{2\, c_4\, g_A^2\, {m^3}}{F^4} \Big\{(d - 5)\,
   I_{A}^{(2)}(t) 
  - t\, I_{A}^{(4)}(t)\Big\}   & \nonumber
\end{flalign} 

\subsubsection*{graphs n+o}
 \begin{flalign} 
D^+_{no} & =  Q(t)\; B^+_{no}  &
B^+_{no} & =  {\frac{8\, g_A^2\, {m^3}}{F^4 
  (m^2-s) }}\big(c_3 - c_4\, ( d-2 )\big ) I^{(2)}(m^2)  & \nonumber \\ 
D^-_{no} & =  D^+_{no}  &
B^-_{no} & =  B^+_{no}  & \nonumber
\end{flalign} 

\subsubsection*{graph s}
 \begin{flalign} 
D^+_{s} & =  Q(t)\; B^+_{s}   & 
B^+_{s} & =  {\frac{6\, g_A^2\, {m^3}\, M^2}{F^4 (m^2-s)^2}} 
  \big({\frac{c_2}{d}} - 2\, c_1 + c_3\big) \, \Delta_\pi  & \nonumber \\ 
D^-_{s} & =  D^+_{s}  & 
B^-_{s} & =  B^+_{s}  & \nonumber
\end{flalign} 

\subsubsection*{graph v}
 \begin{flalign} 
D^+_{v} & =  {\frac{1}{4\, F^4}} \Big\{ -40\, c_1\, M^2 + 2\, c_2\, 
  \big({\frac{4\, M^2}{d}} + \frac{(s-u)^2}{4\,m^2}\big)  + 4\, c_3\, (4\, M^2
  - t) \Big\}   
  \Delta_\pi  & \nonumber \\ 
B^+_{v} & =  0 \cspace
D^-_{v} =  0  \cspace
B^-_{v} =  \frac{2\, c_4\, m}{F^4} \Delta_\pi  & \nonumber 
\end{flalign}

\setcounter{equation}{0}
\section{Renormalization of the effective couplings}
\label{sec:renormalization} The
renormalization of the coupling constants $l_3$ and $l_4$ that occur
in the mesonic part of the effective Lagrangian is discussed in
\cite{Gasser:1984yg}.  These constants enter our amplitude when
expressing the bare quantities $M,F$ in terms of their physical values
$M_\pi,F_\pi$ and in the wave function renormalization of the pion
field. The renormalization of the couplings of ${\cal L}_{\ind
  N}^{(3)}$ is given in \cite{Ecker:1994pi Ecker:1996rk}, where the
HBCHPT formalism is used to extract the divergent part of the one loop
functional.  In infrared regularization, the amplitude is of the form
$M^d f$, where $f$ is a function whose chiral expansion only contains
integer powers of the chiral expansion parameter. Hence the poles at
$d=4$ always appear in the combination $(d-4)^{-1}+\ln M/\mu$. The
same is true for HBCHPT, because the nonrelativistic loop integrals do
not involve the nucleon mass scale. The coefficients of the chiral
logarithms must be the same in the two formulations of the effective
theory, because these are of physical significance and cannot depend
on the regularization scheme.  We conclude that the divergences
encountered in our approach are the same as those occurring in HBCHPT.

As a side remark, we note that in 
dimensional regularization, the
renormalization of the effective coupling constants is different. In that
approach, the infrared regular part of the amplitude is retained. This
part also contains poles at $d=4$, but these are not accompanied by a chiral
logarithm. Instead the pole terms occur in the combination
$(d-4)^{-1}+\ln m/\mu$. Also, some of the divergences contained in the regular
and singular parts cancel -- the
dimensionally regulated amplitude contains chiral logarithms without an
associated pole at $d=4$.  

We define the renormalized coupling constants by
\bea
l_i\al=\al l_i^r(\mu)+\gamma_i\, \lambda\co\hspace{2em}
d_i=d_i^r(\mu)+\frac{ \delta_i}{F^2}\, \lambda\co\hspace{2em} 
e_i=e_i^r(\mu)+\frac{\epsilon_i}{F^2 m}\, \lambda\co\no
\lambda\al=\al\frac{\mu^{d-4}}{(4\pi)^2}\left\{\frac{1}{d-4}-\frac{1}{2}
\left(\rule{0em}{1em}\,\mbox{ln}\, 4\pi +\Gamma'(1)+1\right)\right\}
\fs\nonumber\eea
For the pion couplings $l_3$, $l_4$ the coefficients are
\bea \gamma_3=-\mbox{$\frac{1}{2}$}\co\hspace{2em} 
\gamma_4 = 2\fs\nonumber\eea 
Those relevant for $d_1$, $d_2$, $\ldots$ may be 
taken from 
ref.~\cite{Ecker:1994pi Ecker:1996rk}: 
\begin{flalign*}
  \delta_1&=- \mbox{$\frac{1}{6}$}\,\gAbare^4 & \delta_2 &=
  -\mbox{$\frac{1}{12}$}-\mbox{$\frac{5}{12}$}\,\gAbare^2 & 
\delta_3 &= \mbox{$\frac{1}{2}$} +
  \mbox{$\frac{1}{6}$}\,\gAbare^4 &
  \delta_5 &= \mbox{$\frac{1}{24}$} + \mbox{$\frac{5}{24}$}\,\gAbare^2 \\
  \delta_{14} &= \mbox{$\frac{1}{3}$}\,\gAbare^4& \delta_{15} &=0 & 
\delta_{16} &=
  \mbox{$\frac{1}{2}$}\,\gAbare + \gAbare^3 & \delta_{18} &= 0
\end{flalign*}These expressions differ from those given in
\cite{Fettes:1998ud} because these authors introduce additional
equation of motion terms in the Lagrangian, to arrive at a finite
scattering amplitude also off the mass shell.  

Finally, we list the coefficients occurring in the renormalization of the
coupling constants $e_1$, $e_2$, $\ldots$ that enter the 
amplitude at order $O(q^4)$:
\begin{flalign*}
  \epsilon_1 &=\mbox{$\frac{3}{2}$}\,\gAbare^2 - 
\mbox{$\frac{3}{2}$}\,\left( 8\,c_1 - c_2 -
        4\,c_3 \right) \,m \\
  \epsilon_2 &=- \mbox{$\frac{3}{2}$}\,\gAbare^2 -
  \mbox{$\frac{1}{2}$}\left( c_2 + 6\,c_3 \right) \,m \\
  \epsilon_3 &=-1-3\,\gAbare^2 - \mbox{$\frac{22}{3}$}\,\gAbare^4 - 8\,c_1\,m -
  c_2\,m +
  4\,c_3\,m \\
  \epsilon_4 &=10+ 12\,\gAbare^2 + \mbox{$\frac{52}{3}$}\,\gAbare^4
  + 8\,c_2\,m  \\
  \epsilon_5 &= \mbox{$\frac{1}{2}$} + \mbox{$\frac{7}{2}$}\,\gAbare^2 +
  \mbox{$\frac{13}{3}$}\,\gAbare^4 - 8\,c_1\,m +
  \mbox{$\frac{13}{6}$}\,c_2\,m + 5\,c_3\,m \\
  \epsilon_6 &= -12 - 8\,\gAbare^2 - 8\,\gAbare^4 \\
\epsilon_7 &= -3 - \mbox{$\frac{2}{3}$}\,\gAbare^2 -
 \mbox{$\frac{11}{3}$}\,\gAbare^4 \\ 
\epsilon_8 &= -\gAbare^2 - \mbox{$\frac{1}{3}$}\,\gAbare^4 -
 \mbox{$\frac{1}{3}$}\,\left( c_2 + 6\,c_3 \right) \,m \\ 
\epsilon_9 &=1+ \mbox{$\frac{22}{3}$}\,\gAbare^4 +
 4\,c_4\,m  +
 \gAbare^2\,\left( 13 + 24\,c_4\,m\
   \right)  \\ 
\epsilon_{10} &=- \mbox{$\frac{16}{3}$}\,\gAbare^4 -
 \mbox{$\frac{4}{3}$}\,\gAbare^2\,
    \left( 3 + 8\,c_4\,m \right)  \\ 
\epsilon_{11} &=\mbox{$\frac{1}{6}$}-\mbox{$\frac{7}{6}$}\,\gAbare^2 -
 \mbox{$\frac{2}{3}$}\,\gAbare^4 +
 \mbox{$\frac{2}{3}$}\,c_4\,m 
\end{flalign*}
One readily checks that these renormalizations ensure a finite result for the
quantities $\m$, $\gpiN$, $M_\pi$, $F_\pi$, for which we listed the explicit
expressions in section \ref{nucleon mass and coupling constant}.  The same
holds for the coefficients occurring in the polynomial part of our
representation (see appendix \ref{Subthreshold expansion}), so that the
scattering amplitude approaches a finite limit when the cutoff is removed.

\section{Subthreshold coefficients}
\label{Subthreshold coefficients}
In this appendix, we list the explicit results obtained for the
expansion of the subthreshold coefficients in powers of the quark masses. 
These relations specify the polynomial part of the dispersive representation
(\ref{representation to O4}) in terms of the
effective coupling constants. We express the result in terms of the physical
pion mass. To simplify the formulae,
we set the renormalization scale $\mu$ equal to $M_\pi$ and  
indicate this with a tilde on the renormalized couplings:
\bea \tilde{d}_i\al=\al d_i-\frac{\delta_i}{F^2}\,\lambda_\pi=
d_i^r(\mu)-\frac{\delta_i}{16\,\pi^2\,F^2}\,\ln\frac{\M}{\mu}\co\no
\tilde{e}_i\al=\al e_i-\frac{\epsilon_i}{F^2}\,\lambda_\pi=
e_i^r(\mu)-\frac{\epsilon_i}{16\,\pi^2\,F^2}\,\ln\frac{\M}{\mu}\fs
\nonumber\eea
Note that the chiral logarithms are then contained in the coupling constants.

\subsubsection*{Amplitude {\boldmath$D^+$\unboldmath}}
\begin{multline*} 
d_{00}^+ = - \frac{2\,\left( 2\,c_1 - c_3 \right) \,
       \M^2}{\F^2}  + 
   \frac{\gA^2\,
      \left( 3 + 8\,\gA^2 \right) \,
      \M^3}{64\,\pi \,\F^4} \\ + 
  \M^4\,\left(  \frac{\tilde{e}_{3}}
       {\F^2}  + 
      \frac{8\,c_1\,\tilde{l}_{3}}
       {\F^4}  + 
      \frac{3\,\left( \gA^2 + 
           6\,\gA^4 \right) }{64\,\pi^2\,
         \F^4\,\m}  - 
      \frac{2\,c_1 - c_3}
       {16\,\pi^2\,\F^4}  \right) 
\end{multline*} 
\vspace{0cm}
\begin{multline*} 
d_{10}^+ =  \frac{2\,c_2}{\F^2}  - 
   \frac{\left( 4 + 5\,\gA^4 \right) \,
      \M}{32\,\pi \,\F^4} \\ + 
  \M^2\,\left(  \frac{\tilde{e}_{4}}
       {\F^2}  - 
      \frac{16\,c_1\,c_2}
       {\F^2\,\m}  - 
      \frac{1 + \gA^2}
       {4\,\pi^2\,\F^4\,\m}  - 
      \frac{197\,\gA^4}
       {240\,\pi^2\,\F^4\,\m} \
     \right) 
\end{multline*} 
\vspace{0cm}
\begin{multline*} 
d_{01}^+ = - \frac{c_3}{\F^2}  - 
   \frac{\gA^2\,
      \left( 77 + 48\,\gA^2 \right) \,\M}
      {768\,\pi \,\F^4}  \\ + 
  \M^2\,\left(  \frac{\tilde{e}_{5}}
       {\F^2}  + 
      \frac{52\,c_1 - c_2 - 
         32\,c_3}{192\,\pi^2\,\F^4} 
       -  \frac{\gA^2\,
         \left( 47 + 66\,\gA^2 \right) }{384\,
         \pi^2\,\F^4\,\m}  \right) 
\end{multline*} 
\begin{multline*} 
d_{20}^+ =  \frac{12 + 5\,\gA^4}
    {192\,\F^4\,\M\,\pi }  + 
   \frac{\tilde{e}_{6}}{\F^2}  + 
    \frac{17 + 10\,\gA^2}
     {24\,\pi^2\,\F^4\,\m}  + 
    \frac{173\,\gA^4}
     {280\,\pi^2\,\F^4\,\m} \\ 
\end{multline*} 
\vspace{-1cm}
\begin{multline*} 
d_{11}^+ =  \frac{\gA^4}
    {64\,\F^4\,\M\,\pi }  + 
   \frac{\tilde{e}_{7}}{\F^2}  + 
    \frac{9 + 2\,\gA^2}
     {96\,\pi^2\,\F^4\,\m}  + 
    \frac{67\,\gA^4}
     {240\,\pi^2\,\F^4\,\m} \\ 
\end{multline*} 
\vspace{-1cm}
\begin{multline*} 
d_{02}^+ =  \frac{193\,\gA^2}
    {15360\,\pi \,\F^4\,\M}  + 
   \frac{\tilde{e}_{8}}{\F^2}  - 
   \frac{c_2}
    {8\,\F^2\,\m^2}  + 
   \frac{29\,\gA^2}
    {480\,\pi^2\,\F^4\,\m}  + 
   \frac{\gA^4}
    {64\,\pi^2\,\F^4\,\m} \\  - 
   \frac{19\,c_1}
    {480\,\pi^2\,\F^4}  + 
   \frac{7\,c_2}
    {640\,\pi^2\,\F^4}  + 
   \frac{7\,c_3}{80\,\pi^2\,\F^4}  
\end{multline*} 
\vspace{0cm}
\subsubsection*{Amplitude {\boldmath$D^-$\unboldmath}}
\begin{multline*} 
d_{00}^- =\frac{1}{2\F^2}+\frac{4\,( \tilde{d}_1+\tilde{d}_2 +
      2\,\tilde{d}_{5} ) \, 
      \M^2}{\F^2}  + 
   \frac{\gA^4\,\M^2}
    {48\,\pi^2\,\F^4} \\  - 
  \M^3\,\left( \frac{8 + 
          12\,\gA^2 + 11\,\gA^4}{128\,
          \pi \,\F^4\,\m} - 
      \frac{4\,c_1 + 
         \left( c_3 - c_4 \right) \,
          { \gA^2}}{4\,\pi \,\F^4} \
     \right)
\end{multline*} 
\vspace{0cm}
\begin{multline*} 
d_{10}^- =  \frac{4\,\tilde{d}_{3}}{\F^2}  - 
   \frac{15 + 7\,\gA^4}
    {240\,\pi^2\,\F^4}  + 
    \frac{\M\,\left(168 + 138\,\gA^2 + 
         85\,\gA^4 \right)}{768\,\pi \,\F^4\,
         \m} \\  -
      \frac{ \M\,\left( 8\,c_1 + 8\,c_2 + 
         8\,c_3 + 
         5\,c_3\,\gA^2 - 
         5\,c_4\,\gA^2 \right) }{16\,\pi \,
         \F^4}  
\end{multline*} 
\vspace{0cm}
\begin{multline*} 
d_{01}^- = - \frac{2\,(\tilde{d}_1+\tilde{d}_2)}{\F^2}\, - \, 
   \frac{1 + 7\,\gA^2 + 2\,\gA^4}
    {192\,\pi^2\,\F^4}  \,+\, 
   \frac{\left( 12 + 53\,\gA^2 + 
        24\,\gA^4 \right) \,\M}{384\,\pi\,
      \F^4\,\m } \\ \,-\, 
   \frac{\left( c_3 - c_4 \right) \,
      \gA^2\,\M}{8\,\pi\,
      \F^4 } 
\end{multline*} 
\subsubsection*{Amplitude {\boldmath$B^+$\unboldmath}}
\begin{multline*} 
b_{00}^+ =  \frac{4\,\m\,(\tilde{d}_{14}-\tilde{d}_{15})}
    {\F^2}  - 
   \frac{\gA^4\,\m}
    {8\,\pi^2\,\F^4}  + 
   \frac{\gA^2\,
      \left( 8 + 7\,\gA^2 \right) \,\M}
      {64\,\pi \,\F^4}  - 
   \frac{\left( c_3 - c_4 \right) \,
      \gA^2\,\m\,\M}{2\,
      \pi \,\F^4} \\ 
\end{multline*} 
\vspace{-1cm}
\subsubsection*{Amplitude {\boldmath$B^-$\unboldmath}}
\begin{multline*} 
b_{00}^- =\frac{1}{2\F^2}+ \frac{2\,c_4\,\m}{\F^2}  - 
   \frac{\gA^2\,
      \left( 1 + \gA^2 \right) \,\m\,
      \M}{8\,\pi \,\F^4}  + 
   \frac{\tilde{e}_{9}\,\m\,\M^2}
    {\F^2}  \\ - 
   \frac{\gA^2\,
      \left( 3 + 2\,\gA^2 + 
        9\,c_4\,\m \right) \,
      \M^2}{12\,\pi^2\,\F^4}
\end{multline*} 
\begin{multline*} 
b_{10}^- =  \frac{\gA^4\,\m}
    {32\,\F^4\,\M\,\pi }  + 
   \frac{\tilde{e}_{10}\,\m}
     {\F^2}  + 
    \frac{\gA^2\,
       \left( 25 + 36\,\gA^2 + 
         80\,c_4\,\m \right) }{120\,
       \pi^2\,\F^4} \\ 
\end{multline*} 
\vspace{-1cm}
\begin{multline*} 
b_{01}^- =  \frac{\gA^2\,\m}
    {96\,\F^4\,\M\,\pi }  + 
   \frac{\tilde{e}_{11}\,\m}
     {\F^2}  - 
    \frac{1 - 9\,\gA^2 - 4\,\gA^4 + 
       4\,c_4\,\m}{192\,\pi^2\,
       \F^4} \\ 
\end{multline*} 

\section{{\boldmath$t$\unboldmath}-channel partial waves}
\label{t-channel partial waves}
Only the two $\pi\pi$ partial waves
\begin{align*}
t_0^0&=\frac{2\,t-\M^2}{32\pi\F^2} 
&&\text{and }& t_1^1&=\frac{t-4\M^2}{96\pi\F^2} 
\end{align*}  
enter the $t$-channel unitarity relation (\ref{t-channel unitarity})
at the one loop level.
The imaginary parts of the amplitudes $D$, $B$ can thus be constructed from
the three lowest partial waves $f_+^0$, $f_\pm^1$ of the tree level $\pi N$
amplitude: 
\begin{align*}
\mbox{Im}_t D^+ &=\frac{16\,\pi}{4\m^2-t}\, \im f_+^0 \co &\mbox{Im}_t D^- & =
  \frac{24\,\pi\,\nu}{4\m^2-t} \left\{2\,\m \, \im f_+^1 -
    \frac{t}{2\sqrt{2}}\, \im f_-^1  \right\}\co \\
  \mbox{Im}_t B^+ &=0\co & \mbox{Im}_t B^-& =6\,\pi\,\sqrt{2}\; \im f_-^1\fs
\end{align*}
\noindent
The Born term contribution to the relevant $t$-channel partial waves
reads 
\begin{align*}
{f_B}_+^0(t)&=\frac{g_{\pi N}^2\,\m}{4\pi}\left\{ f(\kappa)-\frac{t}{4\,m^2} 
\right\}\co\\
{f_B}_+^1(t)&=\frac{g_{\pi
    N}^2\,\m}{\pi}\left\{1-\frac{f(\kappa)}{(t-2\,\M^2)\kappa^2}-
\frac{1}{24\,\m^2}\right\}\co \\ 
{f_B}_-^1(t)&=\frac{g_{\pi N}^2}{\sqrt{2}\,\pi}\left\{
  \frac{(1+\kappa^2)\,f(\kappa)-1}{(t-2\,\M^2)\kappa^2}-\frac{1}{12\,\m^2}
\right\}\co
\end{align*}
while the one generated by  ${\cal L}_{\ind N}^{(2)}$ is linear in $t$:
\begin{align*}
f_+^0(t)&=\frac{m^2}{24\,\pi\, F^2}
\!\left\{-24\,M^2\, c_1+(4\,M^2-t)\,c_2 +
6\, (2\,M^2-t)\,c_3 \right\}\!+O(q^3)\co\\
f_+^1(t)&=\frac{m+c_4\,t}{24\pi F^2}
\co\hspace{2cm}f_-^1(t)=\frac{1+4\,\m\,c_4}{12\,\sqrt{2}\,\pi\,\F^2}\fs 
\end{align*}
The function $f(\kappa)$ stands for
\begin{align*}
f(\kappa)&=\frac{\arctan\kappa}{\kappa}\co& \kappa
 &=\frac{\sqrt{(t-4\,\M^2)\,(4\,\m^2-t)}}{t-2\,\M^2}\fs
\end{align*}
Note that the chiral expansion of this function does not converge at
$t=4\M^2$: Since $\kappa$ counts as a quantity of $O(q^{-1})$, the 
$\arctan$ is replaced by its Taylor series in 
inverse powers of $\kappa$. The point 
$t=4\M^2$ corresponds to $\kappa=0$ and is thus outside the radius of 
convergence of that series. The 
infrared singularities occurring in the scalar form factor in the 
vicinity of the $t$-channel threshold are discussed in 
detail in ref.~\cite{Becher Leutwyler 1999} -- the scattering amplitude
exhibits essentially the same structure there.

\section{Imaginary parts of the loop integrals}
\label{Imaginary parts}
In this section we give the exact low energy imaginary part of the loop
integrals defined in Appendix \ref{Loop integrals}. We can distinguish three
different types of singularities:
\begin{itemize}
\item[1.] Singularities in the low energy region,
\item[2.] Physical singularities in the high energy region, 
\item[3.] Unphysical singularities generated by the regularization.
\end{itemize}
The singularities of the first kind are those generated by intermediate states
that involve at most one nucleon. They are identical in dimensional and
infrared regularization. The singularities of type (2) and (3) are absent at
any finite order of the chiral expansion.  An example for type (2) is
the $t$-channel cut due to $\bar{N}N$ intermediate states in the triangle
integral with two nucleon- and one pion-propagator.  This integral develops
an imaginary part for $t>4\,m^2$, which can be obtained from the elastic
unitarity relation in the $t$-channel. In the chiral expansion, the 
corresponding contribution is converted into a Taylor series in $t$. 
In infrared regularization, the loop integrals contain further singularities,
which are not related to physical intermediate states. The
infrared part of the self energy integral $I(s)$, for instance, 
has a cut for $s<0$ and
a pole at $s=0$. Singularities of this type do not occur 
in dimensional regularization. In the following, we only list the
imaginary parts of type (1) -- only these are relevant for the low energy 
structure of the scattering amplitude.

\subsubsection*{1 meson, 1 nucleon}
The dimensionally regularized self energy integral has a right hand cut
starting at $s=(m+M)^2$. Its discontinuity is given by
\begin{align*}
  \im H(s)  & = \frac{\rho(s)}{16\,\pi\, s}\;\theta(s-s_+) \co\;\;\; s_\pm=(m
   \pm M)^2 &\\ 
   \text{and }\rho(s)&=\sqrt{(s-s_+)\,(s-s_-)} = 2 \,m\,
  \sqrt{\omega^2-M^2}\fs 
\end{align*}
The regular part of $H(s)$ has a pole at $s=0$ and a cut for $s<0$. The
imaginary part of $R$ is
\begin{equation*}
 \im R(s) = \frac{\rho(s)}{32\,\pi\, s}\;\theta(-s)
\end{equation*}
The infrared part is given by the difference $I(s)=H(s)-R(s)$ and hence has a
cut on the left as well as on the right. The left hand cut lies outside the
low energy region and is absent to any order of the chiral expansion. Note
that the right hand cut is the same in both regularizations. The imaginary
part of the coefficients of the vector and tensor integrals reads
\begin{align*}
  2\,s\,\im I^{(1)}(s)  &=  \left(s -m^2  + M^2 \right) \,\im I(s) &\\
  12\,s\,\im I^{(2)}(s) &= -\rho(s)^2 \,\im I(s) &\\
  3\,s^2\,\im I^{(3)}(s)&= \left( {\left( s - m^2 + M^2 \right)}^2 - M^2
    \,s \right) \,\im I(s) & 
\end{align*}
\subsubsection*{2 mesons}
\begin{equation*}
  \im J(t) = \frac{1}{16\pi}\,\sqrt{1-\frac{4M^2}{t}}\;\theta(t-4M^2)\co
\end{equation*}
\begin{align*}
  4\,\im J^{(2)}(t) & = \im J(t)\co& 12\,t\,\im J^{(1)}(t) & =  \left(t
  -4\,M^2 \right) \,\im J(t)\fs 
  \end{align*}
\subsubsection*{2 mesons, 1 nucleon}
For $t<4\,m^2$ the imaginary part of the scalar triangle integral is given by
\begin{align*}
  \im I_{21}(t) &= \frac{\theta(t-4
    M^2)}{8\pi\sqrt{\,t\,(4 M^2-t)}}\;\arctan
  \frac{\sqrt{(t-4M^2)(4 m^2-t)}}{t-2M^2}\fs \\
\end{align*}
$\im I_{21}^{(n)}(t)$ can be expressed in terms of $\im
I_{21}(t)$ and $\im J(t)$:
\begin{align*}
  2\,Q^2\,\im I_{21}^{(1)}(t) & = \left( 2\,M^2 - t
  \right)\,\im I_{21}(t) + 2\,\im J(t)\co &\\
  8\,\im I_{21}^{(2)}(t)  & =  \left(4\,M^2-t\right)\,\im I_{21}(t)
  +2\,\left(t -2\,M^2 \right)\,\im I_{21}^{(1)}(t) \co&\\ 
  8\,Q^2\,\im I_{21}^{(3)}(t) & = 6\,\left( 2\,M^2 - t \right) \,\im
  I_{21}^{(1)}(t) +\left(t -4\,M^2 \right)\,\im I_{21}(t)\co&\\
  8\,t\,\im I_{21}^{(4)}(t) & = 2\,\left( 2\,M^2 - t \right)\,\im
  I_{21}^{(1)}(t) +\left(3\,t -4\,M^2 \right)\, \im I_{21}(t)\co&
\end{align*}
where $Q^2=4\,m^2-t$.

\subsubsection*{1 meson, 2 nucleons}
We restrict ourselves to the two special cases
\begin{align*}
 I_A(t)&=I_{12}(m^2,t)\co &
 I_B(s)&=I_{12}(s,M^2)\fs
\end{align*}
The functions $I_A^{(n)}(t)$ are analytic in the low energy region, 
\begin{align*}
\im I_A^{(n)}(t)&=0 & &\mbox{ for }t<4\,m^2 \fs
\end{align*}
For $s>0$ the
absorptive part of $I_B(s)$ is given by
\begin{align*}
\im I_B(s) &=  \frac{\theta(s-s_+)}{16\,\pi\,\rho(s)} \,\ln
 \left\{ \frac{ \left( 2\,m\,\omega -M^2 \right)\,s }{ m^2  \,s- \left( m^2  -
 M^2 
  \right)^2} \right\}& \\
\end{align*}
The imaginary part of the coefficients of the tensorial decomposition can be
obtained from the relations
\begin{multline*}
  2 \,\rho(s)^2\,\im I_{B}^{(1)}(s) = \left( m^2 - M^2
    - s \right)\\ 
\times \left(\left( m^2 - s + 2\,M^2 \right)\, \im I_B(s) +
    \im I(s) \right) \co
\end{multline*} 
\vspace{-0.5cm}
\begin{multline*} 
  2\,\rho(s)^2\,\im I_{B}^{(2)}(s) = \left( m^2 - s
  \right) \,\left( 3\,m^2 + s - 3\,M^2 \right)\,\im
  I_B(s)\\
  + \left( m^2+3\, s - M^2 \right) \,\im I(s) \co
\end{multline*} 
\vspace{-0.5cm}
\begin{multline*} 
  4\,\im I_{B}^{(3)}(s) = 2\,M^2\,\im I_B(s)- \left(2 \,m\, \omega+ 3\,M^2
      \right)\, \im I_{B}^{(1)}(s) \\
  - \left( 2\,m\,\omega+ M^2  \right)\,\im I_{B}^{(2)}(s) \co
\end{multline*}
\vspace{-0.5cm}
\begin{multline*} 
  4 \,\rho(s)^2\,\im I_{B}^{(4)}(s) = 4\,M^2\,\im
  I_{B}^{(3)}(s) - \left( 2\,m\,\omega+ 2\,M^2  \right)\,\im I^{(1)}(s)\\ 
  + 4\,\left( 2\,m^2 \,\omega^2 + M^2
    \,m\,\omega - M^4   \right)\,\im I_{B}^{(1)}(s)  \co
\end{multline*}
\vspace{-0.5cm} 
\begin{multline*} 
  4 \,\rho(s)^2 \,\im I_{B}^{(5)}(s) = -4\,\left( 2\,m^2 + m\,\omega -
M^2 \right) \,\left(  2\,m\,\omega + M^2 \right)\,\im
  I_{B}^{(2)}(s)  \\
  + 4\,\left( 4\,m^2 + 4\,m\,\omega + M^2  \right)\,\im I_{B}^{(3)}(s)+
  2\,\left( 2\,m^2 + 3\,m\,\omega + M^2  \right)\,\im I^{(1)}(s) \co
\end{multline*} 
\vspace{-0.5cm} 
\begin{multline*} 
  \;\; 4\,M^2\,\im I_{B}^{(6)}(s) =  2 \,\left( 
    2\,m\,\omega + M^2  \right)\,\left( \im
    I_{B}^{(1)}(s) - 2\,\im I_{B}^{(4)}(s) \right)-\im I^{(1)}(s) \fs \\
\end{multline*}

\subsubsection*{1 meson, 3 nucleons}
The absorptive part of the box integral $I_{13}(s,t)$ is a function of $t$.
For $t<4\,m^2$, it is given by
\begin{align*}
  \im I_{13}(s,t)&=\frac{1}{4\,\pi\sqrt{4\,\zeta(s)- t\,
      \rho(s)^2}}\,\frac{1}{\sqrt{-t}}\,\arcsinh\left\{\frac{\sqrt{-t}\,
      \rho(s)}{2\, \sqrt{\zeta(s)}}\right\}\co &\\
  \zeta(s)&=\left(2\,m\,\omega-M^2\right)\left(m^2\,s-(m^2-M^2)^2\right)&
\end{align*}
\begin{multline*}
4\,\left( \rho(s)^2  + s\,t \right)\,\im
I_{13}^{(1)}(s,t)  =   \left( m^2  -  M^2  - s \right)\Big\{ 2\,\im
  I_B(s) \\ + \left( 2\,m^2 - 2\,s  +  4\,M^2   - t \right) \,\im
  I_{13}(s,t)\Big\}
\end{multline*}\vspace{-0.5cm} 
\begin{multline*}
  4\,\left( \rho(s)^2 + s\,t \right) \,\im
  I_{13}^{(2)}(s,t) = 2\,\left( m^2 - M^2 + s \right)\,\im I_{B}(s) \\ +
  \left( \left( m^2 - s \right) \,\left( 4\,m^2 - 2\,M^2 - t \right) - M^2
    \,\left( 2\,M^2 - t \right) \right) \,\im I_{13}(s,t)
\end{multline*}
\noindent Note, that the tensor integrals are not singular at
$\rho(s)^2 + s\,t=0$. This can e.g.~be seen by using the relation
\begin{equation*}
\im I_B(s) = \frac{\sqrt{4\,s\,\zeta(\omega ) + 
          \rho(\omega)^4}}{2\,s}\,
   \im I_{13}(s,-\frac{\rho(\omega )^2}{s}) 
\end{equation*}
To the accuracy of our calculation, we need only the first two terms of the
Taylor expansion of these integrals in $t$ (see section \ref{Box graph}).

\section{Low energy expansion of the loop
  integrals}\label{Low energy expansion}
We give the explicit form of the coefficients of the tensor decomposition of
the loop integrals to next-to-leading order. The result is expressed
in terms of the dimensionless variables
\begin{align*}
\Omega&=\frac{s-m^2-M^2}{2\,m\,M}\co & \tau&=\frac{t}{M^2}\co &\text{and } && \alpha=\frac{M}{m}\fs
\end{align*}
It involves the three functions
\begin{align*}
f(\Omega)&=\frac{1}{8\pi^2}\, \sqrt{1-\Omega^2}\, \arccos(-\Omega)\co \\
\bar{J}(\tau)&=J(t)-J(0)=\frac{1}{8 \pi^2}
\left\{1-\sqrt{\frac{4-\tau}{\tau}}\,
\arcsin\,\frac{\sqrt{\tau}}{2}\right\}\co\\
 g(\tau)&= \frac{ 1}{32\,\pi \,{\sqrt{\tau}}} \,\ln  \frac{2\, + 
{\sqrt{\tau}}}{2\, -\sqrt{\tau}} -  \frac{1}{32\,\pi } \,\ln \left\{ 1 +
\frac{\alpha}{{\sqrt{4 - \tau}}}\right\} \\
  &\hspace*{2cm}+\frac{\alpha}{32\,\pi^2} \,\left\{1 +  \frac{\pi}{\sqrt{4  - \tau}}
 +  \frac{2\,\left( 2 - \tau \right)}
{\sqrt{\,\tau\left( 4 - \tau \right) }}\,\arcsin \frac{\sqrt{\tau}}{2}
\right\}\fs 
\end{align*}
The function $f(\Omega)$ is associated with the scalar self energy integral
and $g(\tau)$ is related to the triangle integral with two meson and a single
nucleon propagator. Note that $g(\tau)$ contains arbitrarily high orders of the
expansion parameter $\alpha$. It is constructed in such a way that the
representations for $I_{21}^{(n)}(\tau)$ also cover the region around
$\tau=4$, where the chiral expansion of these integrals breaks down (see
section \ref{Algebraic representation} and ref.~\cite{Becher Leutwyler 1999}).

\subsubsection*{2 mesons:\; $J=I_{20}$}
\begin{multline*}
J(t) = \bar{J}(\tau) - {\frac{1}{16\, \pi^2}} - 2\, \lambda_\pi   \nonumber
\\ \end{multline*}
\vspace{-1cm} 
\begin{multline*}
  J^{(1)}(t) = {\frac{ \tau - 4 }{12\, \tau }}\bar{J}(\tau) -
  {\frac{1}{576\, \pi^2}} +
  ({\frac{1}{\tau }} - {\frac{1}{6}})\,  \lambda_\pi  \nonumber \\
\end{multline*}
\vspace{-1cm} 
\begin{multline*}
  J^{(2)}(t) = {\frac{1}{4}}\bar{J}(\tau) - {\frac{1}{64\, \pi^2}} -
  {\frac{2
      + \tau }{2\, \tau}} \,\lambda_\pi \nonumber \\
\end{multline*}

\subsubsection*{1 meson, 1 nucleon:\; $I=I_{11}$}
\begin{multline*}
  I(s) = -\alpha \,\big(1-2\,\alpha\,\Omega\big)\, f(\Omega) +
  \frac{\alpha\, 
  (\Omega - \alpha )}{16\, \pi^2} - 2\alpha\,\big(\Omega+\alpha-2\, \alpha\,
  \Omega^2\big) \,\lambda_\pi +O(\alpha^3)  \nonumber \\
\end{multline*}
\vspace{-1cm}
\begin{multline*}
  I^{(1)}(s)= -\alpha^2\,\big(\Omega+\alpha - 4\,\alpha\,\Omega^2
  \big)\, 
  f(\Omega) + \frac{{\alpha^2} \Omega^2 \,(1 - 2\, \alpha\, \Omega ) }{16 \,
    \pi^2} 
  \nonumber \\
  + \alpha^2 \,\big(1 - 2\, \Omega^2 - 6\, \alpha\, \Omega + 8\, \alpha\,
  \Omega^3\big) \,\lambda_\pi +O(\alpha^4) \nonumber
\end{multline*}
\begin{multline*}
  I^{(2)}(s) = \frac{m^2\, \alpha^3}{3} \,\big(1 - 4\, \alpha\,
  \Omega \big)\, \big(\Omega^2 - 1\big)\, f(\Omega)+ \frac{m^2\,\alpha^3\,
    \Omega}{144\, \pi^2}\big(6 - 5\, \Omega^2 - 15\, \alpha\, \Omega + 14\,
  \alpha\,  \Omega^3 \big) \nonumber \\
  - {\frac{1}{3}}\, m^2\, {\alpha^3}\,\big(3\, \Omega - 2\, \Omega^3 + 3 \,
  \alpha - 12\, \alpha\, \Omega^2 + 8\, \alpha\, \Omega^4\big) \,\lambda_\pi
  +O(\alpha^5) \nonumber
\end{multline*}
\begin{multline*}
  I^{(3)}(s) =\frac{\alpha^3}{3} \,\big(1 - 4\, \Omega^2 - 12\,
  \alpha \, \Omega + 24\, \alpha\, \Omega^3\big)\, f(\Omega)+ \frac{\alpha^3\,
    \Omega}{72\, \pi^2} \big(7\, \Omega^2
  -3 + 18\, \alpha\, \Omega - 30\, \alpha \, \Omega^3 \big) \nonumber \\
  + \frac{2}{3} \alpha^3 \,\big(3\, \Omega - 4\, \Omega^3 + 3\,\alpha - 24\,
  \alpha\, \Omega^2 + 24\, \alpha\, \Omega^4 \big) \,\lambda_\pi+O(\alpha^5)
  \nonumber
\end{multline*}

\subsubsection*{2 mesons, 1 nucleon}

\begin{multline*}
I_{21}(t) = {\frac{1}{m^2\, \alpha }} \big(g(\tau) + \alpha\,
\lambda_\pi 
\big)+O(\alpha) \\
\end{multline*}
\vspace{-1cm}
\begin{multline*}
I_{21}^{(1)}(t) = {\frac{1}{8\, m^2}} \Big\{\alpha \,(2 - \tau )\,
g(\tau) + 2 
\bar{J}(\tau) - {\frac{1 - \pi\, \alpha }{8\, \pi^2}} - 4\,  \lambda_\pi
\Big\}+O(\alpha^2) \\  
\end{multline*}
\vspace{-1cm}
\begin{multline*}
I_{21}^{(2)}(t) = \frac{\alpha}{16} \Big\{2 \,(4 - \tau )\,  g(\tau) +
\alpha \,(\tau - 2)\, \bar{J}(\tau) + {\frac{4\, \pi + \alpha  \,(\tau - 4)
    }{16 
   \, \pi^2}} \\ + 4\, \alpha \,(4 - \tau ) \,\lambda_\pi \Big\}+O(\alpha^3)
\end{multline*}
\begin{multline*}
  I_{21}^{(3)}(t) = {\frac{1}{64\, m^2}} \Big\{2\, \alpha \,(\tau -
  4)\, g(\tau) + 3\, \alpha^2 \,(2 - \tau )\, \bar{J}(\tau) - \frac{\alpha
    \,(4\, \pi - \alpha\, \tau ) }{16\, \pi^2} \\ + 8\, {\alpha^2} \,(\tau - 4)
  \,\lambda_\pi \Big\}+O(\alpha^3)
\end{multline*}
\begin{multline*}
  I_{21}^{(4)}(t) = \frac{1}{16\, m^2\, \alpha\, \tau } \Big\{2 \,(3\, \tau -
  4)\, g(\tau) + (2 - \tau )\, \alpha\, \bar{J}(\tau) + {\frac{4 \, \pi +
      \alpha \,(8 - \tau ) }{16\, \pi^2}} \\ + 8\, \alpha\, \tau\, \lambda_\pi
  \Big\}+O(\alpha) \nonumber
\end{multline*}

\subsubsection*{1 meson, 2 nucleons}

\begin{multline}
  I_{12}(s,t) = -\frac{1}{4\, m^2\, \Omega^2} \,( 2\,
  \Omega-\alpha - 2\, \alpha\, \Omega^2\big)\, f(\Omega)- \frac{1}{m^2} \big(1
  - \alpha\, \Omega \big) \,\lambda_\pi \nonumber \\+ \frac{1}{64\, m^2\,
    \pi^2 \Omega^2} \big(2\, \pi\, \Omega + 2\, \Omega^2 - \pi \alpha - 2\,
  \alpha\, \Omega + 2\, \alpha\, \Omega^3 \big)+O(\alpha^2) \nonumber
\end{multline} 
\begin{multline}
  I_{12}^{(1)}(s,t) = - \frac{\alpha}{24\, m^2\, \Omega} \big(6\,
  \Omega +
  \alpha - 16\, \alpha\, \Omega^2\big)\, f(\Omega)+\nonumber \\
  {\frac{\alpha}{1152\, m^2\, \pi^2\, \Omega }} \big( 18\, \Omega^2+ 3\, \pi
  \,\alpha - 21\,
  \alpha\, \Omega  - 8\, \alpha\, \Omega^3 \big) \nonumber \\
  - \frac{1}{12\, m^2}\, \alpha \big( 6\, \Omega+9\, \alpha - 16\, \alpha\,
  \Omega^2\big) \,\lambda_\pi+O(\alpha^3) \nonumber
\end{multline}
\begin{multline}
  I_{12}^{(2)}(s,t) = -\frac{\alpha}{12\, m^2\, \Omega^2} \big(2 +
  \Omega^2\big)\, f(\Omega) + \frac{\alpha}{576\, \pi^2\, m^2\, \Omega^2} \big(
  6\, \pi + 12\, \Omega - \Omega^3 \big) \\ -
  \frac{\alpha\, \Omega }{6\, m^2} \,\lambda_\pi  \nonumber+O(\alpha^2) 
\end{multline}
\begin{multline}
  I_{12}^{(3)}(s,t) = \frac{\alpha^2}{12\, \Omega^2}
  \big(\Omega^2 - 1 \big)\, \big(2\, \Omega - \alpha - 6\, \alpha\, \Omega^2
  \big)\, f(\Omega) \nonumber \\ \shoveright{ +\frac{\alpha^2}{576\, \pi^2\,
      \Omega^2} \Big\{ 2\, \Omega \big(3\, \pi + 6\, \Omega - 5\,
    \Omega^3\big) - \alpha \big(3
    \, \pi + 6\, \Omega + 13\, \Omega^3 - 18\, \Omega^5\big) \Big\} } \\
  - \frac{\alpha^2}{6} \big(3 - 2\, \Omega^2 - 8\, \alpha\, \Omega  + 6
  \alpha\, \Omega^3\big) \,\lambda_\pi+O(\alpha^4) \nonumber
\end{multline}

\begin{multline}
  I_{12}^{(4)}(s,t) = \frac{\alpha^2}{48\, m^2\, \Omega^2}
  \Big\{2\, \Omega - 8\, \Omega^3 - \alpha \big(1 + 14\, \Omega^2 - 36\,
  \Omega^4\big) \Big\}\,
  f(\Omega) \nonumber  \\
  \shoveright{- {\frac{\alpha^2}{2304\, m^2\, \pi^2\, \Omega^2}} \Big\{ 2\,
    \Omega \big(3\, \pi + 6\, \Omega - 14\, \Omega^3\big) - \alpha \big(3\,
    \pi + 6\, \Omega + 22\, \Omega^3 - 72 \,
    \Omega^5\big) \Big\}} \nonumber \\
  + \frac{\alpha^2}{12\, m^2} \big(3 - 4\, \Omega^2- 16\, \alpha\, \Omega  +
  18\, \alpha\, \Omega^3\big) \,\lambda_\pi+O(\alpha^4) \nonumber
\end{multline}
\begin{multline}
I_{12}^{(5)}(s,t) = O(\alpha^2) \nonumber \\
\end{multline}
\vspace{-1cm}
\begin{multline}
I_{12}^{(6)}(s,t)= -\frac{\alpha^2}{24\, m^2 \,\Omega} \big(1 + 2\,
 \Omega^2\big)\, f(\Omega) + 
 {\frac{\alpha^2}{1152\, m^2\, \pi^2\, \Omega 
  }} \big(3\, \pi + 6\, \Omega + 4\, \Omega^3
  \big)  \\  - \frac{\alpha^2\, \Omega^2}{6\, m^2} \,\lambda_\pi +O(\alpha^3)
 \nonumber 
\end{multline}
\subsubsection*{1 meson, 3 nucleons}
\begin{multline*}
I_{13}(s,t) =
 \frac{1}{128\, m^4\, \pi^2\, \alpha\, \Omega^3} 
  \Big\{ 2\, \Omega \big(\pi + 2\, \Omega \big)
  - \alpha \big(2 \,\pi + 4\, \Omega - \pi\, \Omega^2 - 4\, \Omega^3\big)
 \Big\}\nonumber\\ 
 -\frac{\Omega -\alpha}{4\, m^4\, \alpha\, \Omega^3}\,  f(\Omega) +O(\alpha)
  \nonumber
\end{multline*}
\begin{multline*}
  I_{13}^{(1)}(s,t) = -\frac{1}{24\, m^4\, \Omega^2} \big(3\,
  \Omega - 2\, \alpha - 4\, \alpha\, \Omega^2\big)\, f(\Omega)-\frac{3 - 4\,
    \alpha\, \Omega }{12 m^4}\, \lambda_\pi \nonumber \\ + \frac{1}{1152\,
    m^4\, \pi^2\, \Omega^2} \Big\{ 9\, \Omega\, \big(\pi + \Omega\big )- 2\,
  \alpha 
  \big(3\, \pi + 6\, \Omega - 5\, \Omega^3\big) \Big\}+O(\alpha^2) \nonumber
\end{multline*}
\begin{multline*}
  I_{13}^{(2)}(s,t) = -\frac{1}{24\, m^4\, \Omega^3}\big(2 +
  \Omega^2\big)\,
  f(\Omega)-\frac{1}{12\, m^4} \, \lambda_\pi \nonumber \\ + \frac{1}{1152\,
  m^4\, \pi^2\, 
  \Omega^3} \big(6\, \pi + 12\, \Omega - \Omega^3 \big)+O(\alpha)  \nonumber
\end{multline*}

\section{Low energy expansion of the amplitude}
The representation of the amplitude given in sections \ref{Dispersive
representation} and \ref{Dispersion integrals} consists of the Born
term, a set of dispersive contributions that account for the cuts in
the $s$-, $t$- and $u$-channels and a polynomial. The dispersive part
is described by the functions $D_1^\pm(s),\ldots, B_2^\pm(t)$, which
can be expressed as integrals over the corresponding imaginary parts,
according to (\ref{disp Ds}) and (\ref{disp Dt}). As discussed in
section \ref{Algebraic representation}, the first few terms of the
chiral expansion of these integrals can be evaluated in closed
form. In the present appendix, we give the corresponding explicit
expressions.

The chiral expansion of the scattering amplitude involves a choice of 
kinematic variables to be kept fixed. 
As discussed in section \ref{Algebraic representation}, the
convergence of the expansion improves considerably if the lab.~energy
$\Omega=\omega/M_\pi$ is replaced by the c.m.~energy $\Omq=\omega_q/M_\pi$.
At the same time, however, the change of variables generates bookkeeping
problems, because polynomials in $\Omq$ do not represent polynomials in
$\nu,t$. For this reason, we work with the
variables $\omega,t$ in the $s$-channel and use $\nu,t$ for the
$t$-channel. The corresponding expansion at fixed $\Omq$ is readily obtained
from the expressions given below: It suffices to express $\omega$ in 
terms of $\Omq$ and to expand the result in powers of $\alpha$ 
at fixed $\Omq$. 

When combined with the chirally expanded polynomial part specified in
appendix \ref{Subthreshold coefficients} the expressions given below
can be used to work out the chiral expansion of the scattering lengths
and effective range parameters, for instance.  As discussed in section
\ref{Algebraic representation}, the expansion of observables that
involve derivatives of the amplitude converges only extremely slowly
because of the strong infrared singularities generated by the $s$- and
$u$-channel cuts. For numerical analysis, we strongly recommend the use
of the dispersive representation specified in sections \ref{Dispersive
representation} and \ref{Dispersion integrals}.

\subsection*{The functions {\boldmath$D^{\pm}_1(s),D^{\pm}_2(s),
B^{\pm}_1(s)$\unboldmath}}
For all of the expanded one-loop graphs that contain a cut in the $s$-channel, 
the discontinuity across the cut is given by a polynomial in $\omega$ and 
$t$ times the factor
$\sqrt{\omega^2-\M^2}/\omega^3$, where $\omega$ is the pion lab
momentum. The factor $\sqrt{\omega^2-\M^2}$ represents the imaginary
part of the elementary function
\bea
f(\omega)=\frac{1}{8\pi^2}\sqrt{1-\frac{\omega^2}{\M^2}} \,
\arccos\left(-\frac{\omega}{\M}\right)\co\nonumber\eea
which does develop the required square root discontinuity:
Approaching the real axis from above, we have
\begin{equation}
f(\omega)=\frac{1}{8\pi^2} \sqrt{\Big|1-\frac{\omega^2}{\M^2}\Big|} \begin{cases}
 -\arccosh(-\frac{\omega}{\M}) & \text{
 for }  \omega < -\M \\
 \phantom{-}\arccos(-\frac{\omega}{\M}) & \text{
 for } -\M< \omega < \M \\
 \phantom{-}\arccosh(\frac{\omega}{\M})-i\pi & \text{
 for } \omega > \M  
\end{cases}  
\nonumber\end{equation}
The imaginary part determines the real part up to a polynomial.
In fact, in the normalization specified in section 
\ref{Dispersion integrals},
the real part is fully determined by the imaginary part, through the condition
that the Taylor series expansion of the functions $D^\pm_1(s)$, $D^\pm_2(s)$, 
$B^\pm_1(s)$ in powers of $\omega$ does not contribute to 
the subthreshold coefficients listed in eq.~(\ref{eq:Tpoly}).
It therefore suffices to list the
contributions proportional to $f(\omega)$, which we denote by 
$\hat{D}^\pm_1(s),\hat{D}^\pm_2(s),\hat{B}^\pm_1(s)$, respectively. 
The explicit expressions read:
\begin{multline*}
\hat{D}_1^+(s)  =  \frac{\M}{12 \F^4 \, \m\,\omega^3} 
  \big\{ \Delta^+_1 + g_A^2 \,\Delta^+_2 + g_A^4 \,\Delta^+_3\big\}  
f(\omega) \\
\end{multline*}
\vspace{-1.6cm}
\begin{flalign} 
\dspace \Delta^+_1 & =  12\, \omega^4 (- \m \,\omega  + 3\, \omega^2  -
  \M^2)  \nonumber \\ 
\dspace \Delta^+_2 & =  24\, \omega^4\, (\omega^2 -
  {\M^2})  & \nonumber \\ 
\dspace \Delta^+_3 & = 8(\omega^2 - \M^2)^2  \left(-\m\,\omega+3\,\omega^2+
\M^2 \right)  & \nonumber
 \end{flalign}
\vspace{-1em}
\begin{multline*}
\hat{D}_2^+(s)  =  \frac{\M \,(\omega^2-\M^2)}{12 \F^4 \, \m\,\omega^3} 
  \big\{2\,g_A^2\,\omega^2  + g_A^4 \,(- 4\, \m\, \omega 
 +5\, \omega^2+ 4\,\M^2)\big\}  f(\omega) 
\end{multline*}
\vspace*{-1.3em}
\begin{multline*}
\hat{B}_1^+(s)  =  \frac{g_A^2\, \M\, (\omega^2 - \M^2)
  }{3\, \F^4\, \omega^3 } \big\{2\, \omega^2+ g_A^2 (-  \m \,\omega+2
  \omega^2  + \M^2)  \\ 
- 8\, (c_3 - c_4)  \m\, \omega^2\big\}
   f(\omega) 
\end{multline*}
\vspace{-1.3cm}
\begin{multline*}
  \hat{D}_1^-(s) = \frac{\M}{12\, \F^4\, \m\, \omega^3}
  \big\{\Delta^-_1 + g_A^2 \,\Delta^-_2 + g_A^4 \,\Delta^-_3 + \Delta^-_4\big\}  f(\omega)\\
\end{multline*}
\vspace{-1.3cm}
\begin{flalign} 
  \dspace \Delta^-_1 & = 6\, \omega^4 (- \m\, \omega +3\, \omega^2 -
  \M^2)  & \nonumber \\
  \dspace \Delta^-_2 & = 6\, \omega^2 (2\,\omega^4-2\,
\omega^2\,\M^2
      +  \M^4 ) & \nonumber \\
  \dspace \Delta^-_3 & = 2\,(- \m \, \omega+3\, \omega^2 + \M^2)\,  (\omega^2 -
  \M^2)^2   & \nonumber \\
  \dspace \Delta^-_4 & = -48\, \m\, \omega^4 \left\{(c_2 + c_3) \omega^2 - 2\,
    c_1\, \M^2\right\}& \nonumber\\
&\hspace{3em}- 16\, g_A^2\, (c_3 -
  c_4)\, \m\, \omega^2\, (\omega^2 - \M^2)^2 &
  \nonumber
\end{flalign}
\vspace{-1em}
\begin{multline*}
  \hat{D}_2^-(s) = \frac{g_A^2\,\M\,(\omega^2-\M^2)}{12\, \F^4\, \m\, \omega^3}
  \big\{3\,\omega^2 + g_A^2 \, (- \m \, \omega+3\, \omega^2 + \M^2)\\ 
-8\,(c_3 -
  c_4)\, \m \omega^2\big\}  f(\omega)
\end{multline*}
\vspace{-1.3cm}
\begin{multline*} 
\hat{B}_1^-(s) =  \frac{g_A^2\, \M\, (\omega^2 - \M^2) }{3\,
  \F^4\, \omega^3 } \big\{ 2\, g_A^2\, (- \m
  \omega+2\, \omega^2  + \M^2)+3\, \omega^2   \\ 
  + 8\, c_4\, \m\, \omega^2\big\}  f(\omega)
\end{multline*}
Note that the expressions contain fictitious singularities at $\omega=0$.
These drop out when evaluating 
the functions $D_1^\pm(s)$, $D_2^\pm(s)$, $B_1^\pm(s)$, which are
obtained by simply removing the leading terms in the expansion in powers
of $\omega$: 
\bea \hat{D}^\pm_1(s)\al=\al\sum_{n=-3}^4 d_{1,n}^\pm\,\omega^n
+D^\pm_1(s) \co\no
\hat{D}^\pm_2(s)\al=\al\sum_{n=-3}^2 d_{2,n}^\pm\,\omega^n
+D^\pm_2(s) \co\no
\hat{B}^\pm_1(s)\al=\al\sum_{n=-3}^2 b_{1,n}^\pm \omega^n
+B^\pm_1(s) \fs\nonumber\eea

\subsection*{The functions {\boldmath$D_3^\pm(t)$, $B_2^\pm(t)$\unboldmath}}

Only the diagrams belonging to the topologies (k) and (l) have a
branch point at $t=4\M^2$. Note that in the case of (k), the
two-pion-vertices from ${\cal L}_{\ind N}^{(1)}$ are relevant as well
as those from ${\cal L}_{\ind N}^{(2)}$, which are proportional to
$c_1$, $c_2$, $c_3$, $c_4$. Explicit representations for the
contributions from these graphs are given in appendix \ref{Loop
graphs}.  Those of type (k) may be expressed in terms of the familiar
loop integral $\bar{J}(t)$. The low energy structure of the vertex
diagrams (l) is discussed in detail in \cite{Becher Leutwyler
1999}. To the order we are considering here, the corresponding
integrals may be expressed in terms of $\bar{J}(t)$ and of the
function $g(t)$ which is the expanded version of the triangle integral
$I_{21}(t)$ with two mesons and one nucleon. In the $t$-channel, we
thus encounter two functions that play the same role as the quantity
$f(\omega)$ used for the $s$-channel.  Again, it suffices to list the
contributions proportional to these functions (the explicit expressions for
$\bar{J}(t)$ and $g(t)$ are given in section \ref{Algebraic
representation}):
\begin{multline} 
\hat{D}_3^+(t) = \frac{\M^2 - 2\, t}{12\, \F^4} \Big\{\Big(24\, c_1\,
\M^2 + c_2\, (t - 4\, \M^2) \\ + 6\, c_3\, (t - 2\, \M^2)+ \frac{3\,
g_A^2\, t}{2\, \m}\Big) \bar{J}(t) + \frac{3\, g_A^2\, (2\,
\M^2-t)}{\M} g(t) \Big\} \nonumber
\end{multline}
\begin{multline} 
B_2^+(t) =  0 \nonumber \\ 
\end{multline}
\vspace{-1cm}
\begin{multline} 
  \hat{D}_3^-(t) = \frac{1}{12\, \F^4}\left\{t\, -\, 4\, \M^2 + g_A^2\,
    (5\, t -
    8\, \M^2)\right\} \bar{J}(t) \\
  - \frac{g_A^2}{8\, \F^4\,\m\, \M} \left(3\, t^2 - 12\, t\, \M^2 +
    8\, \M^4\right)g(t) \nonumber
\end{multline}
\begin{multline} 
\hat{B}_2^-(t) =  \frac{1}{12\, \F^4} \left\{2\, g_A^2\, (5 \M^2-2 t )  +
  (1 + 4\, c_4\, \m)  (t - 4\, \M^2) \right\} \bar{J}(t) \\
 + \frac{g_A^2 \m
  }{2\, \F^4 \M}\left (t - 4\, \M^2 \right) g(t)
  \nonumber
\end{multline}
The functions $D_3^\pm(t)$ and $B_2^-(t)$ are then obtained by
subtracting from the above expressions the first few terms of their
expansion in $t$:
\begin{align}
\hat{D}_3^+(t) &=d_{3,0}^+ + d_{3,1}^+\, t +d_{3,1}^+\, t^2 + D_3^+(t)\no
\hat{D}_3^-(t) &=d_{3,0}^- + d_{3,1}^-\, t + D_3^-(t)\no
\hat{B}_2^-(t) &=b_{2,0}^- + B_2^-(t)\nonumber
\end{align}

\setcounter{equation}{0}
\section{Integral equations for \boldmath{$\pi N$} scattering}
\label{Royappendix}
In section \ref{sec:Roy}, we briefly described a system of integral
equations for the partial waves of $\pi N$ scattering, which is
analogous to the Roy equations for the $\pi\pi$ scattering
amplitude. The present appendix provides the technical details needed
to explicitly work out the kernels that occur in these equations.

Our starting point is the hypothesis that, at low energies, 
the properties of the scattering
amplitude are governed 
by the pole from the Born term and by the unitarity cuts in the $s$-, 
$t$- and $u$-channels and that only the $S$- and $P$-waves generate a 
significant contribution to the latter. In the $s$-channel partial wave
decomposition, the terms from these waves read
\bea\label{eq:PWs} A^\pm\al =\al 
4\pi \left\{\frac{(\sqrt{s}+\m)\,(f^\pm_{0+}+3\,z\,f^\pm_{1+})}{E+\m}-
\frac{(\sqrt{s}-\m)\,(f^\pm_{1-}-f^\pm_{1+})}{E-\m}+\ldots\right\}\co\no
B^\pm\al =\al 
4\pi \left\{\frac{f^\pm_{0+}+3\,z\,f^\pm_{1+}}{E+\m}+
\frac{f^\pm_{1-}-f^\pm_{1+}}{E-\m}+\ldots\right\}\co\eea
where $z=\cos\theta$ is related to the momentum transfer, $t=2q^2(z-1)$. The
variables $q$ and $E$ denote the momentum and the energy of the nucleon 
in the centre of mass system, respectively. 
The analogous terms from the 
lowest $t$-channel partial waves read
\bea\label{eq:PWt} A^+\al=\al \frac{16\pi\,f^0_+}{4\m^2-t}+\ldots\co
\hspace{2em}
A^-=-\frac{24 \pi\, \m\, \nu\,(\sqrt{2}\,\m\,f^1_--2f^1_+)}
{4\m^2-t}+\ldots\,\co\no
B^+\al=\al 0+\ldots\co\hspace{5.1em}
B^-=6\sqrt{2}\,\pi\,f^1_-+
\ldots\,\co\eea
where the dots stand for contributions with $J\geq2$.

This shows that the relevant contributions from the singularities in
the $s$- and $u$-channels are linear in $t$, while those from the
$t$-channel are linear in $\nu$.  The above hypothesis is thus
equivalent to the assumption that the invariant amplitudes $A^\pm$ and
$B^\pm$ may be described in terms of functions of a single variable,
which moreover only have a right hand cut:\footnote{In the present
context, the invariant amplitudes $A^\pm$ are more convenient to work
with than $D^\pm$, because the $P$-waves $f^\pm_{1+}$ give rise to a
quadratic $t$-dependence in $D^\pm$.}  
\bea
\label{eq:AB}\hspace{-3em}A^+\al=\al A_{pv}^++ A_1^+(s)+
A_1^+(u)+t\,A_2^+(s)+ t\,A_2^+(u) +A^+_3(t)+A^+_p\co\no
\hspace{-3em}A^-\al=\al A_{pv}^-+
A_1^-(s)- A_1^-(u)+t\,A_2^-(s)- t\,A_2^-(u) +\nu\,A^-_3(t)+A^-_p\co\\
\hspace{-3em} B^+\al=\al B_{pv}^++
B_1^+(s)- B_1^+(u)+t\,B_2^+(s)- t\,B_2^+(u) +B^+_p\co\no
\hspace{-3em} B^-\al=\al B_{pv}^-+
B_1^-(s)+ B_1^-(u)+t\,B_2^-(s)+ t\,B_2^-(u)+B_3^-(t) +B^-_p\fs
\nonumber\eea
The first terms on the right denote the pseudovector Born contributions, while
$A^\pm_p$ and $B^\pm_p$ stand for polynomials in the Mandelstam 
variables. The imaginary parts of the functions $A_1^+(s),\ldots,B_3^-(t)$
are determined by those of the 9 partial waves occurring in (\ref{eq:PWs}),
(\ref{eq:PWt}): In the notation of eqs.~(\ref{eq:fs}), (\ref{eq:ft}),
the discontinuities of $A^\pm_1(s),A^\pm_2(s),
B^\pm_1(s),B^\pm_2(s)$ are given by
\bea \mbox{Im} A_1^\pm\al=\al 4\pi\,
\frac{( \sqrt{s}+\m)\,(\mbox{Im} f_1^\pm+3\,\mbox{Im} f_3^\pm)}{E+\m}
-4\pi\,\frac{(\sqrt{s}-\m)\,(\mbox{Im} f_2^\pm-\,\mbox{Im} f_3^\pm)}
{E-\m}\co\no
\mbox{Im} A_2^\pm\al=\al 6\pi\,\frac{(\sqrt{s}+\m)\,\mbox{Im} f_3^\pm}
{(E+\m)^2(E-\m)}
\co\\
\mbox{Im}B^\pm_1\al=\al 4\pi\,
\frac{\mbox{Im} f_1^\pm+3\,\mbox{Im} f_3^\pm}{E+\m}+4\pi\
\frac{\mbox{Im} f_2^\pm-\mbox{Im} f_3^\pm}
{E-\m}\co\no
\mbox{Im}B^\pm_2\al=\al 6\pi\,\frac{\mbox{Im} f_3^\pm}{(E+\m)^2(E-\m)}
\co\nonumber\eea
while those of $A^\pm_3(t),B^\pm_3(t)$ are determined by the $t$-channel
partial waves:
\bea 
\mbox{Im}A_3^+\al=\al 16\pi\,\frac{\mbox{Im} f_4^+}{4\m^2-t}\co
\hspace{1em}
\mbox{Im}A_3^-= -24\pi\,\m\,\frac{\sqrt{2}\,\m\, \mbox{Im}f_4^-
-2\, \mbox{Im}f_5^-}{4\m^2-t}\co\no
\mbox{Im}B_3^+\al=\al 0\co \hspace{5.95em}
\mbox{Im}B_3^-= 6\sqrt{2}\,\pi\, \mbox{Im}f_4^-\fs 
\eea

We can now set up the analogon of the Roy equations. In order not to burden
the discussion with problems of technical nature, we first disregard
subtractions. The dispersion relations obeyed by the functions associated with
the $s$-channel cuts then read
\bea G(s)\al=\al\frac{1}{\pi}\int_{(\m+\M)^2}^\infty
\frac{ds'\,\mbox{Im}G(s')}{s'-s-i\epsilon} 
\co\hspace{2em}G=A_1^\pm,A_2^\pm,B_1^\pm,B_2^\pm\co \label{unsubtracted Disp}\eea
while those related to the $t$-channel singularities are of the form
\bea H(t)\al=\al\frac{1}{\pi}\int_{4\M^2}^\infty
\frac{dt'\,\mbox{Im}H(t')}{t'-t-i\epsilon}\co
\hspace{2em}H=A_3^\pm,B_3^\pm \fs\eea The above equations specify the
amplitude in the low energy region, in terms of the imaginary parts of
the 9 relevant partial waves and of the constants occurring in the
polynomial parts $A_p^\pm,B_p^\pm$ of the representation
(\ref{eq:AB}). In particular, the real parts of those partial waves
may be worked out from the $s$- and $t$-channel partial wave
decompositions of that representation. 

To extract the $s$-channel partial waves, we need to perform the integrals 
\bea\label{abeta} \alpha^\pm_\ell(s)=
\int_{-1}^{+1} dz\,A^\pm(s,t)\,P_\ell(z)\co\hspace{1em}
\beta^\pm_\ell(s)=\int_{-1}^{+1}
dz\,B^\pm(s,t)\,P_\ell(z)\co\eea where $t$ is to be expressed
in terms of $s$ and $z$, with $t=2\,q^2(z-1)$.  The partial
waves of interest are then given by 
\bea\label{fs}
f^\pm_1(s)\al=\al\frac{E+\m}{16\pi\sqrt{s}}\left\{
\alpha^\pm_0(s)+(\sqrt{s}-\m)\,\beta^\pm_0(s)\right\}\no \al\al
+\frac{E-\m}{16\pi\sqrt{s}}\left\{
-\alpha^\pm_1(s)+(\sqrt{s}+\m)\,\beta^\pm_1(s) \right\}\co\no
f^\pm_2(s)\al=\al\frac{E+\m}{16\pi\sqrt{s}}\left\{
\alpha^\pm_1(s)+(\sqrt{s}-\m)\,\beta^\pm_1(s)\right\}\\ \al\al
+\frac{E-\m}{16\pi\sqrt{s}}\left\{
-\alpha^\pm_0(s)+(\sqrt{s}+\m)\,\beta^\pm_0(s) \right\}\co\no
f^\pm_3(s)\al=\al\frac{E+\m}{16\pi\sqrt{s}}\left\{
\alpha^\pm_1(s)+(\sqrt{s}-\m)\,\beta^\pm_1(s)\right\}\no \al\al
+\frac{E-\m}{16\pi\sqrt{s}}\left\{
-\alpha^\pm_2(s)+(\sqrt{s}+\m)\,\beta^\pm_2(s) \right\}
\fs\nonumber\eea

In order to extract the
$t$-channel partial waves, we need to expand the amplitude 
in powers of $\nu$ at fixed $t$. Concerning the contribution from the  
$t$-channel cuts, the operation is trivial, because these are linear in
$\nu$. Those arising from the cuts in the $s$-channel represent a 
superposition of terms of the form
\bdm \frac{1}{s'-s}
=\frac{1}{s'-\m^2-\M^2- 2\m\,\nu+\frac{1}{2}\,t}\edm
with $s'\geq(\m+\M)^2$. The relevant angular variable is given by 
\bea Z=\frac{\m\,\nu}{p_-q_-}=-\cos\theta_t\co
\hspace{2em}p_-=\sqrt{\m^2-\mbox{$\frac{1}{4}$}\,t}
\co\hspace{2em}q_-=\sqrt{\M^2-\mbox{$\frac{1}{4}$}\,t}\co\nonumber\eea
so that the partial wave expansion amounts to the series
\bea\label{PWseries} \frac{1}{s'-s}\al=\al \frac{1}{2\,p_-q_-}\,
\sum_{\ell=0}^\infty
(2\ell+1)\,P_\ell(Z)\,Q_\ell(Z')\co\\
Z'\al\equiv\al \frac{2s'-2\m^2-2\M^2+t}{4\,p_-q_-}\fs\nonumber\eea
In the $Z-$plane, the expansion converges in an ellipse that has the 
foci at $Z=\pm1$ and passes through the point $Z=Z'$. 
In the context of our system of
integral equations, we need the $t-$channel partial waves for $t>4\M^2$.
For real values of $Z$ and $t<4\m^2$, the variable $\nu$ is then purely
imaginary and the denominator $s'-s$ is different from zero, in the entire
range of integration over $s'$. Hence the partial wave expansion 
converges and can be inverted without problems. 

In fact, in the case of the amplitude $C$, we may apply the formula
(\ref{tchannelPW}) to convert the dispersive representation into a series of
Legendre polynomials of the form (\ref{tchannelPW}) and read off the
coefficients.  For the amplitude $B$, however, the partial wave
decomposition involves the first derivative of the Legendre polynomials.
The corrolary of (\ref{PWseries}) that is relevant in this case
reads:
\bea\frac{1}{s'-s}= \frac{1}{2\,p_-q_-}\,
\sum_{\ell=1}^\infty P_\ell'(Z)\,\{Q_{\ell-1}(Z')-
Q_{\ell+1}(Z')\}\fs\eea
This then allows us to read off the contributions to 
$t-$channel partial waves that are generated by the $s$- and $u$-channel cuts.

For the specific partial waves of
interest, we may also write down explicit projection formulae analogous to
(\ref{abeta}) and (\ref{fs}):
\bea f_4^+(t) \al=\al \frac{p_-}{8\pi}\int_{-1}^{1}dZ
\left\{p_-\,A^+(\nu,t)+ q_-\, \m\,Z\, B^+(\nu,t)\right\} \co\no
f_4^-(t)\al=\al\frac{\sqrt{2}}{16\pi}\int_{-1}^{1}dZ\,
(1-Z^2)\,B^-(\nu,t)\co\\ f_5^-(t)\al=\al\frac{1}{8\pi
q_-}\int_{-1}^{1}dZ\,Z \left\{ p_-\,A^-(\nu,t)+q_-\,\m\, Z\,
B^-(\nu,t)\right\}\co\nonumber\eea
with $\nu=Z\, p_-q_-/\m$.  
Note that, for $4\M^2<t<4\m^2$, the integral extends over purely imaginary
values of $\nu$. 

Collecting the various pieces, the system of equations takes
the form (\ref{integral equation}): The $S$- and $P$-waves are
expressed as linear superpositions of contributions from the Born
term, from the polynomial and from the imaginary parts of these waves.
The advantage of working in the $t$-channel isospin basis is that the
system does not intertwine the partial waves with even and odd
isospin.  For the $t$-channel partial waves, the unitarity condition
is also diagonal in these variables: \bea\label{eq:unit}
\mbox{Im}f^+_4(t)\al=\al
\{1-4\M^2/t\}^{-\frac{1}{2}}\,t_0^0(t)^{\star}\,f^+_4(t)\co\\
\mbox{Im}f^-_i(t)\al=\al\{1-4\M^2/t\}^{-\frac{1}{2}}\,t_1^1(t)^{\star}\,
f^-_i(t)\co\hspace{2em} i=4,5\nonumber\eea In the $s$-channel,
however, unitarity does connect the two sets of amplitudes. The
relation (\ref{elastic unitarity}) applies to the partial waves of a
given $s$-channel isospin: \bea f_i^{\frac{1}{2}}\al=\al
f^+_i+2f^-_i\co\hspace{2em} f_i^{\frac{3}{2}}=f^+_i-f^-_i
\co\hspace{2em}i=1,2,3\fs\eea Expressed in terms of the amplitudes
$f^\pm_1$, $f^\pm_2$, $f^\pm_3$, elastic unitarity requires
\bea\label{eq:unis} \mbox{Im}f^+_i(s)\al=\al
q\,|f^+_i(s)|^2+2\,q\,|f^-_i(s)|^2\co\\ \mbox{Im}f^-_i(s)\al=\al q\,
f^+_i(s)\,f^-_i(s)^\star+q\,f^-_i(s)\,f^+_i(s)^\star +
q\,|f^-_i(s)|^2\fs\nonumber\eea

Finally, we need to specify the polynomial part of the representation 
(\ref{eq:AB}), which accounts for the contributions
from the higher partial waves. This part is intimately connected with the
issue of subtractions, because the contributions from the subtraction constants
amount to a polynomial. Note also that the decomposition of the amplitude
in eq.~(\ref{eq:AB}) is not unique. We may, for instance, add a constant to
the functions $A_1^-(s),A^-_2(s),B_1^+(s),B_2^+(s)$, 
without changing the scattering amplitude. 

While the imaginary parts of the functions $B_i^\pm(s)$ fall off
sufficiently fast at high energies for the integrals
in eq.~(\ref{unsubtracted Disp}) to converge, the
dispersion relations for the functions $A_i^\pm(s)$ need to be
subtracted at least once.  In order to suppress the
contributions of the higher partial waves we oversubtract and
introduce two subtractions for the amplitudes $A_1^\pm(s)$ and one for
$B_1^\pm(s)$. Since the contributions of $A_2^\pm(s)$ and $B_2^\pm(s)$
are suppressed by a factor of $t$, we use the minimal number of
subtractions for these. Crossing symmetry reduces the number of
subtraction constants by a factor of two, so that the eight
subtractions lead to four subtraction constants. We identify these
with the subthreshold parameters $d_{00}^+$, $d_{10}^+$, $d_{00}^-$,
$b_{00}^-$, so that we can write the polynomial part of the
amplitude as 
\begin{align} A^+_p &= d_{00}^+ + t\,d_{01}^+ \co &
B^+_p &= 0\\ 
A^-_p &= \nu\,(d_{00}^--b_{00}^-)\co &B^-_p &= b_{00}^-\co \nonumber 
\end{align}

Since the amplitude $B^+$ is
odd under crossing, we can perform an additional subtraction in
$B_2^+(s)$, without introducing a new subtraction constant. 
Subtracting at $s_0=\m^2+\M^2$, the dispersion relations in the variable $s$
take the form 
\bea G_n(s)\al\equiv 
\al\frac{(s-s_0)^n}{\pi}\int_{(\m+\M)^2}^\infty\frac{ds'\,\mbox{Im}G(s')}
{(s'-s_0)^n(s'-s)}\co \eea  
with $n=2$ for $A_1^\pm(s)$ and $n=1$ for $A_2^\pm(s), B_1^\pm(s),B_2^+(s)$.  
Only the one for $B_2^-(s)$ remains unsubtracted, $n=0$. The analogous 
dispersion relations in the variable $t$ read
\bea
H_n(t)\al=\al\frac{t^n}{\pi}\int_{4\M^2}^\infty\frac{dt'\,\mbox{Im}H(t')}
{t^{\prime\,n}(t'-t)}\co
\eea with $n=2$ for $A_3^+(t)$ and $n=1$ for
$A_3^-(t),B_3^-(t)$.

This then completes our system of equations. In order to solve it, 
the system must be iterated.
Starting with a given input for $\mbox{Im}f^\pm_n$,
the relations (\ref{integral equation}) determine the corresponding real 
parts. The unitarity 
conditions (\ref {eq:unit}), (\ref{eq:unis}) then yield a new set of
imaginary parts, etc. Note that the outcome is meaningful only
at low energies, whereas the dispersion integrals extend to infinity. Also,
elastic unitarity is valid only at low energies -- we need experimental
information for the behaviour of the imaginary parts of the partial waves
in the inelastic region, $s>(\m+2\M)^2$, $t>16\M^2$.

In the case of $\pi\pi$ scattering, the mathematics of the Roy equations has
been explored in detail and it has been shown that two subtractions suffice to
arrive at a remarkably accurate representation in the elastic region
\cite{ACGL}. For $\pi N$ scattering, the properties of the corresponding
system of equations yet need to be explored. Concerning the behaviour of the
imaginary parts at higher energies, as well as the contributions from
higher partial waves, the phenomenological information is much better
than in the case of $\pi\pi$ scattering. Also, the $\pi\pi$ phase shifts,
which enter the $t$-channel unitarity condition, are now known very 
accurately \cite{CGL}. We are confident that the
proposed system of equations will indeed provide the required link between the
physical region and the Cheng-Dashen point and thus allow an accurate 
determination of the $\sigma$-term.

\end{appendix}

\end{document}